\documentclass{article}[12pt]

\usepackage{amssymb}
\usepackage{amsmath}
\usepackage{enumerate}
\usepackage{latexsym}
\usepackage{mathrsfs}
\usepackage{amsthm}
\usepackage{verbatim}
\usepackage{graphicx}
\usepackage{epstopdf}
\usepackage{epsfig}
\usepackage{color}
\usepackage{subfig}
\usepackage{float}
\usepackage{color}
\usepackage{titlesec}
\usepackage{bm}
\usepackage[colorlinks,linkcolor=blue,anchorcolor=green,citecolor=red]{hyperref}
\usepackage[T1]{fontenc}

\usepackage{charter}
\usepackage{pythonhighlight}
\usepackage{fancyhdr}
\usepackage{amscd}
\usepackage{mathtools}

\usepackage{booktabs}

\ifdefined\showcaptionsetup
 \PassOptionsToPackage{caption=false}{subfig}
\fi
\makeatother
\numberwithin{figure}{section}
\theoremstyle{plain}
\newtheorem{thm}{\protect\theoremname}
\theoremstyle{remark}

\theoremstyle{plain}

\theoremstyle{plain}
\newtheorem{prop}[thm]{\protect\propositionname}
\theoremstyle{plain}
\newtheorem{assum}[thm]{\protect\assumptionname}

\usepackage{caption}
\usepackage{algorithm} 
\usepackage{algorithmic} 

\theoremstyle{definition}

\providecommand{\lemmaname}{Lemma}
\providecommand{\propositionname}{Proposition}
\providecommand{\remarkname}{Remark}
\providecommand{\theoremname}{Theorem}
\providecommand{\assumptionname}{Assumption}

\theoremstyle{remark}
\newtheorem{remark}{Remark}

\makeatletter
\@addtoreset{equation}{section}
\makeatother

\makeatletter 
\@addtoreset{equation}{section}
\makeatother  

\usepackage{geometry}
\usepackage{listings}
\usepackage{setspace}
\definecolor{Code}{rgb}{0,0,0}
\definecolor{Decorators}{rgb}{0.5,0.5,0.5}
\definecolor{Numbers}{rgb}{0.5,0,0}
\definecolor{MatchingBrackets}{rgb}{0.25,0.5,0.5}
\definecolor{Keywords}{rgb}{0,0,1}
\definecolor{self}{rgb}{0,0,0}
\definecolor{Strings}{rgb}{0,0.63,0}
\definecolor{Comments}{rgb}{0,0.63,1}
\definecolor{Backquotes}{rgb}{0,0,0}
\definecolor{Classname}{rgb}{0,0,0}
\definecolor{FunctionName}{rgb}{0,0,0}
\definecolor{Operators}{rgb}{0,0,0}
\definecolor{Background}{rgb}{0.98,0.98,0.98}
\lstdefinelanguage{Python}{
	numbers=left,
	numberstyle=\footnotesize,
	numbersep=1em,
	xleftmargin=1em,
	framextopmargin=2em,
	framexbottommargin=2em,
	showspaces=false,
	showtabs=false,
	showstringspaces=false,
	frame=l,
	tabsize=4,
	basicstyle=\ttfamily\small\setstretch{1},
	backgroundcolor=\color{Background},
	commentstyle=\color{Comments}\slshape,
	stringstyle=\color{Strings},
	morecomment=[s][\color{Strings}]{"""}{"""},
	morecomment=[s][\color{Strings}]{'''}{'''},
	morekeywords={import,from,class,def,for,while,if,is,in,elif,else,not,and,or,print,break,continue,return,True,False,None,access,as,,del,except,exec,finally,global,import,lambda,pass,print,raise,try,assert},
	keywordstyle={\color{Keywords}\bfseries},
	morekeywords={[2]@invariant,pylab,numpy,np,scipy},
	keywordstyle={[2]\color{Decorators}\slshape},
	emph={self},
	emphstyle={\color{self}\slshape},
}

\begin{document}
	
\title{An Asymptotic Approach for Modeling Multiscale Complex Fluids at the Fast Relaxation Limit}
\author{Xuenan Li \thanks{Department of Applied Physics and Applied Mathematics, Columbia University, xl3383@columbia.edu} \qquad Chun Liu \thanks{Department of Applied Mathematics, Illinois Institute of Technology, cliu124@illinoistech.edu} \qquad Di Qi \thanks{Department of Mathematics, Purdue University, qi117@purdue.edu}}
\date{}
\maketitle

\begin{abstract}
We present a new asymptotic strategy for general micro-macro models which analyze complex viscoelastic fluids governed by coupled multiscale dynamics. In such models, the elastic stress appearing in the macroscopic continuum equation is derived from the microscopic kinetic theory, which makes direct numerical simulations computationally expensive. To address this challenge, we introduce a formal asymptotic scheme that expands the density function around an equilibrium distribution, thereby reducing the high computational cost associated with the fully coupled microscopic processes  while still maintaining the dynamic microscopic feedback in explicit expressions.
The proposed asymptotic expansion is based on a detailed physical scaling law which characterizes the multiscale balance at the fast relaxation limit of the microscopic state. An asymptotic closure model for the macroscopic fluid equation is then derived according to the explicit asymptotic density expansion. Furthermore, the resulting closure model preserves the energy-dissipation law inherited from the original fully coupled multiscale system. Numerical experiments are performed to validate the asymptotic density formula and the corresponding flow velocity equations in several micro-macro models. This new asymptotic strategy offers a promising approach for efficient computations of a wide range of multiscale complex fluids.
\end{abstract}

\section{Introduction}\label{sec:intro}

Complex multiscale fluids demonstrating non-Newtonian viscoelastic and rheological phenomena are ubiquitous in nature and engineering, which comprises a wide variety of materials, including polymeric and fiber solutions, liquid crystals, magneto-hydrodynamical fluids, and blood suspensions \cite{bird1987dynamics,doi1988theory,anderson1998diffuse,owens2002computational,giga2017variational,wang2021two}. 
The multiphysics and multiscale phenomena exhibited in the complex fluids can be attributed to the two-way coupling between the microscopic kinetic particles and the macroscopic fluid dynamics \cite{gurtin1982introduction,kroger2000structure,yue2004diffuse}. 

The interacting coupling mechanism can be described by the macroscopic deformation in terms of certain internal microscopic variables through kinematic transport \cite{bird1987dynamics,bird1987dynamics2,doi1988theory}. 
The macroscopic elastic stress thus can be induced by the molecular configuration in the microscopic level. 
Understanding these elastic effects is crucial for accurately modeling the complicated yet essential behaviors exhibited by complex fluids. 
Precise mathematical strategies are still needed to describe and predict the macroscopic behaviors in the multiscale complex fluids with consideration of the microscopic kinetic theory.

The macroscopic motion of the viscoelastic fluids is modeled by the Navier-Stokes equation 
\begin{equation}
	\rho\left(\partial_t u+u\cdot\nabla u\right)+\nabla p=\eta\Delta u+\nabla\cdot \tau,\label{eq:NS}
\end{equation}
with suitable boundary and initial conditions. Here, $u$ is the velocity field, $\rho$ is the mass, $p$ is the hydrostatic pressure, $\eta$ is the fluid viscosity, and $\tau$ is the elastic stress tensor.
The main task in modeling complex fluids from \eqref{eq:NS} is to construct a \emph{constitutive law} that connects the stress tensor $\tau$ and the strain tensor $s=\frac{1}{2}\left(\nabla u+\nabla u^{T}\right)$. In Newtonian fluids, the stress $\tau$ and strain $s$ are related by a linear relation with $\tau=2\eta s$, where $\eta$ is the viscosity constant \cite{bris2009multiscale}. However, the stress in non-Newtonian fluids is related to the flow deformation in order to capture the underlying elastic effects.

There are several macroscopic models in the existing literature which try to directly solve the non-Newtonian stress $\tau$ from the continuum mechanics in a differential constitutive equation  such as the Oldroyd-B models \cite{bird1987dynamics,bris2009multiscale,lions2000global}. The heterogeneous multiscale method \cite{engquist2007heterogeneous,weinan2009general} allows seamless simulation of the macroscopic state using the microscopic process in equilibrium enforcing an extreme scale separation of the macro and micro state evolution. However, directly building the constitutive equation requires quantitative and generalizable formulation of the underlying microscopic physical processes that are usually untractable due to the complex nature of the fluids. Thus, the purely macroscopic description often becomes insufficient to capture the detailed multiscale properties.

An alternative strategy to circumvent the difficulty is to recover the constitutive law based on the explicit microscopic energy law through an \emph{energetic variational approach} introduced in \cite{liu2009introduction,giga2017variational}. The governing equation can be derived according to the competition between the kinetic and free energy together where the energy dissipation comes from the Least Action Principle (LAP) \cite{abraham2008foundations,arnol2013mathematical} and Maximum Dissipation Principle (MDP) \cite{onsager1931reciprocal1,onsager1931reciprocal2}. 
In particular, the free energy $\mathcal{F}$ depends on the full deformation gradient $F\left(X,t\right)$ given by the general form
\begin{equation}
	\mathcal{F}=\int_{\Omega_0} W\left(F\right)\mathrm{d}X, \quad F\left(X,t\right)=\frac{\partial x\left(X,t\right)}{\partial X},\label{eq:free-ene}
\end{equation}
where $x\in\Omega_x\subset\mathbb{R}^d$ and $X\in\Omega_0$ are the corresponding Eulerian and Lagrangian coordinate connected by the flow map $\frac{\partial x}{\partial t}=u\left(x,t\right)$ with initial state $x\left(X,0\right)=X$. The elastic stress  $\tau$ in \eqref{eq:NS} is derived according to the LAP  by taking the variation of the free energy \eqref{eq:free-ene} with respect to $x$ such that
\begin{equation}
	-\frac{\delta}{\delta x}\int_{0}^{T}\mathcal{F}\mathrm{d}t=-\frac{\delta}{\delta x}\left(\int_{0}^{T}\int_{\Omega_{x}}\frac{W\left(\tilde{F}\right)}{\det\tilde{F}}\mathrm{d}x\:\mathrm{d}t\right)=\nabla\cdot\left(\frac{W_{F}\left(\tilde{F}\right)\tilde{F}^{T}}{\det\tilde{F}}\right)\coloneqq\nabla\cdot\tau,\label{eq:stress-deform}
\end{equation}
where $\tilde{F}\left(x\left(X,t\right),t\right)=F\left(X,t\right)$ is the deformation tensor under Eulerian coordinate. 

Since the deformation tensor $F$ carries all information from  the microscopic patterns, the remaining task is to construct the explicit form of the macroscopic potential $W$. For example,  the free energy in Newtonian fluids depends only on the determinant of $F$, that is, $W\left(F\right)=\omega\left(\frac{\rho_0}{\det F}\right)\det F$ \cite{gurtin1982introduction}. In the case of linear isotropic elasticity, the stress tensor in \eqref{eq:stress-deform} gets reduces to $\tau=\frac{FF^T}{\det F}$ following the relation $W_F=F$ \cite{temam2005mathematical}. Still, in order to capture the multiscale coupling mechanisms in the more general situations, it becomes necessary to derive the constitutive laws by taking into account the detailed microscopic structures through the explicit kinetic formulation. The development of effective multiscale strategies then becomes necessary to bridge the microscopic kinetic theory and the macroscopic continuum mechanics.

The micro-macro model  \cite{bird1987dynamics,doi1988theory,jourdain2004variance,jourdain2004existence} emerges as an effective approach by characterizing behaviors in the induced microscopic elastic stress to the macroscopic fluids in terms of the distribution of molecular configurations. 
In this setting, the molecular configuration can be described by the microscopic configuration variable $q\in\mathbb{R}^n$ (such as the end-to-end vector describing the molecular orientation in the elastic dumbell model in \cite{bird1987dynamics2}). An additional Fokker-Planck equation is introduced to describe the evolution of the probability distribution $f\left(x,q,t\right)$ containing microscopic elastic effects. 
In this way, a direct link can be built between the macroscopic stress derived from the free energy dependent on the deformation tensor \eqref{eq:stress-deform} and the kinetic equation representing a wide class of microscopic processes. 

However, the resulting micro-macro system involves a fully coupled multiscale density function in a high dimensional space, thus leads to a computationally demanding task especially when complicated microstructures are involved (even with the more complicated lattice materials with $q\in\mathbb{R}^{n\times n}$ as a tensor field such as  \cite{reese2003meso}). 
A common practice is to use moment closures to approximate the microscopic contributions in the leading-order moments relying on near-Gaussian assumptions and specific microscopic potentials \cite{yu2005micro,hyon2008enhanced,hyon2010some}. Developments of an accurate modeling framework  preserving precise energy conservation structures and efficient computational strategies that are applicable to more general fluids and distribution functions with non-negligible higher-order moments remain a long lasting challenge \cite{hyon2008enhanced,wang2021two,bao2025deterministic}. 

In this paper, we propose a new modeling strategy that enables efficient computation of the general micro-macro equations thus to achieve a better understanding of the detailed properties found in various complex fluids. Our main contributions are summarized as follows: 
\begin{itemize}
    \item We derived an explicit asymptotic formula for the multiscale density function  (Proposition~\ref{prop:asymp_density}) to establish a precise constitutive relation for the multiscale coupling mechanism directly from the macroscopic strain tensor; 
    \item We constructed hierarchical asymptotic closure models for the macroscopic fluid equations (Proposition~\ref{prop:special_solns}) according to the explicit high-order asymptotic expansions;
    \item We showed an associated energy-dissipation law for the asymptotic closure models  (Proposition~\ref{prop:energy-dissip}), preserving key structural properties from the energetic variational principle.
\end{itemize}

The modeling idea starts with tracking the limiting performance of the density dynamics at a fast relaxation limit, where the microscopic density function quickly converges to its equilibrium state satisfying a Gibbs distribution. Then, the asymptotic models can be naturally constructed by introducing higher-order corrections to the equilibrium distribution in a step-by-step fashion based on a careful calibration of the balanced microscopic advection and diffusion effects. This new approach enables efficient prediction of the macroscopic flow field avoiding the most expensive computation in solving the high-dimensional density equations involving multiscale coupling. The precise energy-dissipation equation can be further used to lay out solid theoretical foundation for the asymptotic closure modeling strategy and guarantee the stability in the proposed computational strategy.


The new asymptotic closure models are easy to implement and the order of accuracy can be designed flexibly according to the requirement of the problem. The inheritance of the  energy dissipation law indicates that the proposed approach is energetically stable and enables a systematic analysis for the existence and uniqueness of the solution.
Furthermore, the modeling strategy can be adaptive to a wider class of general multiscale complex fluids, with minimum constraints to the flow structures and elastic potential function. To demonstrate the scope of skills in the new asymptotic closure models, we test the performance in a series of test examples, starting from various density functions with a steady flow field, then a simply coupled fluid system with a dominant potential flow in the leading order, and finally to the fully coupled micro-macro system subject to different microscopic potential structures. The numerical tests give both qualitative and quantitative confirmations of high-order accuracy in the much more efficient closure models. Especially, the results demonstrate robust performance even far away from the limiting near equilibrium regime with small parameters, implying further generalization of the strategy to be applied to many  other general complex fluids.

The outline of the paper is organized as follows. A quick review on the energetic variational principle to derive the micro-macro model is given in Section~\ref{sec:formulation}. The intuition and derivation of the asymptotic formula for the high-dimensional density function follow in Section~\ref{sec:asymp-density} based on the energy dissipation relation of the original equations. The corresponding macroscopic fluid equations under this asymptotic expansion are derived in Section~\ref{sec:macro-closure} together with the associated energy equation. Detailed numerical experiments are performed in Section~\ref{sec:nums}. Finally, we give the concluding discussion in Section~\ref{sec:conclusion} followed by an Appendix showing detailed computations of the related formulas.

\section{Energetic variational formulation of the micro-macro model}\label{sec:formulation}
In this section, we review the energetic variational formulation for multiscale polymeric fluids.  
In addition, we establish a link between the macroscopic potential $W(F)$ in \eqref{eq:stress-deform} and the micromechanical dynamics consisting of elastic particles $q$  subject to a general microscopic potential $U\left(q\right)$. 

\subsection{The micro-macro model from energetic variational principle}\label{sec:model1}
We start with the micro-macro model for simple polymeric fluids as an example which consist of beads joined by springs or elastic rods. 
The (microscopic) molecule configuration can be described by the microscopic coordinate $q\in \Omega_q\subset \mathbb{R}^{n}$ (order parameter) and molecules are distributed in an underlying (macroscopic) incompressible fluid which can be described by a velocity field $u(x,t):\Omega_x \times \mathbb{R}_{+} \rightarrow \mathbb{R}^n$ under the Eulerian coordinates $x \in\Omega_x\subset \mathbb{R}^{n}$. 
The following energy-dissipation law \cite{giga2017variational, wang2021two} is proposed for the micro-macro system of polymeric fluids
\begin{equation}
	\begin{aligned}
		& \frac{\mathrm{d}}{\mathrm{d}t}\int_{\Omega_{x}} \left\{\frac{1}{2}\rho|u|^{2}  + \lambda\int_{\Omega_{q}}\left[\gamma^2f\log f+fU\left(q\right)\right]\mathrm{d}q\right\}\mathrm{d}x \\
		& = -\int_{\Omega_{x}}\left\{2\eta\left\Vert\frac{\nabla u+\nabla u^{T}}{2}\right\Vert^{2}+\lambda\int_{\Omega_{q}} \frac{1}{D}f\left\vert V-\left(\nabla u\right)q\right\vert^{2}  \mathrm{d}q\right\}\mathrm{d}x.
	\end{aligned}\label{eq:energy-dissipation-law}
\end{equation}
Above, the first terms on the left and right hand side of the equation represent the kinetic energy and the energy dissipation due to the macroscopic effect. 
To describe the microscopic processes,  the microscopic velocity $V(x,q,t)=\frac{\partial q}{\partial t}$ is introduced for molecules in the configuration space, and $f(x,q,t)$ is the probability density function (PDF) based on the macroscopic location $x$ and the microscopic molecular orientation $q$ which  satisfies the conservation law (as the multiscale Fokker-Planck equation)
\begin{equation}
	\partial_t f + \nabla_x \cdot (uf) + \nabla_q \cdot (Vf) = 0. \label{eq:conservation-density}
\end{equation}
The model contains three scaling parameters: $\lambda$ represents the ratio between the macroscopic kinetic energy and the microscopic free energy; $\gamma$ characterizes the rate of microscopic diffusion; and $D$ is the inverse of the \emph{Deborah number} \cite{poole2012deborah} that characterizes the time scale ratio between the material relaxation time and the macroscopic observation time.

In \eqref{eq:energy-dissipation-law}, we define the kinetic energy $\mathcal{K}$ and the Helmholtz free energy $\mathcal{F}$, and the energy dissipation $\mathcal{D}_1$ and $\mathcal{D}_2$ at the macro and microscopic level respectively, that is 
\begin{equation}
	\begin{aligned}
		&\mathcal{K} = \int_{\Omega_{x}}\frac{1}{2}\rho|u|^{2} \: \mathrm{d}x, \qquad\qquad\quad \mathcal{F} = \lambda\int_{\Omega_q\times\Omega_x} \gamma^2f\log f + Uf \: \mathrm{d}q \mathrm{d}x.\\
		&\mathcal{D}_1 = \eta\int_{\Omega_{x}}\left\Vert\frac{\nabla u+\nabla u^{T}}{2}\right\Vert^{2}\:\mathrm{d}x, \quad \mathcal{D}_2 = \frac{\lambda}{2D}\int_{\Omega_{q}\times\Omega_x} f\left\vert V-\left(\nabla u\right)q\right\vert^{2}  \mathrm{d}q\mathrm{d}x.
	\end{aligned}\label{eq:def-energy}
\end{equation}
The micro-macro equations can be derived based on LAP and MDP according to the above energy \eqref{eq:def-energy} in macro- and micro-scales respectively. 
The macroscopic momentum equation for $u$ and the microscopic velocity $V= (\nabla u)q - D\nabla_q\left(\gamma^2 \log f +U\right)$ can be found by performing variational principle on the macroscopic and microscopic state as
\[
\frac{\delta \int_0^{T}\left(\mathcal{K}-\mathcal{F}\right)\mathrm{d}t }{\delta x} = \frac{\delta \left(\mathcal{D}_1+\mathcal{D}_2\right)}{\delta u},\qquad -\frac{\delta\int_0^{T}\mathcal{F}\mathrm{d}t}{\delta q} = \frac{\delta \mathcal{D}_2}{\delta V}.
\]
The equation for the probability density $f$ can be reached by combining the result with the conservation law of density \eqref{eq:conservation-density}.
Therefore, the dynamics of this micro-macro model for polymeric fluids corresponding to the energy-dissipation law \eqref{eq:energy-dissipation-law} is given by the following PDE system
\addtocounter{equation}{0}\begin{subequations}\label{eq:micro-macro}
	\begin{align}
		&\rho(\partial_t u + u\cdot \nabla u)  = -\nabla p +\eta \Delta u + \lambda \nabla_x \cdot \Bigg(\int_{\Omega_q} f \nabla_q U \otimes q \: \mathrm{d}q - \gamma^2N(x,t)I\Bigg), \label{eq:macro-u}\\
		&\partial_t f + \nabla_x \cdot (uf)  = - \nabla_q \cdot \left((\nabla u q)f\right) + D\nabla_q\cdot \big(\gamma^2 \nabla_q f  +  f\nabla_q U\big), \label{eq:micro-f}\\
		&\partial_t \rho + \nabla \cdot\left(u\rho\right) = 0, \label{eq:macro-F}
	\end{align}
\end{subequations}
where $p(x,t)$ is the pressure and $N(x,t) = \int_{\Omega_q} f(x,q,t)\mathrm{d}q$ is the macroscopic particle density, $I$ is the identity matrix, and the tensor product is the matrix defined as $(a\otimes b)_{ij}=a_i b_j$. 
Especially, the micro-macro model \eqref{eq:micro-macro} can accept different potential function $U$ characterizing the microscopic structure. Typical examples include the spring potential from the Hookean law, $U\left(q\right)=\frac{1}{2}H\left|q\right|^{2}$ with $H$ the elastic constant; and the finite-extensible-nonlinear-elastic (FENE) potential, $U\left(q\right)=-\frac{1}{2}Hq_{0}^{2}\log\left(1-\left|q\right|^{2}/q_{0}^{2}\right)$, which provides a more practical formulation with detailed analysis and  model closure strategies  \cite{bird1987dynamics2,yu2005micro,hyon2008maximum,hyon2010some}. 
In addition, a direct generation to lattice materials (such as reticulated polymers in \cite{gao1998numerical,reese2003meso}) can be made based on the energetic variational formulation of the micro-macro model \eqref{eq:micro-macro} (see the discussions in Section~\ref{sec:conclusion}). 

Still, solving the coupled micro-macro equations numerically is a demanding task, where the most computationally expensive part comes from resolving the density function $f\left(x,q,t\right)$ at both physical points $x$ and configuration space of $q$. Efficient computational strategies for a general potential function applicable to a wider class of viscoelastic fluids without an exact macroscopic constitutive law is needed and requires a better understanding of particular dynamical structures in the micro-macro model \eqref{eq:micro-macro}.

\subsection{A connection to the deformation tensor in the macroscopic model}
The energy-dissipation law \eqref{eq:energy-dissipation-law} establishes a direct link between the microscopic process and the macroscopic deformation. 
The Cauchy-Born rule \cite{lin2007micro} indicates that the microscopic configuration is transported with the flow on the macroscopic level, that is, $q=FQ$ where $F$ is the deformation tensor in \eqref{eq:free-ene} and $Q$ is the Lagrangian coordinate in the configuration space. Thus the microscopic velocity can be found from direct computation, $V=\frac{\mathrm{d}}{\mathrm{d}t}(FQ)=(\nabla u)q$. For simplicity, we consider the incompressible flow with $\det F=1$. The free energy in \eqref{eq:def-energy} can be written under the Lagrangian coordinate as
    \begin{equation}
		\mathcal{F}=\lambda\int_{\Omega_{0}}\left[\int\gamma^{2}f_0\log f_0+U\left(FQ\right)f_0\:\mathrm{d}Q\right]\mathrm{d}X,\label{eq:ene-lagrangian}
	\end{equation}
where $f_0\left(X,Q\right)=f\left(x(X,0),q(X,Q,0),0\right)$ is the initial probability distribution function. The same macroscopic equation with the elastic stress $\tau$ in \eqref{eq:macro-u1} can be derived by taking the variation of the above free energy with respect to $x$ \cite{wang2021two}. 
Comparing \eqref{eq:ene-lagrangian} to the free energy defined in \eqref{eq:free-ene}, we find the explicit expression for the macroscopic potential according to the micro-macro model
	\begin{equation}
		W\left(F\right)=\lambda\int U\left(FQ\right)f_{0}\left(X,Q\right)\:\mathrm{d}Q.\label{eq:macro_potential}
	\end{equation}
Then, consistent elastic stress in \eqref{eq:macro-u1} can be found directly from the definition in \eqref{eq:stress-deform} using the explicit form of $W$ in \eqref{eq:macro_potential} and the chain rule under the Eulerian coordinate
	\[
	\begin{aligned}\tau_{ij}= & \left(W_{F}\left(\tilde{F}\right)\tilde{F}^{T}\right)_{ij} 
		=  \lambda\int\sum_{m,l}\frac{\partial U}{\partial q_{m}}\left(\tilde{F}Q\right)\frac{\partial\left(\tilde{F}Q\right)_{m}}{\partial F_{il}}\tilde{F}_{jl}f_{0}\mathrm{d}Q\\
		= & \lambda\int\sum_{l}\frac{\partial U}{\partial q_{i}}\left(\tilde{F}Q\right)Q_{l}\tilde{F}_{jl}f_{0}\mathrm{d}Q 
		=  \lambda\int\left(\nabla_{q}U\left(\tilde{F}Q\right)\otimes\left(\tilde{F}Q\right)\right)_{ij}f_{0}\mathrm{d}Q\\
		= & \lambda\int\left(\nabla_{q}U\left(q\right)\otimes q\right)_{ij}f\left(x,q,t\right)\mathrm{d}q.
	\end{aligned}
	\]
The last equality uses the change of coordinate between Eulerian and Lagrangian coordinates and the relation in the incompressible case $f\left(X,Q,0\right)=f\left(x(X,t),q(X,Q,t),t\right)$.

\begin{remark}
	In the case of incompressible flow, the mass conservation law \eqref{eq:macro-F} get reduced to $\nabla\cdot u=0$ and the density becomes a constant $\rho=\rho_0$. 
	The last term in the stress becomes a gradient, $\nabla\cdot(NI)=\nabla N$, thus can be absorbed by the pressure. In particular, Integrating the  equation \eqref{eq:conservation-density} about $q$ gives
	\[
	\partial_{t}N+u\cdot\nabla N=0.
	\]
	This implies that $N(x,t)=\mathrm{const.}$ from a uniform initial distribution. Therefore, the momentum equation \eqref{eq:macro-u} gets simplified to
	\begin{equation}
		\partial_t u + u\cdot \nabla u +\nabla p =  \eta \Delta u + \lambda \nabla \cdot\tau,\quad \tau=\int_{\Omega_q} f \left(\nabla_q U\right) \otimes q \: \mathrm{d}q. \label{eq:macro-u1}
	\end{equation}
\end{remark}

\section{Asymptotic solution of the probability density at the fast relaxation limit}\label{sec:asymp-density}
To develop efficient computational strategies for the coupled macro-micro
model, we propose an asymptotic approximation for the joint probability density function $f(x,q,t)$ defined in the coupled state and configuration space. In particular, we consider a mild separation of scales around the limit where the microscopic kinetic process possesses a fast relaxation time returning to the equilibrium described by the invariant Gibbs distribution. Below, we first demonstrate a parameter scaling relation revealing a physical interpretation of the detailed restoring balance between the dominant micro- and macro-scale coupling effects at the corresponding approximation limit.  Then, an explicit asymptotic formula for the multiscale density function is derived in terms of the equilibrium distribution.

\subsection{Physical scaling law with detailed micro and macro deformation balance}\label{subsec:scaling_params}
To first gain some intuition on the macro- and micro-scale coupling mechanism with the scaling parameters, we focus on the fluctuation-dissipation balance in the microscales according to the energy $\mathcal{F}$ and $\mathcal{D}_2$ in \eqref{eq:def-energy}, that is
\begin{equation}
	\frac{\mathrm{d}}{\mathrm{d}t} \int_{} \left(\gamma^2f\log f + Uf\right) \: \mathrm{d}q \mathrm{d}x = -\frac{1}{2D}\int_{} f\left\vert V-\left(\nabla u\right)q\right\vert^{2}  \mathrm{d}q\mathrm{d}x.\label{eq:FDT_spring}
\end{equation}
The corresponding multiscale Fokker-Planck equation for the probability density $f$ with no macroscopic velocity $u=0$ then satisfies 
\begin{equation}
	\begin{aligned}
		\partial_t f &+\nabla_q\cdot(Vf) = 0, \\
		V\left(x,q,t\right) &= (\nabla u)q - D\nabla_q\left(\gamma^2 \log f(x,q,t) +U(q)\right),
	\end{aligned}\label{eq:micro_transport}
\end{equation}
where the microscopic velocity $V$ is recovered by applying the energetic variational principle according to \eqref{eq:FDT_spring}. 
In \eqref{eq:micro_transport}, the first term $(\nabla u)q$ in the micro-velocity $V$ accounts for macroscopic flow deformation at the microscopic scale, while the rest part of $V$ represents the microscopic molecular dynamics which includes  advection due to the microscopic potential $U$ and the molecular diffusion effect with strength $\gamma$. The parameter $D$ then controls the balance between these two competing effects. The equation for $f$ thus describes the combined effects between the macroscopic flow and microscopic dynamics in the microscopic level of the density function.
This motivates us to conduct a detailed study of the limiting performance considering the balanced contributions from the macro and micro scales.

We then focus on the microscopic motion described by \eqref{eq:micro_transport} which can be formulated by the stochastic differential equation (SDE) based on the following microscopic configuration $Q$
\[
\mathrm{d}Q=\left(\nabla uQ-D\nabla_{q}U\right)\mathrm{d}t+\sqrt{2D}\gamma\mathrm{d}B_{t}.\label{eq:micro_sde}
\]
Above, $D^{-1}=\mathrm{De}$ can be viewed as the
\emph{Deborah number} (also known as the \emph{Weissenberg number} in \cite{bris2009multiscale}) that characterizes the ratio of time scales
between the material elastic relaxation time and the macroscopic flow observation
time. As  the Deborah number $D^{-1}\rightarrow0$,
it implies that the microscopic elastic structure relaxes to its equilibrium
distribution in a much faster rate compared to the deformation due to the observed macroscopic
motion $\nabla u$. According to the microscopic velocity in \eqref{eq:micro_transport}, the evolution of the probability equation is dominated by the term $\nabla_q\left(\gamma^2 \log f +U(q)\right)$ as $D\rightarrow\infty$.  This implies that an 
equilibrium Gibbs distribution will be quickly reached in the microscopic configuration of the density distribution 
\begin{equation}
f_{\mathrm{eq}}(q)=\frac{1}{Z_\mathrm{eq}}\exp\left(-U(q)/\gamma^2\right).\label{eq:equili_gibbs}
\end{equation}
Therefore, the parameter can be interpreted as $\gamma^2=k_{B}T$ where $k_B$ is the Boltzmann constant and $T$ is the temperature \cite{lin2007micro,yu2005micro}.

Based on the multiscale coupling dynamics, we can introduce the deformation effects due to the macroscopic flow to the full density function $f$ as an additional perturbation on top of the microscopic equilibrium distribution $f_{\mathrm{eq}}(q)$. In addition, we take another limit $\gamma\rightarrow 0$ around temperature zero, that is, the microscopic equilibrium $f_{\mathrm{eq}}$ will concentrate near the local minima of $U(q)$. Therefore, the following leading-order expansion of the probability density function can be introduced
\begin{equation}
f  = \overline{f}(x,q,t)f_{\mathrm{eq}}= \left[f_0(x,t)+\gamma^2f_1(x,q,t)\right]\exp(-U(q)/\gamma^2),\label{eq:quasi-seamless}
\end{equation}
where the leading term $f_0$ is only dependent on the macro state $(x,t)$. This leads to the simplified microscopic velocity \eqref{eq:micro_transport} where the  fluctuation-dissipation balance in the first order between $U$ and $f_{\mathrm{eq}}$ is filtered as
\[
V = (\nabla u)q - D\gamma^2\nabla_q \log\overline{f} =(\nabla u)q - D\gamma^4(\nabla_q f_1)/\overline{f}.
\]
Notice that on the right hand side of the above identity, the additional factor $\gamma^2$ is introduced due to the ansatz for the leading-order distribution near equilibrium $f\sim f_0(x,t)f_{\mathrm{eq}}(q)$. Matching the terms on the right hand side due to the macroscopic flow deformation and the microscopic restoration, we find that the leading-order balance due to the microscopic advection $\nabla_q\cdot(Vf)$ in \eqref{eq:micro_transport} using the above velocity becomes
\[
\nabla_q\cdot\left[f_0(\nabla uq) e^{-U/\gamma^2}\right] \sim D\gamma^4\nabla_q\cdot\left[(\nabla_q f_1)e^{-U/\gamma^2}\right].
\]
Therefore, we introduce the \emph{micro-macro deformation balance ratio} $c$ according to the above multiscale balance relation
\begin{equation}
	c = D\gamma^4=\frac{k_{B}^2T^2}{\mathrm{De}}.\label{eq:balance_param}
\end{equation}
Under this scaling relation $D\gamma^4=O(1)$ as the parameters $\gamma\rightarrow0$ and $D\rightarrow\infty$, it indicates that the strong restoring force generated by the macroscopic flow deformation from the microscopic equilibrium (due to large $D$ as fast relaxation) is balanced by the quick relaxation to the locally stable microscopic state (due to small $\gamma^2$ as low temperature). This limit region illustrates an interesting scenario when the system approaches to the near equilibrium state in a fast rate with a microscopic configuration dominated by its extrema. 
According to this detailed scaling law \eqref{eq:balance_param}, we are able to derive an explicit asymptotic representation of the density function.
Notice that the proposed density expansion \eqref{eq:quasi-seamless} can be viewed as a `quasi-seamless' approximation that adds a detailed microscopic correction $f_1$ to the seamless multiscale method in \cite{weinan2009general}.

\subsection{Explicit asymptotic expansion for the multiscale density function}
Based on the parameter scaling law \eqref{eq:balance_param}, we consider the performance of the probability density equation \eqref{eq:micro-f} near the  limit regime of the micro-macro model 
\begin{equation}
	\partial_t f + \nabla_x \cdot (uf) + \nabla_q \cdot (\nabla u qf) = D \nabla_{q}\cdot \left[f \nabla_q
	\left(\gamma^2\log f+U\right)\right],\label{eq:asymp_den_spring}
\end{equation}
where the model parameters satisfy the scaling $c=O(1)$ in \eqref{eq:balance_param} with the Deborah number $D \rightarrow \infty$ and $\gamma \rightarrow 0$ and $D\gamma^4 = c$. According to the micro- and macro-scale balance discussed in Section~\ref{subsec:scaling_params}, we seek the asymptotic expansion of the density function according to the following ansatz in terms of orders of the scaling parameter $\gamma^2$
\begin{equation}
	f(x,q,t) = \overline{f}(x,q,t) e^{-\frac{U(q)}{\gamma^2}} = \left[ f_0(x,t) + \gamma^2 f_1(x,q,t)+ \gamma^4 f_2(x,q,t)+O\left(\gamma^6\right)\right] e^{-\frac{U(q)}{\gamma^2}}.\label{eq:ansatz-spring}
\end{equation}
Above,  the coefficient function $\overline{f}(x,q,t)=f_0+\gamma^2\overline{f}_1$ is separated from the original  full density function.  We assume that macroscopic density $f_0(x,t)$ varies near the equilibrium microscopic distribution $f_{\mathrm{eq}}\sim \exp{(-\frac{U(q)}{\gamma^2})}$ in the leading order, and $\overline{f}_1(x,q,t)=f_1+\gamma^2 f_2 +O(\gamma^4)$ includes the fluctuation corrections from all the higher-order terms.

In order to precisely characterize the balance in the density equation at each order,  we  introduce the following assumptions that add constraints on either the macroscopic flow velocity field $u$ or the microscopic potential structure $U$.
\begin{assum}\label{assump}
	Assume that one of the following two conditions is satisfied in the micro-macro model \eqref{eq:micro-macro}:
	\begin{description}
		\item[i) ]  The macroscopic velocity field $u$ satisfies the Helmholtz decomposition that is dominated by a potential flow at the leading order
		\begin{equation}
			u = \tilde{u}+\gamma^4v,\quad \tilde{u}=\nabla \phi,\label{eq:flow-potential}
		\end{equation}
		where $v$ is a rotational vortical flow as a high-order perturbation to the potential flow $\tilde{u}$.\\
        
		\item[ii) ]  The microscopic potential function $U$ satisfies the following structural property
		\begin{equation}
            \nabla_q U\left(q\right)=\Psi(q)q,\label{eq:micro-potential1}
		\end{equation}
		where $\Psi$ is a scalar function such that $\nabla \Psi$ is parallel to configuration vector $q$.
	\end{description}
\end{assum}

\begin{remark}
	The two cases in Assumption \ref{assump} are both representative situations in fluid fields. \eqref{eq:flow-potential} represents a dominant potential flow field with the rotational flow emerged as a high-order term that works for all potential function $U$. The particular scale $\gamma^4$ in the vortical flow will be revealed more clearly in the leading-order macroscopic flow model \eqref{eq:closure1} in Proposition~\ref{prop:special_solns}.  \eqref{eq:micro-potential} includes a wide variety of radial symmetric microscopic models  including the Hookean spring model and the FENE dumbbell model. 
	In particular,  the constraint on $U$ is satisfied if the potential is radially symmetric  
	\begin{equation}U\left(q\right)=\tilde{U}\left(\left|q\right|\right),\label{eq:micro-potential}
	\end{equation}
	where $\tilde{U}$ is a scalar function only dependent on the length $\left|q\right|$ of the microstate vector. The radially symmetric potential \eqref{eq:micro-potential} guarantees the general structural assumption \eqref{eq:micro-potential1} with $\nabla_q U=\frac{\tilde{U}^{\prime}\left(|q|\right)}{|q|} q$.
\end{remark}

Under the above assumption, we propose to find the asymptotic solution of the probability density equation \eqref{eq:ansatz-spring} according to the following recurrent relation
\begin{equation}
	\nabla_q U \cdot (\nabla u q)f_{n} = c\nabla_q U \cdot \nabla f_{n+1},\quad n=0,1,2,\cdots\label{eq:f_order}
\end{equation}
The difficulty in solving the above equation \eqref{eq:f_order} is that it involves the general potential function $U$. Thus we introduce the following necessary conditions to solve the easier problems according to the assumptions in \eqref{eq:flow-potential} and \eqref{eq:micro-potential} respectively
\begin{equation}
	\mathrm{i)}\;\left(\nabla\tilde{u}q\right)f_{n}=c\nabla_{q}f_{n+1},\qquad \mathrm{ii)}\; \left(q^{T}\nabla u q\right)f_{n}=cq^{T}\nabla_{q}f_{n+1}.\label{eq:f_order1}
\end{equation}
Notice that the above relation i) implicitly requires $\tilde{u}$ to be a potential flow, which is satisfied by the velocity assumption \eqref{eq:flow-potential} in the leading order. Without the potential flow constraints, the potential constraint \eqref{eq:micro-potential1} enables an easier projection on $q$ as in relation ii). 
Importantly, both situations in \eqref{eq:f_order1} lead to the explicit solutions   
\begin{equation}
f_n=\frac{f_0}{\left(2c\right)^{n}n!}\left(q^{T}\nabla uq\right)^{n},\label{eq:recurr_soln}
\end{equation}
for $n=1,2,\cdots$ by solving the equations  recurrently starting from the leading-order solution of $f_0$. 

Using the above results satisfying \eqref{eq:f_order1}, we are able to compute the asymptotic expansions in \eqref{eq:asymp_den_spring} recurrently according to each order terms. We now provide the effective equations obtained at different orders:

\noindent\textbf{At order $O(\gamma^{-2})$:} The leading order term gives the identity
\[
\nabla_q U \cdot (\nabla u q)f_{0} = c\nabla_q U \cdot \nabla f_{1}.
\]
in \eqref{eq:f_order} with $n=0$, which leads to the explicit relation between $f_0$ and $f_1$ from \eqref{eq:recurr_soln};\\

\noindent\textbf{At order $O(\gamma^0)$:} The next order term gives the transport equation for $f_0$
\[
\partial_{t}f_{0} +\nabla\cdot\left(uf_{0}\right)  =0,
\]
together with the identity 
\[
\nabla_q U \cdot (\nabla u q)f_{1} = c\nabla_q U \cdot \nabla f_{2}.
\]
in \eqref{eq:f_order} with $n=1$ to give the explicit solution of $f_2$ from $f_1$;\\

\noindent \textbf{At order $O(\gamma^2)$:} the higher-order term adds to next order solution for $f_3$ in \eqref{eq:f_order} with $n=2$. In addition, there will be a dynamical equation for $f_1$ that gives an additional correction for $f_3$ (see \eqref{eq:asymp-f2} in Appendix~\ref{subsec:asymp-deri}). \\

Summarizing the above results, we get the following proposition by combining the asymptotic representation \eqref{eq:ansatz-spring} and the explicit hierarchical solutions of \eqref{eq:f_order}. 
\begin{prop}\label{prop:asymp_density}
	If either of the conditions in Assumption \ref{assump} is satisfied, the probability density function in the following expansion up to $O\left(\gamma^{6}\right)$
	\begin{equation}
		\begin{aligned}f\left(x,q,t\right)= & C_{\gamma}\left[1+\frac{\gamma^{2}}{2c}q^{T}\nabla u\left(x,t\right)q+\frac{\gamma^{4}}{8c^{2}}\left[q^{T}\nabla u\left(x,t\right)q\right]^{2}+\right.\\
			& \qquad\left.+\frac{\gamma^{6}}{48c^{3}}\left[q^{T}\nabla u\left(x,t\right)q\right]^{3}\right]f_{0}\left(x,t\right)e^{-\frac{U\left(q\right)}{\gamma^{2}}},
		\end{aligned}
		\label{eq:density_asympt}
	\end{equation}
	gives an asymptotic solution to the Fokker-Planck equation \eqref{eq:asymp_den_spring} at the limit $\gamma\rightarrow 0$, $D\rightarrow\infty$ and $\gamma^4 D=c$, with $C_{\gamma}$ the normalizing constant. In addition, the leading-order macroscopic density function $f_0$ can be solved by the conservation equation
	\begin{equation}
		\partial_{t}f_{0} +\nabla\cdot\left(uf_{0}\right)  =0.\label{eq:dyn_density}
	\end{equation}
\end{prop}

Detailed proof of this proposition can be found in Appendix~\ref{subsec:asymp-deri} including additional correction equations from higher-order asymptotic expansions. It is found that the explicit expressions of $f_n$ in the density function expansion can be computed recurrently by comparing each order terms in the density equation \eqref{eq:asymp_den_spring} according to the asymptotic expansion \eqref{eq:ansatz-spring} and the relations from \eqref{eq:f_order}.

\section{Macroscopic closure models with the asymptotic microscopic feedback}\label{sec:macro-closure}
Next, we consider closure strategies for the macroscopic flow equation \eqref{eq:macro-u} in the coupled micro-macro model without solving the high-dimensional density $f$
\begin{equation}
	\rho(\partial_t u + u\cdot \nabla u) +\nabla p = \eta \Delta u + \lambda \nabla \cdot \left[\int_{\Omega_q} f \nabla_q U \otimes q \: dq - \gamma^2N(x,t)I\right].\label{eq:flow_spring}
\end{equation}
Directly following the asymptotic representation of the multiscale density function \eqref{eq:density_asympt}, we can build an explicit link between the macroscopic elastic stress $\tau$ and the flow gradient $\nabla u$ taking into account information around the equilibrium microscopic distribution.

\subsection{Explicit macroscopic equation with asymptotic microstress feedback}
The asymptotic solution of the density function \eqref{eq:density_asympt} provides an efficient way to compute the explicit microscopic feedback terms depending on the structure of microscopic potential $U$.
Correspondingly, we can represent the elastic stress $\tau$ also in the leading-order asymptotic expansion as
\begin{equation}
	\tau	= \frac{f_0}{m_0}\left[\gamma^2\left(\gamma^{-2}G_0-m_0I\right)+\frac{\gamma^4}{2c}\left(\gamma^{-2}G_1- m_1I\right)+\frac{\gamma^6}{8c^2}\left(\gamma^{-2}G_2- m_2I\right)+O\left(\gamma^8\right)\right].\label{eq:stress_asymp}
\end{equation}
In the above expression, we introduce a sequence of microscopic moments 
\begin{equation}
	\begin{aligned}
		&m_{0}=\int_{\Omega_{q}}e^{-U/\gamma^2}\mathrm{d}q, \quad &G_{0}=\int_{\Omega_{q}}\left(\nabla_{q}U\otimes q\right)e^{-U/\gamma^2}\mathrm{d}q, \\
		&m_{1}=\int_{\Omega_{q}}\left(q^{T}\nabla uq\right)e^{-U/\gamma^2}\mathrm{d}q, \quad &G_{1}=\int_{\Omega_{q}}\left(\nabla_{q}U\otimes q\right)\left(q^{T}\nabla uq\right)e^{-U/\gamma^2}\mathrm{d}q, \\
		&m_{2}=\int_{\Omega_{q}}\left(q^{T}\nabla uq\right)^{2}e^{-U/\gamma^2}\mathrm{d}q,  \quad &G_{2}=\int_{\Omega_{q}}\left(\nabla_{q}U\otimes q\right)\left(q^{T}\nabla uq\right)^{2}e^{-U/\gamma^2}\mathrm{d}q.
	\end{aligned}\label{eq:moments}
\end{equation}
These moment terms are found by directly substituting the density expansion \eqref{eq:density_asympt} into the stress expression on the right hand side of \eqref{eq:flow_spring}: $G_0, G_1, G_2$ are matrices from the first term in the stress, $\int f \nabla_q U \otimes q \: dq$; $m_0, m_1, m_2$ are scalars from the pressure term, $\gamma^2N(x,t)I$. Especially, $m_0$ serves as the normalization factor due to the non-normalized equilibrium density $\exp(-U(\gamma)/\gamma^2)$, so that the leading-order macroscopic density $f_0\sim O(1)$. 
In addition, notice that these moment terms contains the scale factor $\gamma$ due to the integration with respect to the equilibrium density. In fact, using integration by parts, we find that the leading-order term in \eqref{eq:stress_asymp}  gets canceled, so that $G_0 =\gamma^2 m_0I$, that is 
\[
\begin{aligned}
\int_{\Omega_{q}}\left(\nabla_{q}U\otimes q\right)e^{-U/\gamma^{2}}\mathrm{d}q&=\gamma^{2}\int_{\Omega_{q}}\nabla_{q}e^{-U/\gamma^{2}}\otimes\left(-q\right)\mathrm{d}q\\
&=\gamma^{2}\int_{\Omega_{q}}e^{-U/\gamma^{2}}\otimes\nabla_{q}q\mathrm{d}q=\gamma^{2}\left(\int_{\Omega_{q}}e^{-U/\gamma^{2}}\mathrm{d}q\right)I.
\end{aligned}
\]
This identity is indicated by the leading-order balance from the scaling relation \eqref{eq:balance_param}, where the macroscopic flow deformation of the density is balanced by the microscopic restoration at the equilibrium.  In a similar way, the higher-order terms for $n=1, 2\cdots$ get simplified  as
\[
\begin{aligned}G_{n}= & \int_{\Omega_{q}}e^{-U/\gamma^{2}}\left(\nabla_{q}U\otimes q\right)\left(q^{T}\nabla uq\right)^{n}\mathrm{d}q
	= \gamma^{2}\int_{\Omega_{q}}\nabla_{q}e^{-U/\gamma^{2}}\otimes\left(-q\right)\left(q^{T}\nabla uq\right)^{n}\mathrm{d}q\\
	=&\gamma^{2}\int_{\Omega_{q}}e^{-U/\gamma^{2}}\nabla_{q}\left[\left(q^{T}\nabla uq\right)^{n}\right]\otimes q\:\mathrm{d}q+\gamma^{2}\int_{\Omega_{q}}e^{-U/\gamma^{2}}\left(q^{T}\nabla uq\right)^{n}I\:\mathrm{d}q\\
	= & \gamma^{2}n\int_{\Omega_{q}}e^{-U/\gamma^{2}}\left(q^{T}\nabla uq\right)^{n-1}\left(\nabla u+\nabla u^{T}\right)q\otimes q\mathrm\:{d}q+\gamma^{2}m_{n}I.
\end{aligned}
\]
Therefore, we find the asymptotic expansion for the elastic stress $\tau$ in the leading order as
\begin{equation}
	\lambda\nabla\cdot\tau=\frac{\lambda\gamma^{4}}{c m_0}\nabla\cdot\left[f_0\int_{\Omega_{q}}\left(\frac{\nabla u+\nabla u^{T}}{2}\right)q\otimes qe^{-U\left(q\right)/\gamma^{2}}\mathrm{d}q\right]+O\left(\lambda\gamma^{6}\right).\label{eq:stress_asymp_leading}
\end{equation}

This leading-order expression \eqref{eq:stress_asymp_leading} for the microscopic stress feedback to the macroscopic flow provides clear guidance for the scaling relation in the multiscale coupling parameter depending on $\lambda\gamma^{4}$. This relation can be also implied from the microscopic dissipation energy $\mathcal{D}_2$ in \eqref{eq:def-energy}, where the coefficient gives $\frac{\lambda}{D}=c\lambda \gamma^4$ combining with \eqref{eq:balance_param}. According to the two situations in Assumption~\ref{assump}: if we have $\lambda=O(1)$, the leading-order potential flow structure is maintained and the vorticity emerges at the scale $O(\gamma^4)$ consistent with the decomposition in \eqref{eq:flow-potential}; alternatively, if we introduce the multiscaling coupling scaling $\tilde{\lambda}=\lambda\gamma^{4}=O(1)$, the vorticity flow will be induced at the leading order from the microscopic stress that follows the general flow structure with potential function in \eqref{eq:micro-potential1}. We summarize the resulting special solutions due to the leading-order closure models in the following proposition.
\begin{prop}\label{prop:special_solns}
	If the normalized second moments according to the microscopic potential satisfy that
	\begin{equation}
		\tilde{M}_{2}=\frac{1}{m_{0}}\int_{\Omega_{q}}q\otimes qe^{-U\left(q\right)/\gamma^{2}}\mathrm{d}q = O\left(1\right),\quad as\;\gamma\rightarrow 0.\label{eq:2mom}
	\end{equation}
	Explicit macroscopic closure models for the incompressible flow solution $u$ can be developed according to the two conditions in Assumption~\ref{assump} and the corresponding scales in  the multiscale coupling parameter $\lambda$. In details, we have the leading-order equations as $\gamma\rightarrow 0$
	\begin{description}
		\item[i) ] The macroscopic flow decomposition $u=\tilde{u}+\gamma^4 v$ in \eqref{eq:flow-potential} can be maintained during the flow evolution for $\lambda\lesssim O(1)$. The leading-order steady potential flow $\tilde{u}\left(x\right)=\nabla \phi$ can be determined by  $\Delta \phi=0$ with its boundary condition, while the fluctuation vortical flow will remain in the small amplitude not proceeding $O(\gamma^4)$ which can be solved by the closure equation
		\begin{equation}
			\partial_{t}v+\tilde{u}\cdot\nabla v+v\cdot\nabla\tilde{u}+\nabla p=\eta\Delta v+\frac{\lambda}{c}\nabla\cdot \left(f_{0} \nabla\tilde{u} \tilde{M}_2\right).\label{eq:closure1}
		\end{equation}
		\item[ii) ] Vortical flow $u$ will be induced from the microscopic potential function \eqref{eq:micro-potential} satisfying the following closure equation for $\lambda\lesssim O(\gamma^{-4})$ and $\tilde{\lambda}=\lambda\gamma^{4}$
		\begin{equation}
			\partial_{t}u+u\cdot\nabla u+\nabla p=\eta\Delta u+\frac{\tilde{\lambda}}{c}\nabla\cdot\left[f_{0} \left(\frac{\nabla u+\nabla u^{T}}{2}\right)\tilde{M}_2\right].\label{eq:closure2}
		\end{equation}
	\end{description}
	The above equations \eqref{eq:closure1} and \eqref{eq:closure2} are completed with the macroscopic density equation \eqref{eq:dyn_density} for $f_0$ and the incompressibility condition $\nabla\cdot u=0$.
\end{prop}

\begin{remark}
	The closure models \eqref{eq:closure1} and \eqref{eq:closure2} characterize the limiting performance of the macroscopic flow from the microscopic coupling as $\gamma\rightarrow 0$. If the leading-order contribution in the second moments in \eqref{eq:2mom} becomes $O(\gamma^2)$ (such as the case $q_{*}=0$ in \eqref{eq:tensor_asympt}), similar closure equations can be derived by bringing up the next order terms in the elastic stress \eqref{eq:stress_asymp}. 
	In practical numerical simulations with finite $\gamma$, higher-order correction terms in \eqref{eq:stress_asymp} can be added to both closure models for improved accuracy. It is demonstrated that the higher-order corrections can effectively improve the model accuracy and extend the applications to a wide range of parameter values up to $\gamma\sim 1$. Detailed numerical simulations for model \eqref{eq:closure1} are tested in Section~\ref{subsec:flow-test1}, and numerical tests for model \eqref{eq:closure2} are shown in Section~\ref{subsec:flow-test2}.
\end{remark}

\noindent \textbf{Example: quadratic microscopic potential energy $U$.}\\
We illustrate the closure strategy using a simple example of the spring model with a quadratic potential 
\begin{equation}
	U\left(q\right)=\frac{1}{2}\left(q-q_{*}\right)^{T}A^{-1}\left(q-q_{*}\right), \label{eq:quad_potential}
\end{equation}
which is centered at the point $q^*\in\mathbb{R}^2$ and coefficient $A^T=A$. With the microscopic potential in this special case, we are able to integrate the microscopic moments in \eqref{eq:moments} explicitly. The normalization factor can be first found through integration with the normal distribution with $\tilde{q}=q-q_{*}$
\[
m_0=\int e^{-\frac{1}{2}\tilde{q}^{T}\left(\gamma^{2}A\right)^{-1}\tilde{q}}\mathrm{d}q=2\pi\left(\det A\right)^{1/2}\gamma^2.
\]
The higher-order moments moments in \eqref{eq:moments} can be then computed using the second and fourth moments of the normal distribution. 
Therefore, the microscopic tensor stress to the macroscopic flow for the spring model \eqref{eq:quad_potential} can be found in the following leading-order expansion with the scaling factor $c=1$
\begin{equation}
	\begin{aligned}\tau & =\gamma^{4}f_{0}\left(\frac{\nabla u+\nabla u^{T}}{2}\right)q_{*}\otimes q_{*}
		+\frac{\gamma^{6}}{2}f_{0}\left[\left(\nabla u+\nabla u^{T}\right)A+\left(\frac{\nabla u+\nabla u^{T}}{2}\right)q_{*}\otimes q_{*}\left(q_{*}^{T}\nabla uq^{*}\right)\right]\\
		& +\frac{\gamma^{8}}{4}f_{0}\left[\left(\nabla u+\nabla u^{T}\right)q_{*}\otimes q_{*}\left(\nabla u:A\right)+\left(\nabla u+\nabla u^{T}\right)A\left(q_{*}^{T}\nabla uq^{*}\right)\right.\\
		& +\left.\left(\nabla u+\nabla u^{T}\right)q_{*}\otimes q_{*}\left(\nabla u+\nabla u^{T}\right)A+\left(\nabla u+\nabla u^{T}\right)A\left(\nabla u+\nabla u^{T}\right)q_{*}\otimes q_{*}\right]
		+O\left(\gamma^{10}\right),
	\end{aligned}
	\label{eq:tensor_asympt}
\end{equation}
where we define $A:B=\sum a_{ij}b_{ij}$. In this way, a direct relation between the stress $\tau$ and the strain tensor $\frac{1}{2}(\nabla u+\nabla u^{T})$ is established.

\begin{remark}
	In the general potential function $U$ with a local minimum at $q=q_{*}$, $U$ can be expanded around the point $q_*$ such that 
	\[
	U\left(q\right)=U\left(q_{*}\right)+\frac{1}{2}\left(q-q_{*}\right)^{T}A^{-1}\left(q_{*}\right)\left(q-q_{*}\right)+O\left(\left|q-q_{*}\right|^{3}\right),
	\]
	where $A^{-1}=\nabla_{q}^{2}U\left(q_{*}\right)$. A direct application of the Laplace method \cite{wong2001asymptotic} provides the estimate of the integrals in \eqref{eq:moments} for the domain enclosing $q_{*}$ as $\gamma\rightarrow 0$
	\begin{equation}
		\int g\left(q\right)e^{-U\left(q\right)/\gamma^{2}}\mathrm{d}q\cong e^{-\gamma^{-2}U\left(q_{*}\right)}\int g\left(q\right)e^{-\frac{1}{2}\left(q-q^{*}\right)^{T}\left(\gamma^{2}A\right)^{-1}\left(q-q^{*}\right)}\mathrm{d}q.\label{eq:laplace}
	\end{equation}
	In this way, the explicit formula \eqref{eq:tensor_asympt} together with \eqref{eq:laplace} can be used for further efficient computation in the general closure models such as the double well potential tested in Section~\ref{sec:nums} as well as the more computationally demanding lattice potential case in \eqref{eq:pde_incompr} of Section~\ref{sec:conclusion}.
\end{remark}

\subsection{Structure-preserving energy equation for the asymptotic closure models}
One distinctive advantage of the new asymptotic closure models \eqref{eq:closure1} and \eqref{eq:closure2} is that the new asymptotic closure models preserves a corresponding  energy dissipation law under the asymptotic expansion of the density function \eqref{eq:density_asympt}. 
Using the expression \eqref{eq:micro_transport} for the microscopic velocity $v$, the total energy equation \eqref{eq:energy-dissipation-law} can be written as
\begin{equation}
	\begin{aligned}
		& \frac{\mathrm{d}}{\mathrm{d}t}\int_{\Omega_{x}} \left\{\frac{1}{2}\rho|u|^{2}  + \lambda\int_{\Omega_{q}}\left[\gamma^2\log f+U\left(q\right)\right]f\left(x,q,t\right)\mathrm{d}q\right\}\mathrm{d}x \\
		& = -\int_{\Omega_{x}}\left\{2\eta\left\Vert\nabla u\right\Vert^{2}+\lambda D\int_{\Omega_{q}} \left\vert \nabla_q\left(\gamma^2\log f+U\left(q\right)\right)\right\vert^{2}f\left(x,q,t\right)  \mathrm{d}q\right\}\mathrm{d}x.
	\end{aligned}\label{eq:energy-asymp}
\end{equation}
We focus on contributions from the microscopic coupling terms in the energy equation using the asymptotic approximation of the density function
\begin{equation}
	f(x,q,t) = \overline{f}(x,q,t) e^{-\frac{U(q)}{\gamma^2}} \approx \left[ f_0(x,t) + \gamma^2 f_1(x,q,t)+ \gamma^4 f_2(x,q,t)\right] \frac{e^{-U(q)/\gamma^2}}{m_0},\label{eq:den-asymp}
\end{equation}
where $m_0=\int_{\Omega_{q}}e^{-U/\gamma^{2}}\mathrm{d}q$ is the normalization constant. 
Substituting \eqref{eq:den-asymp} into \eqref{eq:energy-asymp} with the explicit solutions  yields the following expressions for the microscopic free energy and dissipation energy
\begin{align}
	\tilde{\mathcal{F}}= & \int_{\Omega_{q}}\left[\gamma^{2}\log f+U\left(q\right)\right]f\mathrm{d}q\\
    = &\frac{\gamma^{4}}{2c}\left(f_{0}\log\frac{f_{0}}{m_{0}}\right)F_{1}+\frac{\gamma^{8}}{8c^{2}}f_{0}\left(1+\log\frac{f_{0}}{m_{0}}\right)F_{2}+O\left(\gamma^8\right),\label{eq:energy-free-asymp}\\
	\tilde{\mathcal{D}}_{2}= & D\int_{\Omega_{q}}\left\vert \nabla_{q}\left(\gamma^{2}\log f+U(q)\right)\right\vert ^{2}f\mathrm{d}q\\
    = &\frac{\gamma^{4}}{c}f_{0}D_{1}+\frac{\gamma^{6}}{2c^{2}}f_{0}D_{2}+O\left(\gamma^{8}\right).
	\label{eq:energy-dissip-asymp}
\end{align}
Above, we define the following normalized moments with respect to the equilibrium microscopic distribution
\begin{equation}
	\begin{aligned}F_{1}\left(x,t\right)= & \sum_{i,j}\frac{\partial u_{i}}{\partial x_{j}}\tilde{M}_{2}^{ij},\qquad\\
		F_{2}\left(x,t\right)= & \sum_{i,j,k,l}\frac{\partial u_{i}}{\partial x_{j}}\frac{\partial u_{k}}{\partial x_{l}}\tilde{M}_{4}^{ijkl},\\
        D_{1}\left(x,t\right)= & \sum_{m,i,j}\left(\frac{\partial u_{m}}{\partial x_{i}}+\frac{\partial u_{i}}{\partial x_{m}}\right)\left(\frac{\partial u_{m}}{\partial x_{j}}+\frac{\partial u_{j}}{\partial x_{m}}\right)\tilde{M}_{2}^{ij},\\
        D_{2}\left(x,t\right)= & \sum_{m,i,j,k,l}\left(\frac{\partial u_{m}}{\partial x_{i}}+\frac{\partial u_{i}}{\partial x_{m}}\right)\left(\frac{\partial u_{m}}{\partial x_{j}}+\frac{\partial u_{j}}{\partial x_{m}}\right)\frac{\partial u_{k}}{\partial x_{l}}\tilde{M}_{4}^{ijkl},
	\end{aligned}
	\label{eq:closure_mom}
\end{equation}
where  $\tilde{M}_{2}^{ij}=\frac{1}{m_{0}}\int q_{i}q_{j}e^{-U/\gamma^{2}}\mathrm{d}q$ and $\tilde{M}_{4}^{ijkl}=\frac{1}{m_{0}}\int q_{i}q_{j}q_{k}q_{l}e^{-U/\gamma^{2}}\mathrm{d}q$ are the normalized second- and fourth-order moments with respect to the microscopic equilibrium distribution. Notice that $\tilde{M}_2$ and $\tilde{M}_4$ may still contain higher-order terms of $\gamma$ due to the integration with the factor $\exp(-U/\gamma^2)$.

We summarize the resulting energy equation in the leading-order asymptotic approximation from the above derivation in the following result. Detailed derivations of the energy principle as well as the equations recovered from the variational approach can be found in Appendix~\ref{subsec:ene-variation}.
\begin{prop}\label{prop:energy-dissip}
	Assume that the probability density function can be written in the asymptotic expansion as in \eqref{eq:density_asympt} and the second-order moments have the order one contribution in \eqref{eq:2mom}.  Then, the closure model \eqref{eq:closure2} from the leading-order representation as $\gamma\rightarrow0$ satisfies the following total energy equation given the scaling relation $\lambda\gamma^4=\tilde{\lambda}$ and $\gamma^4D=c$ in the micro-macro coupling coefficients 
	\begin{equation}
		\begin{aligned}
			\frac{\mathrm{d}}{\mathrm{d}t} &\int_{\Omega_{x}}\left\{ \frac{1}{2}\rho|u|^{2}+\frac{\tilde{\lambda}}{2c} \left(f_{0}\log \frac{f_{0}}{m_0}\right)\left(\nabla u:\tilde{M}_{2}\right)\right\} \mathrm{d}x\\
			& =-2\int_{\Omega_{x}}\left\{ \eta\left\Vert \nabla u\right\Vert ^{2}+\frac{\tilde{\lambda}}{8c}f_{0}\mathrm{tr}\left[\left(\nabla u+\nabla u^T\right)\tilde{M}_{2}\left(\nabla u+\nabla u^T\right)\right]\right\} \mathrm{d}x,
		\end{aligned}
		\label{eq:ene_asymp_leading}
	\end{equation}
	where $\tilde{M}_2=\frac{1}{m_{0}}\int q\otimes q e^{-U/\gamma^{2}} \mathrm{d}q$ is the second moment matrix with respect to the equilibrium microscopic distribution in \eqref{eq:2mom}.
\end{prop}

The second term on the left hand side of the energy equation \eqref{eq:ene_asymp_leading} represents the free energy $\mathcal{F}$ from microscopic coupling, and the second term on the right hand side gives the microscale dissipation energy $\mathcal{D}_2$. Notice that the closure model \eqref{eq:closure1} also satisfies this energy equation by replacing the potential flow $\tilde{u}$ in the microscopic energy $\mathcal{F}$ and $\mathcal{D}_2$.
Importantly, it can be shown that the closure equations can be directly derived through the energetic variational principle of the new energy-dissipation equation (see Appendix~\ref{subsubsec:ene-vari-deri}). The elastic stress forcing can be recovered by the MDP on $\mathcal{D}_2$, while the LAP on $\mathcal{F}$ indicates the structural constraints imposed in Assumption~\ref{assump}.
Inversely, the corresponding energy equation \eqref{eq:ene_asymp_leading} in the asymptotic closure model can be derived directly from the original energy dissipation equation \eqref{eq:energy-asymp} according to the asymptotic expansion of the full density function. The detailed derivation is listed in Appendix~\ref{subsubsec: ene-eqn-deri}.

\section{Numerical results with the asymptotic closure models}\label{sec:nums}

In this section, we perform a detailed numerical study of the asymptotic closure models \eqref{eq:closure1} and \eqref{eq:closure2} compared to direct simulations of the micro-macro solutions \eqref{eq:micro-macro} under different
microscopic potentials. In particular, we consider the following two representative potential
functions: 
\begin{itemize}
    \item the finite-extensible-nonlinear-elastic (FENE) potential, $U\left(q\right)=-\frac{HQ_{0}^{2}}{2}\log\left(1-\frac{\left|q\right|^{2}}{\left|Q_{0}\right|^{2}}\right)$, with a global minimum at $q=\left(0,0\right)$;
    \item the double well potential, $U\left(q\right)=H_{1}\left(q_{1}^{2}-1\right)^{2}+H_{2}q_{2}^{2}$, with two local minima at $q=\left(\pm1,0\right)$.
\end{itemize}
The FENE potential satisfies the structure in \eqref{eq:micro-potential},
while the double well potential provides another typical test case that supports
potential flow in leading order as in \eqref{eq:flow-potential}.



\subsection{Approximation of the density distribution functions}\label{subsec:density-simu}

In the first test, we check accuracy in the asymptotic approximation to
the density function using \eqref{eq:density_asympt}. Using a clean
setup, a steady potential flow field 
\begin{equation}\label{eq:pf_test}
	u=\left(\kappa x,-\kappa y\right),
\end{equation}
is used as the macroscopic flow field with $\kappa$ a positive constant.
Assuming uniform distribution $N\equiv\mathrm{const.}$ in the macroscopic state, the density function $f$ purely depends on the micro state $q$.
Thus, we are able to focus on the approximation in the density distribution 
functions subject to different microscopic potentials.

To recover the true density function, we adopt the direct Monte-Carlo (MC) approach
to solve the corresponding stochastic differential equation in the spatially homogeneous setting by running $M=5\times 10^4$ independent samples
$\left\{ Q^{i}\right\} _{i=1}^{M}$
\begin{equation}
	\mathrm{d}Q_{t}^{i}=\left(\kappa Q_{t}^{i}-D\nabla_{q}U\left(Q_{t}^{i}\right)\right)\mathrm{d}t+\sqrt{2D}\gamma\mathrm{d}W_{t}^{i}.\label{eq:density_mc}
\end{equation}
Above, using the steady potential flow \eqref{eq:pf_test}, we have the flow strain $\frac{1}{2}(\nabla u+\nabla u^T)=kI$ and $W^{i}$ is an
independent 2-dimensional Brownian motion for each sample. The initial
values of the samples $Q_{t}^{i}$ are drawn independently from a
normal distribution with mean zero and small variance. The truth is then 
generated using both the FENE and double well potential. We choose
the parameters for FENE as $H=0.1,Q_{0}=1.5$, and the parameters
for the double well model $H_{1}=0.05,H_{2}=0.1$. The numerical solution is integrated with the standard Euler-Maruyama scheme up to time $T=1$ with a small time step $\Delta t=1\times 10^{-3}$. A wide range of parameter values are tested ranging from $\kappa\in [1,10]$ and $\gamma\in [0.1,1]$. 
Notice that the asymptotic expression \eqref{eq:density_asympt} is
valid near the limit $\gamma\rightarrow0$ and $D\rightarrow\infty$, thus the
density function will quickly converge to the final equilibrium distribution.
Therefore, we focus on comparing the approximation skill in the equilibrium density
distribution of the microscopic state.

We choose the elastic stress $\tau$
and the internal elastic energy as the typical macroscopic quantities to compare
\begin{equation}
	\tau=\int f\nabla_{q}U\otimes q\mathrm{d}q\approx\frac{1}{N}\sum_{i=1}^{N}\nabla_{q}U\left(Q^{i}\right)\otimes Q^{i},\quad E=\int fU\mathrm{d}q\approx\frac{1}{N}\sum_{i=1}^{N}U\left(Q^{i}\right).\label{eq:macro_quan}
\end{equation}
The truth is generated by taking the empirical average of the ensemble
$\left\{ Q^{i}\right\} $.
To achieve an accurate estimate of the truth, we gather the samples
also in $N_{t}$ different time instants after the equilibrium state
is reached, then there will be in total $N=MN_{t}=5\times10^{4}\times500=2.5\times10^{7}$
samples in computing the averages in \eqref{eq:macro_quan}. Using
the asymptotic approximation \eqref{eq:density_asympt}, we get the
expression for the microscopic equilibrium distribution 
directly from the velocity gradient simplified as $\nabla u=\mathrm{diag}(\kappa,-\kappa)$ and $f_{0}$ only acting as a normalizing constant. Notice that a larger value of $\kappa$ implies a more important contribution from the higher-order terms in the density. 
In particular, we compare the accuracy of approximating the above quantities \eqref{eq:macro_quan}
truncated at orders $O\left(\gamma^{4}\right)$ and $O\left(\gamma^{6}\right)$.

\begin{figure}
	\subfloat[$\gamma=0.5,\kappa=10$]{\includegraphics[scale=0.22]{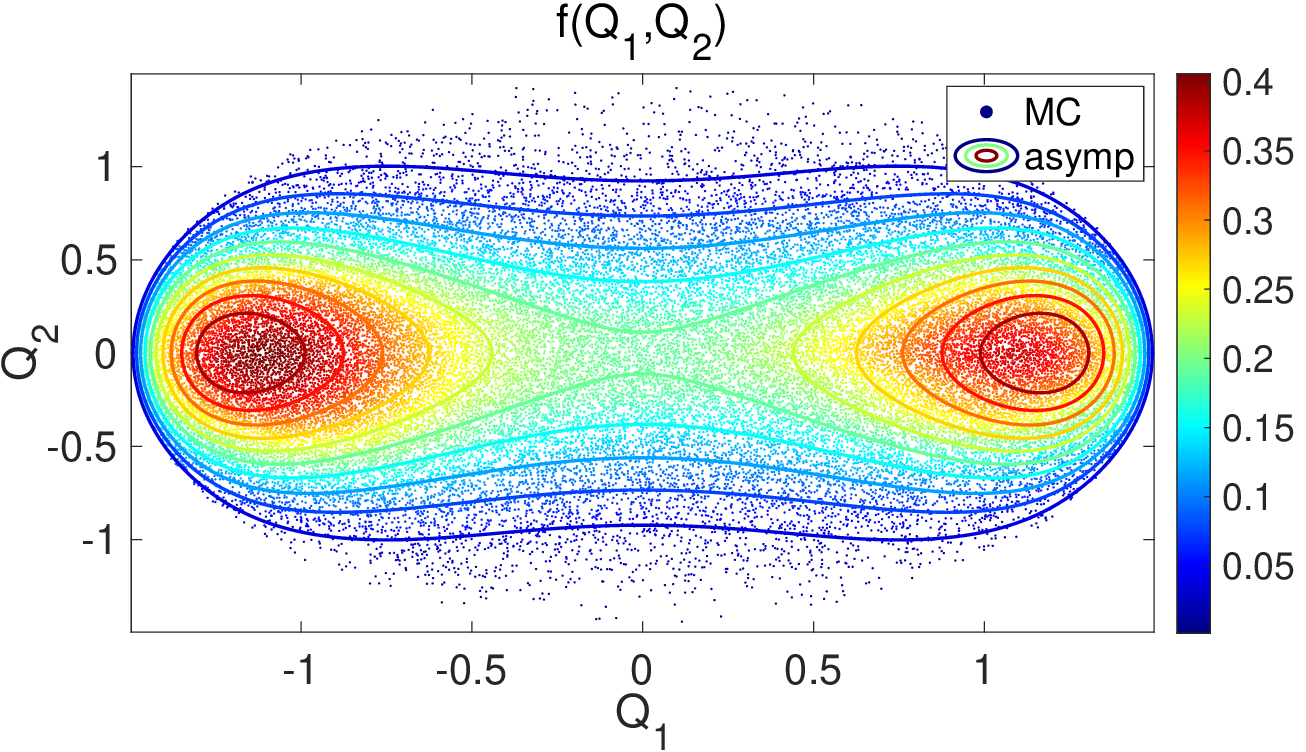}\includegraphics[scale=0.22]{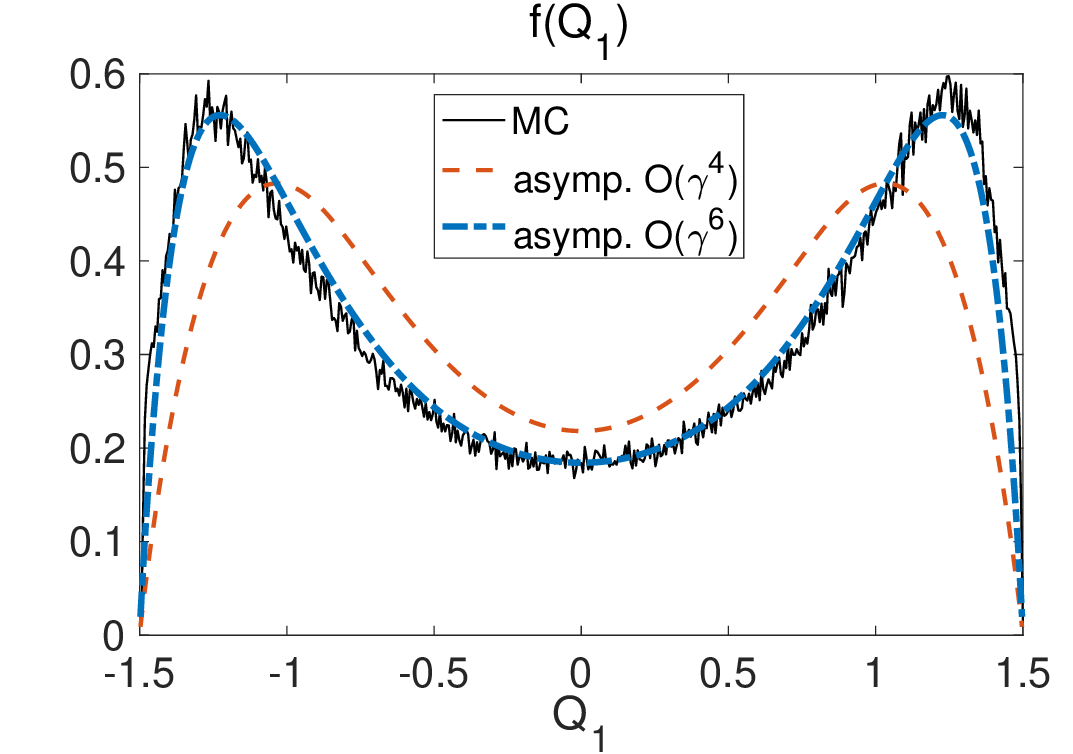}\includegraphics[scale=0.22]{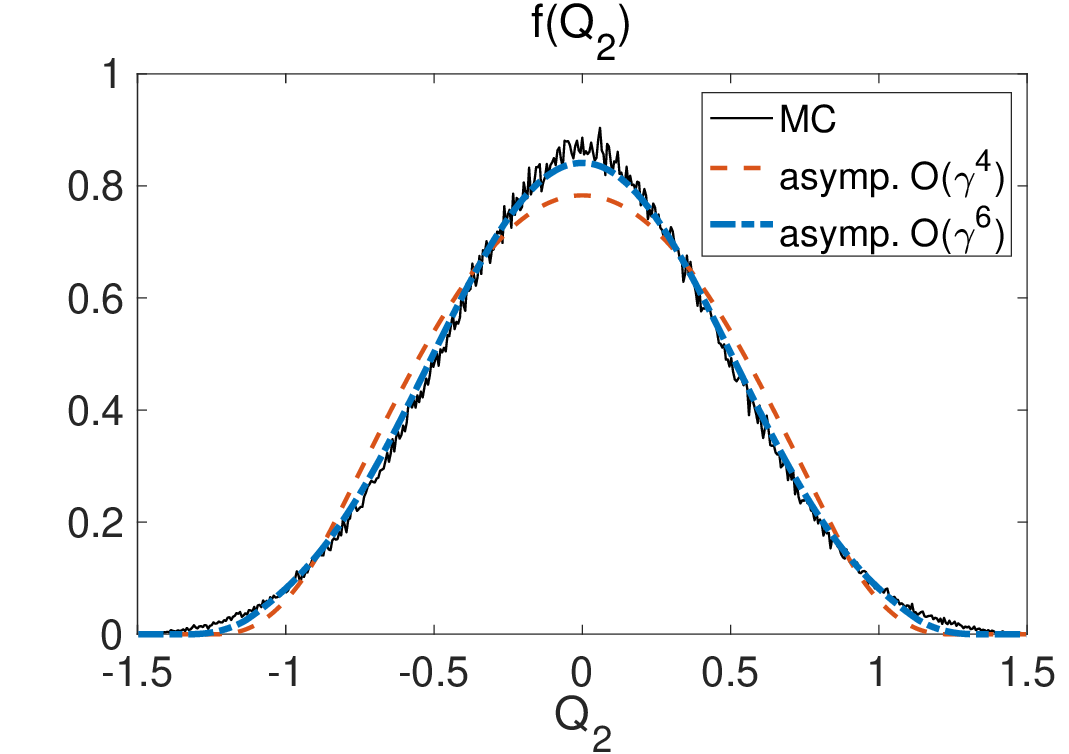}
		
	}
	
	\subfloat[$\gamma=0.2,\kappa=10$]{\includegraphics[scale=0.22]{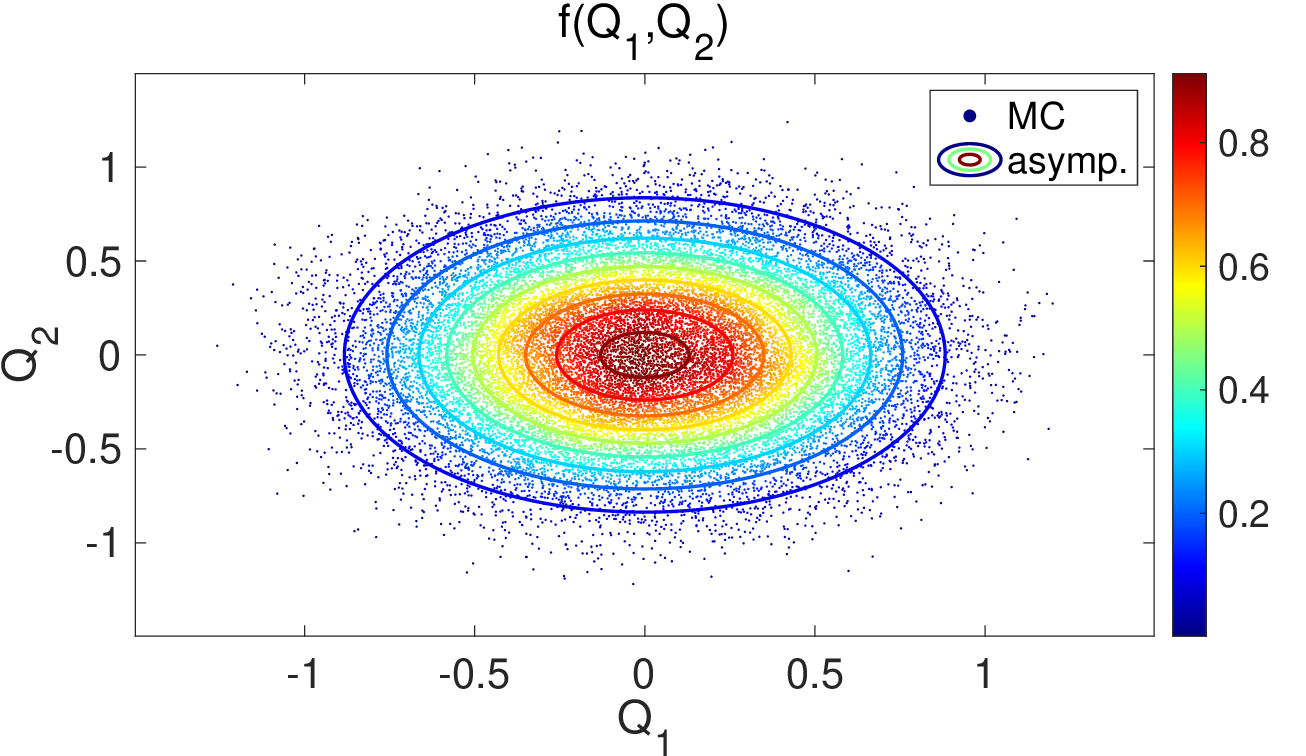}\includegraphics[scale=0.22]{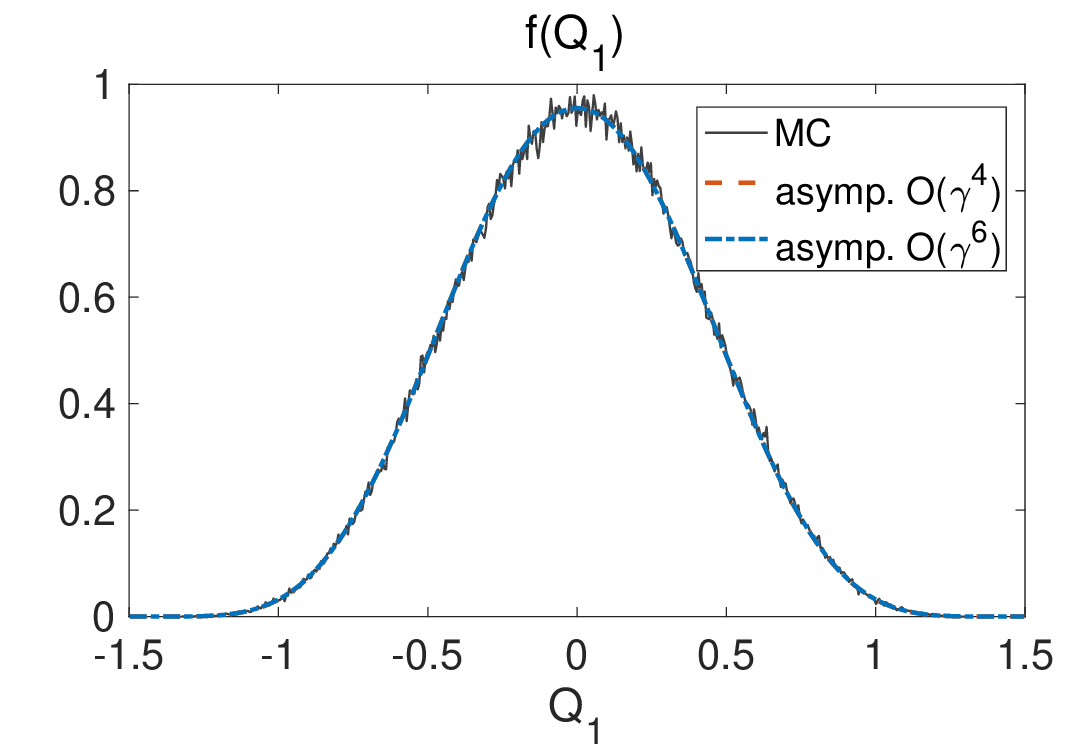}\includegraphics[scale=0.22]{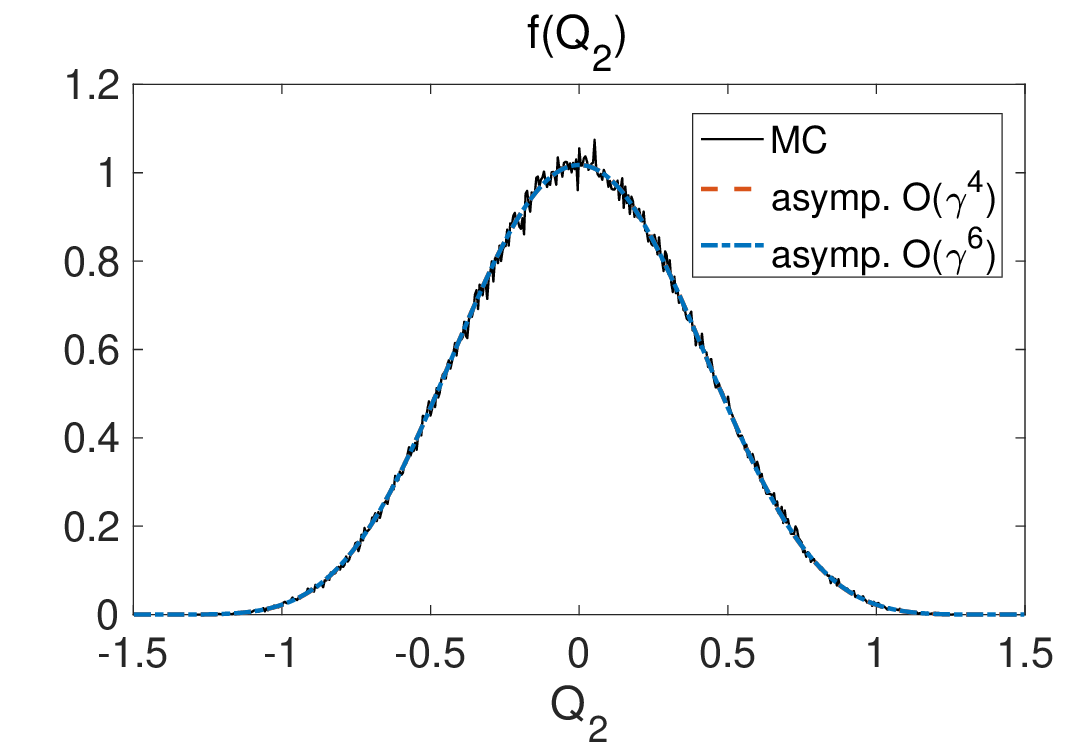}
		
	}
	\caption{Approximation of the microscopic density function in the FENE model.
		Colors in the 2d scatter plot indicate the density of the MC  samples.\protect\label{fig:den_fene}}
	
\end{figure}

\begin{figure}
	\subfloat[$\gamma=0.5,\kappa=5$]{\includegraphics[scale=0.22]{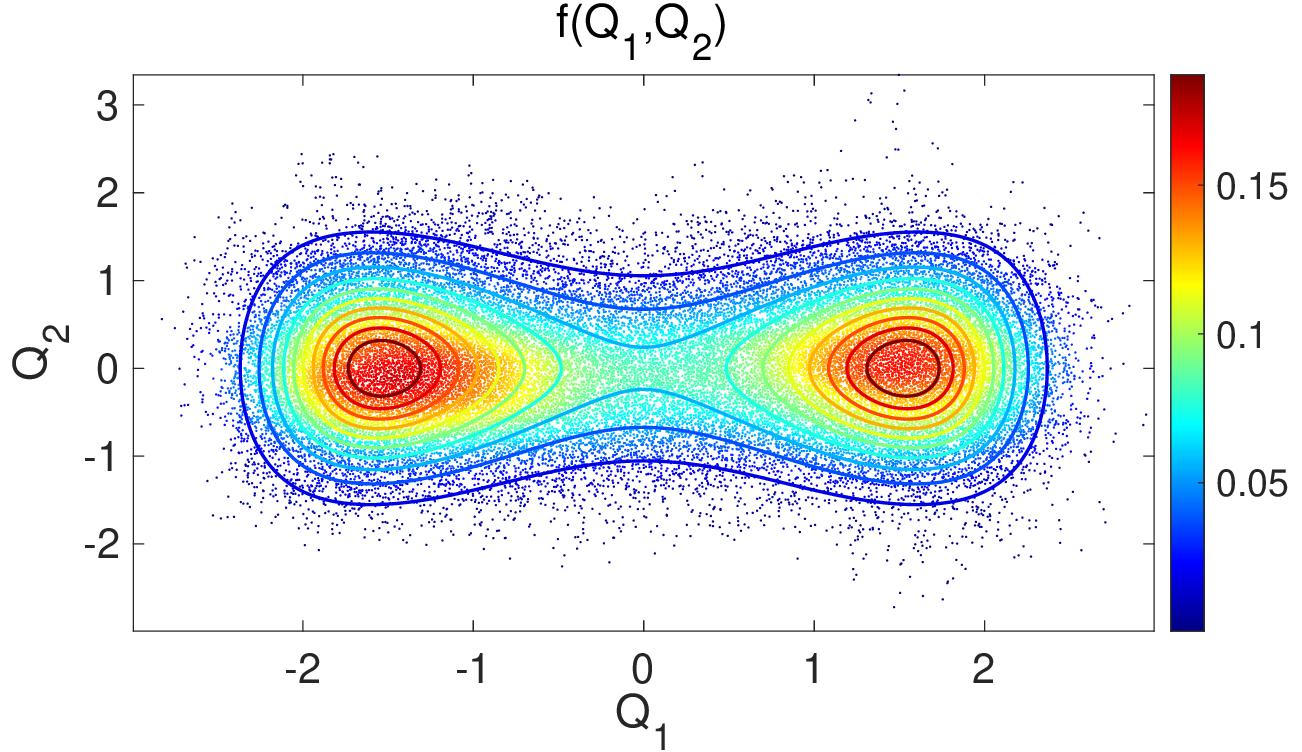}\includegraphics[scale=0.22]{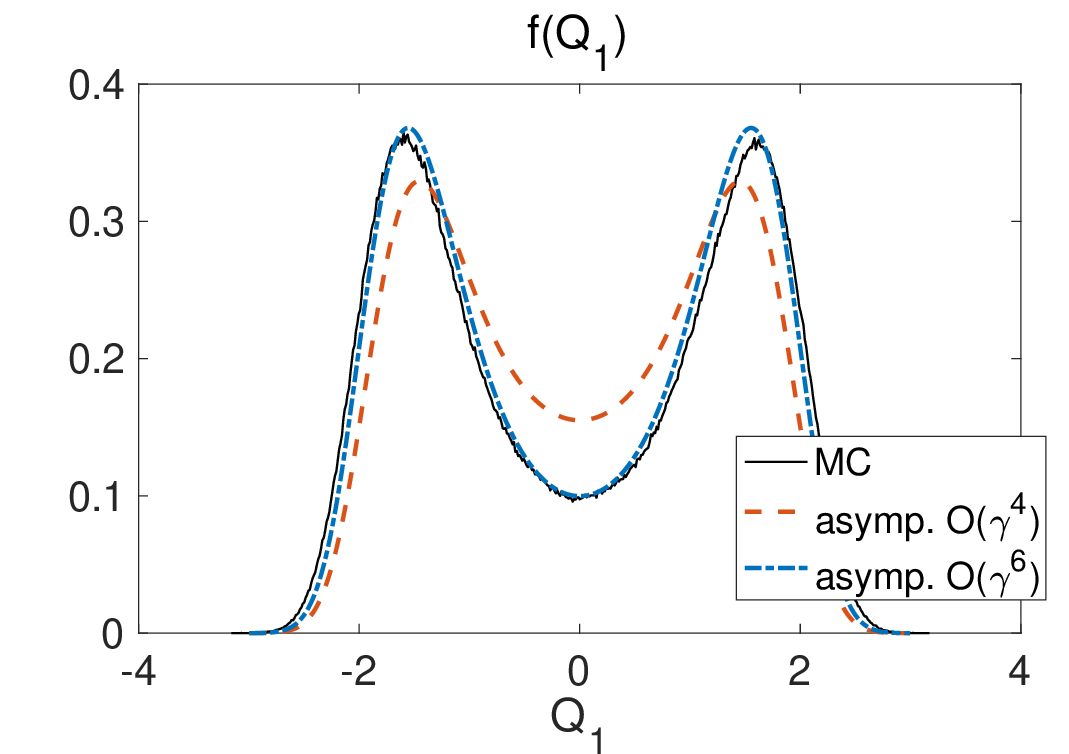}\includegraphics[scale=0.22]{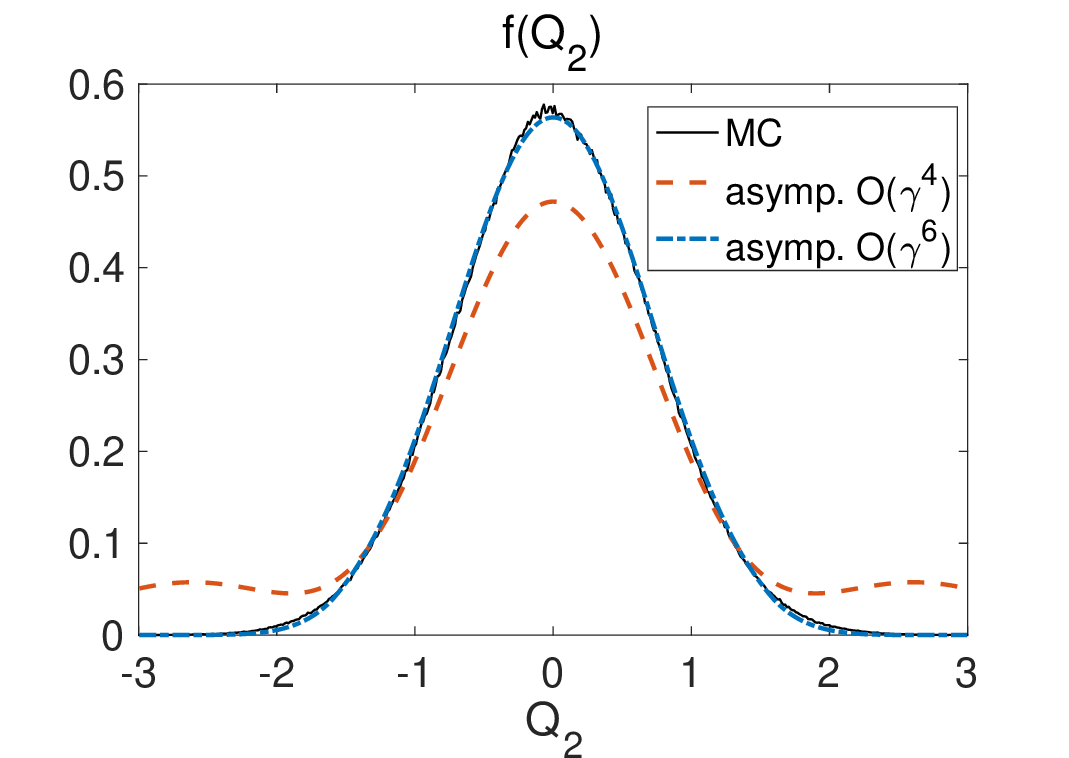}
		
	}
	
	\subfloat[$\gamma=0.1,\kappa=5$]{\includegraphics[scale=0.22]{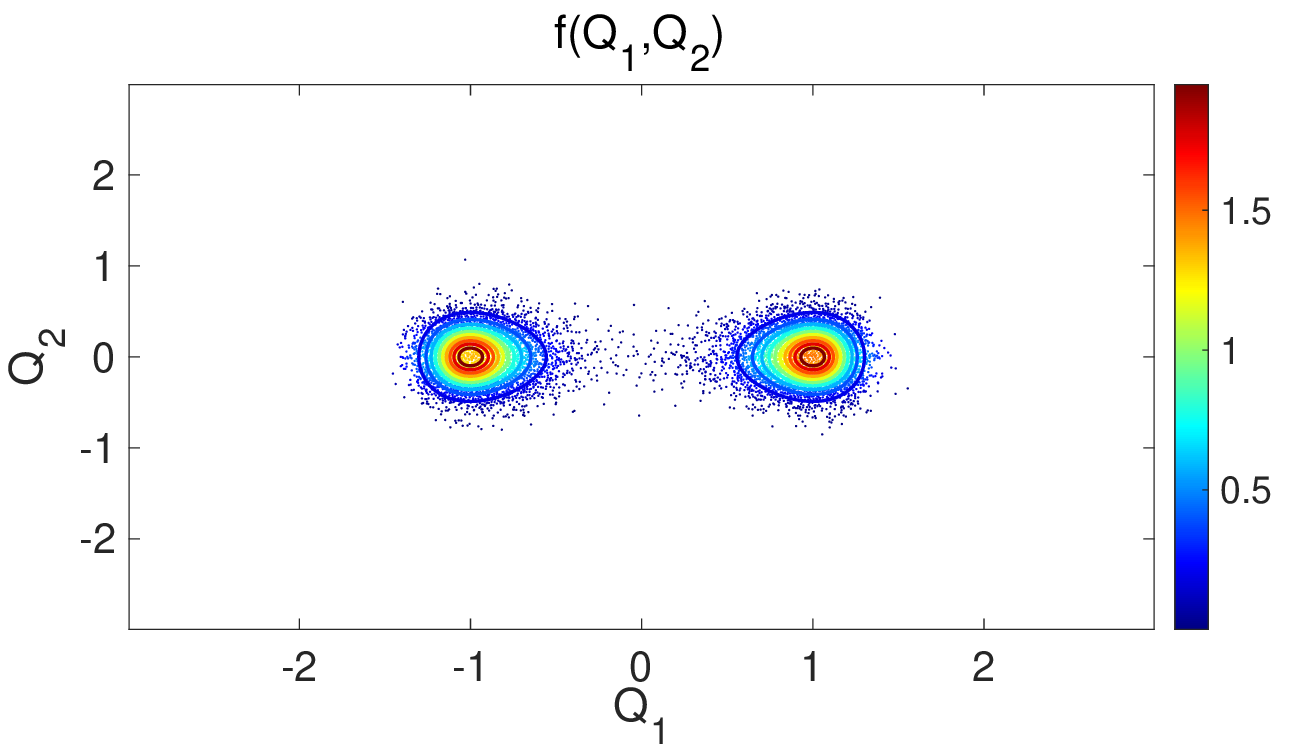}\includegraphics[scale=0.22]{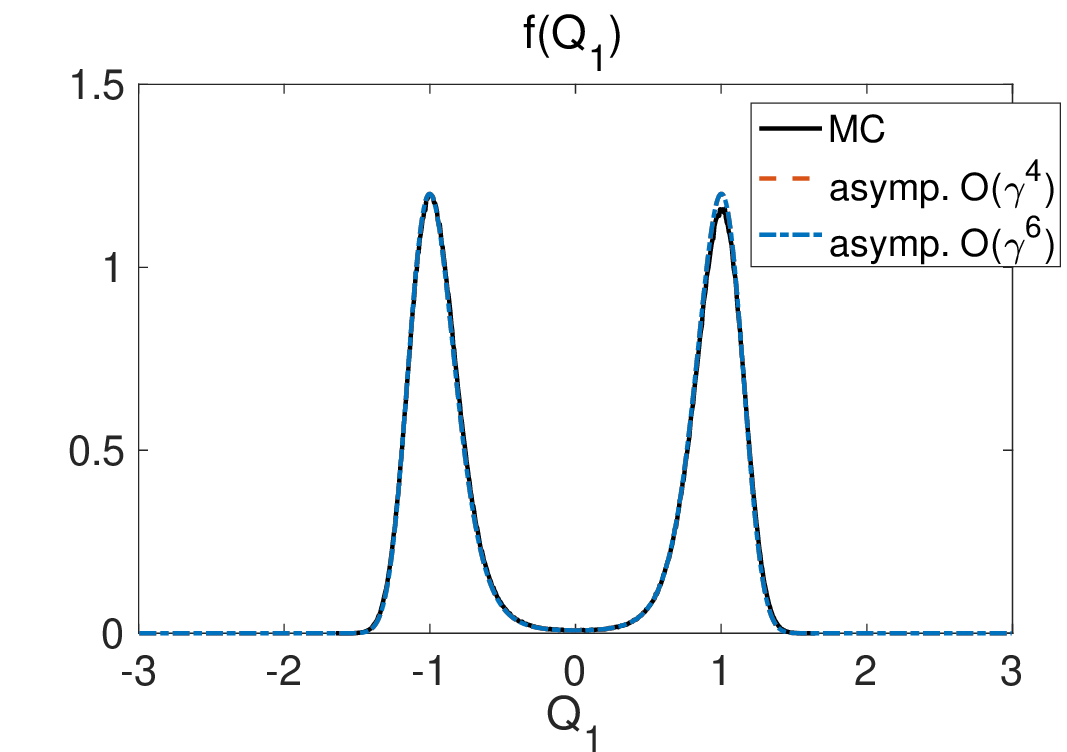}\includegraphics[scale=0.22]{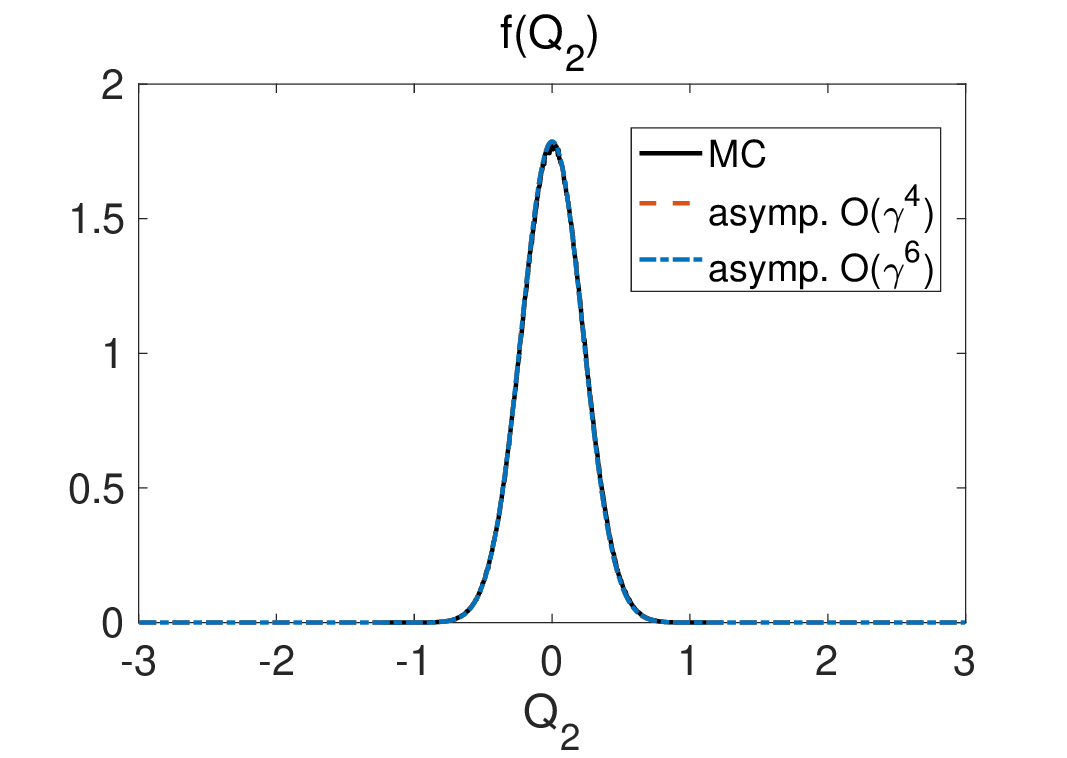}
		
	}
	\caption{Approximation of the microscopic density function in the double well
		model. Colors in the 2d scatter plot indicate  the density of the MC 
		samples.\protect\label{fig:den_quad}}
\end{figure}

Figure \ref{fig:den_fene} and \ref{fig:den_quad} demonstrate the
equilibrium density distributions according to the two potential functions respectively. The asymptotic approximations with different orders of accuracy are
compared with the MC result as the truth in both the 2d joint distribution as well
as the 1d marginal distributions. With a moderate value of $\gamma=0.2$,
both $O\left(\gamma^{4}\right)$ and $O\left(\gamma^{6}\right)$ approximations
capture the density function dominated by the Gibbs  distribution
$f_{\mathrm{eq}}\sim\exp\left(-U/\gamma^{2}\right)$. However, as the value increases to $\gamma=0.5$, the $O\left(\gamma^{4}\right)$
approximation captures the general structure of the density function but misses
the accurate locations of the peaks in the density. With a finer correction
in the $O\left(\gamma^{6}\right)$ approximation, the precise structure
of the density function is also captured even with large
values of $\gamma$ away from the asymptotic limit. In particular, the asymptotic representations show high skill in successfully capturing the bi-modal density structures typically observed in the FENE potential with a large value of $\kappa$.

\begin{figure}
	\subfloat{\includegraphics[scale=0.28]{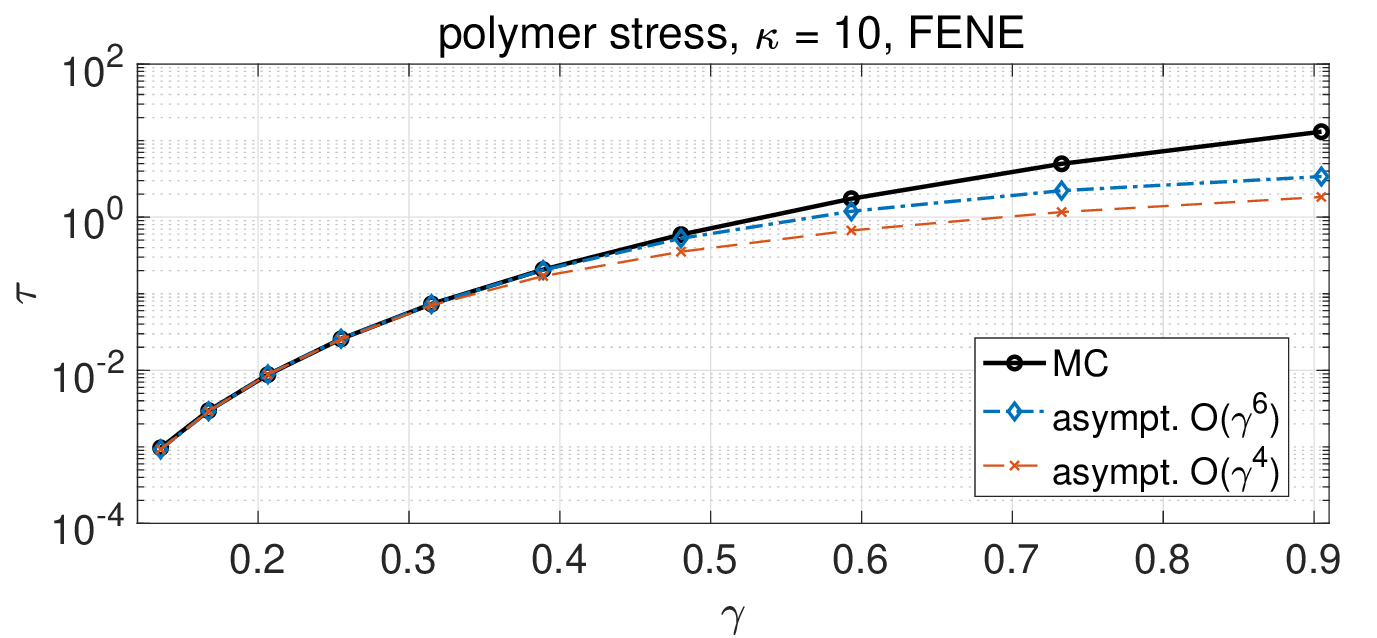}\includegraphics[scale=0.28]{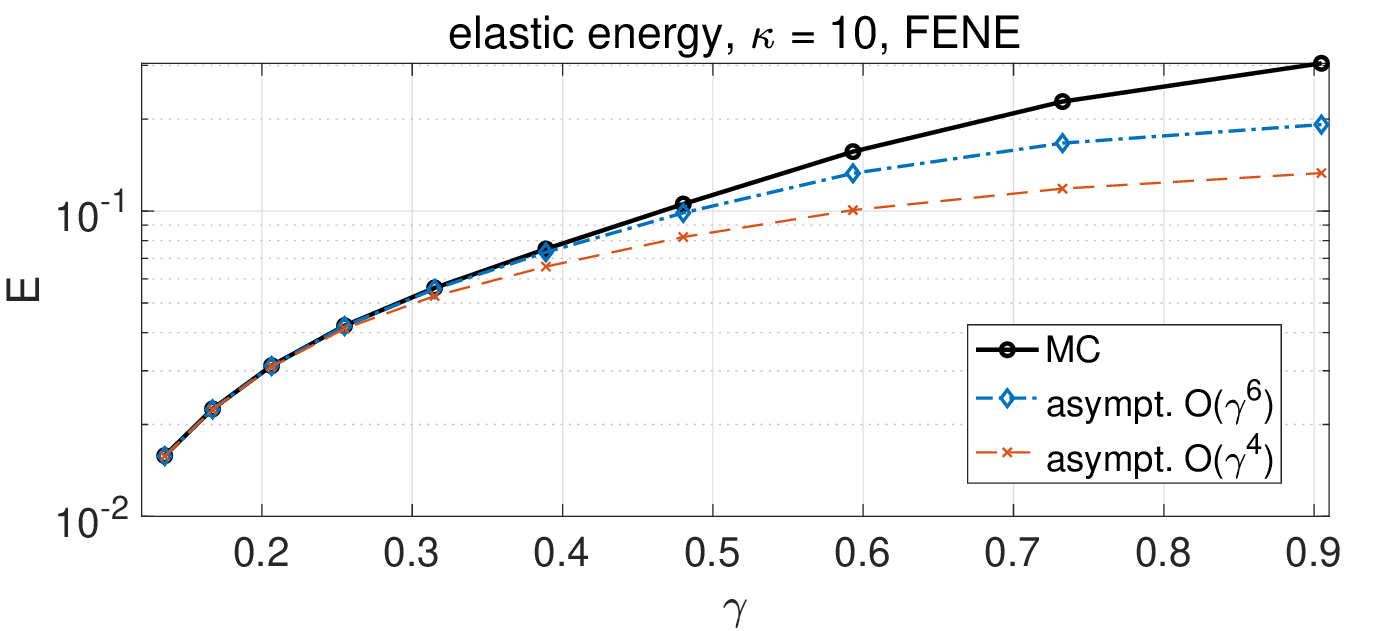}
		
	}
	
	\subfloat{\includegraphics[scale=0.28]{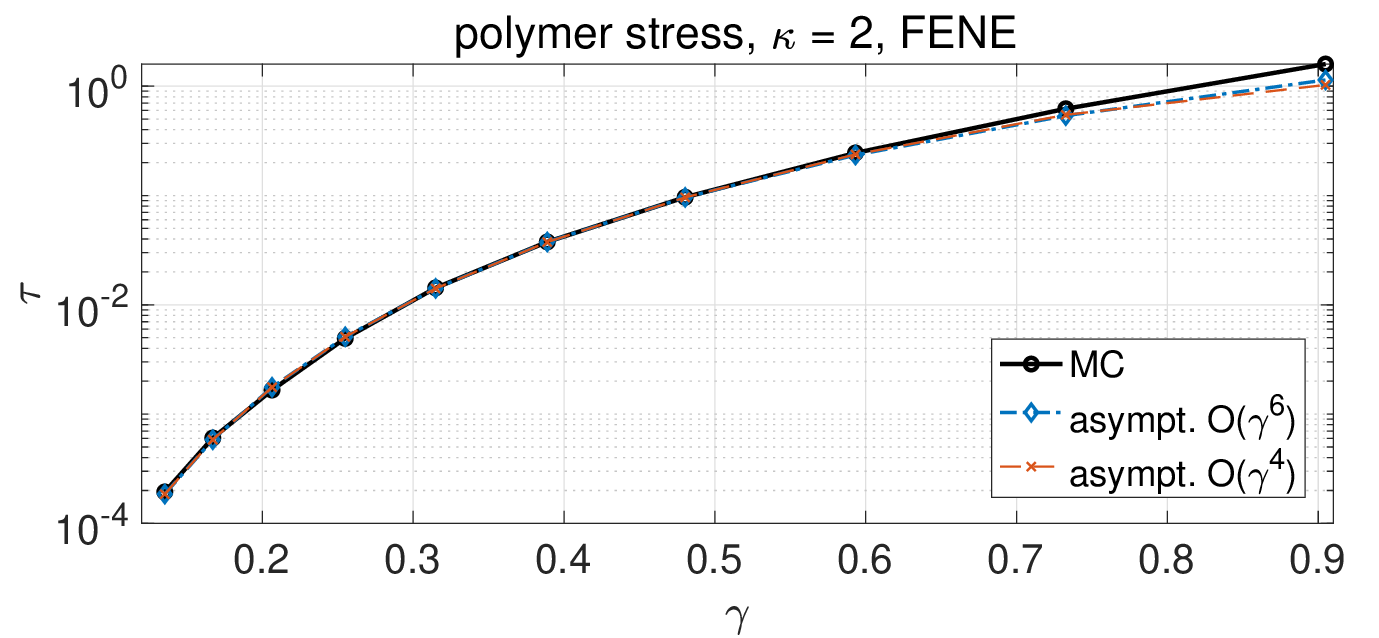}\includegraphics[scale=0.28]{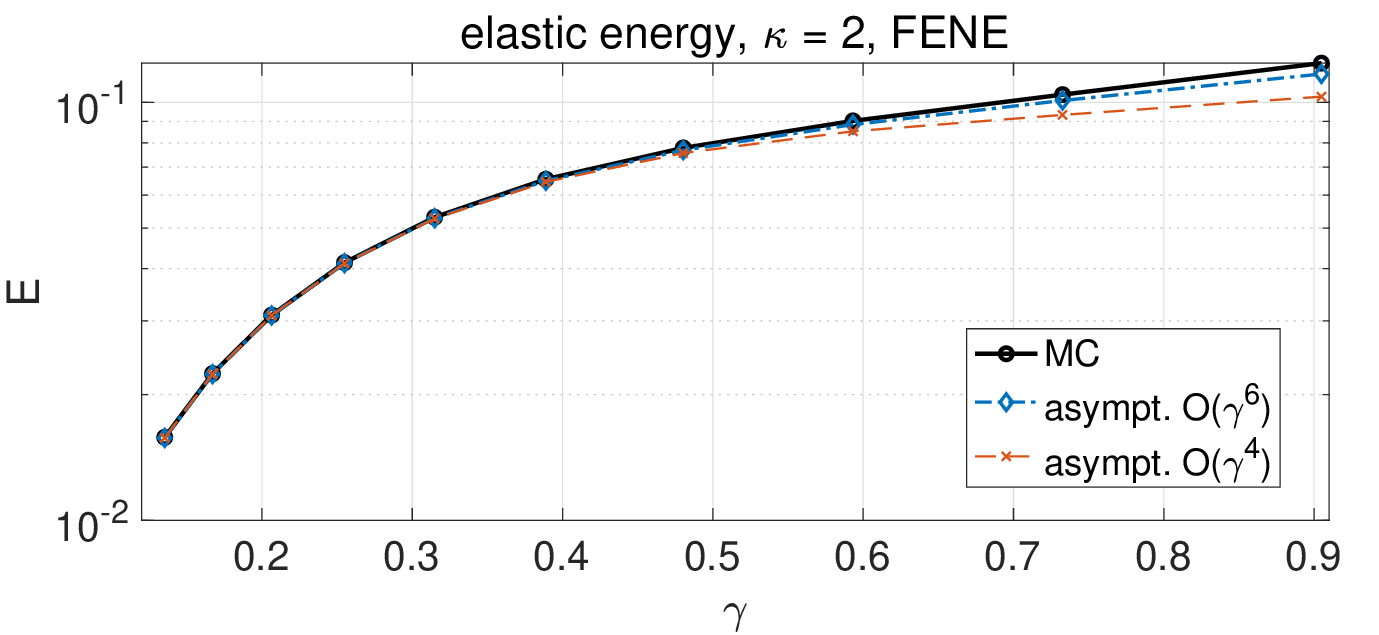}}
	
	\caption{Comparison of the polymer stress $\tau_{11}-\tau_{22}$ and elastic
		energy $E$ from the MC solution and asymptotic approximations in
		the FENE model.\protect\label{fig:Comparison-stress-fene}}
	
\end{figure}

\begin{figure}
	\subfloat{\includegraphics[scale=0.28]{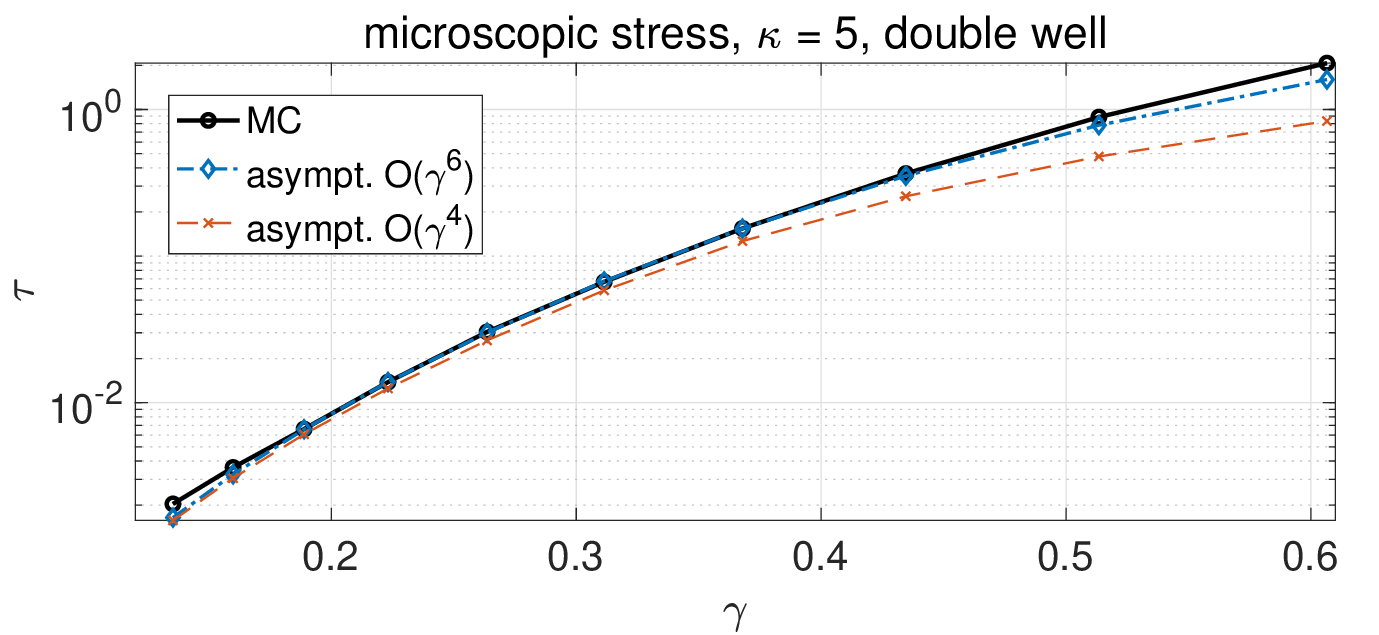}\includegraphics[scale=0.28]{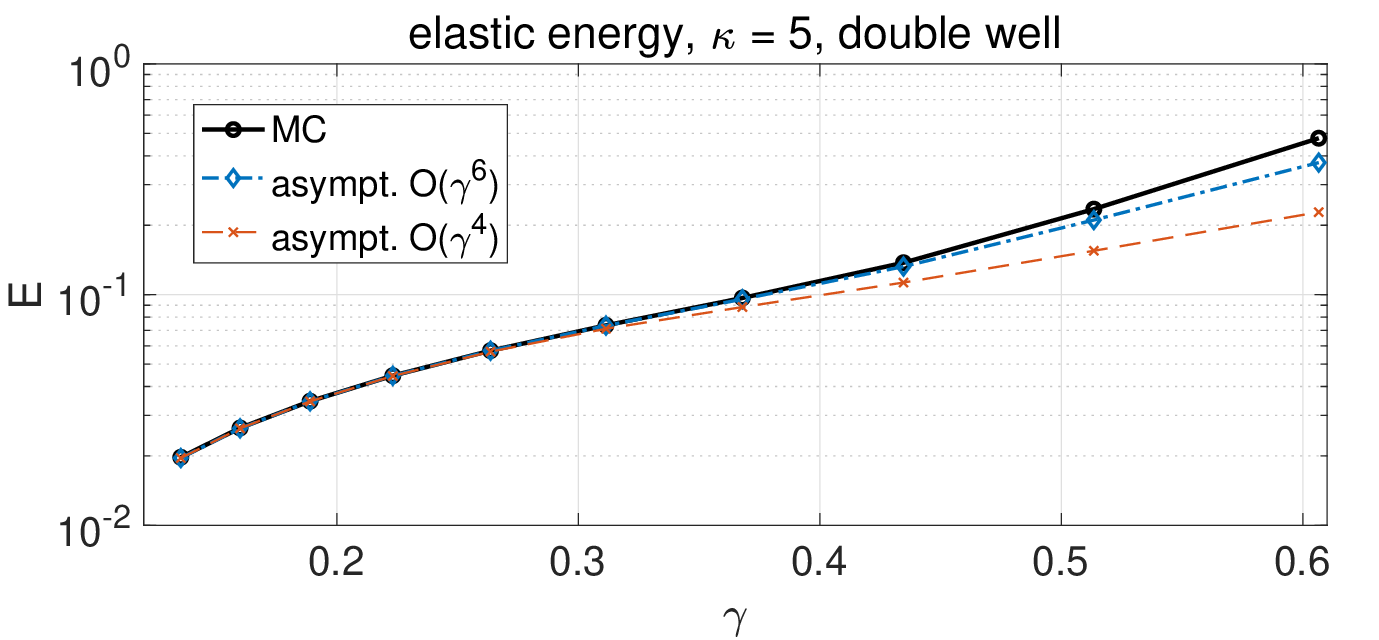}}
	
	\subfloat{\includegraphics[scale=0.28]{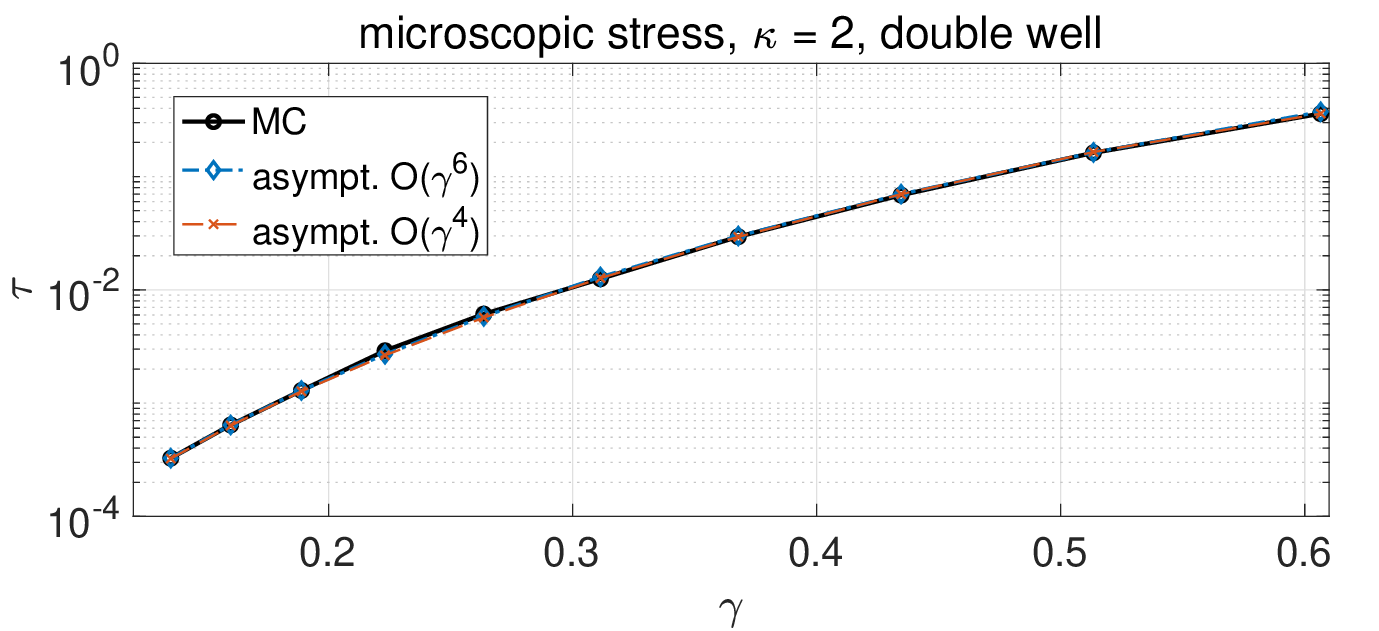}\includegraphics[scale=0.28]{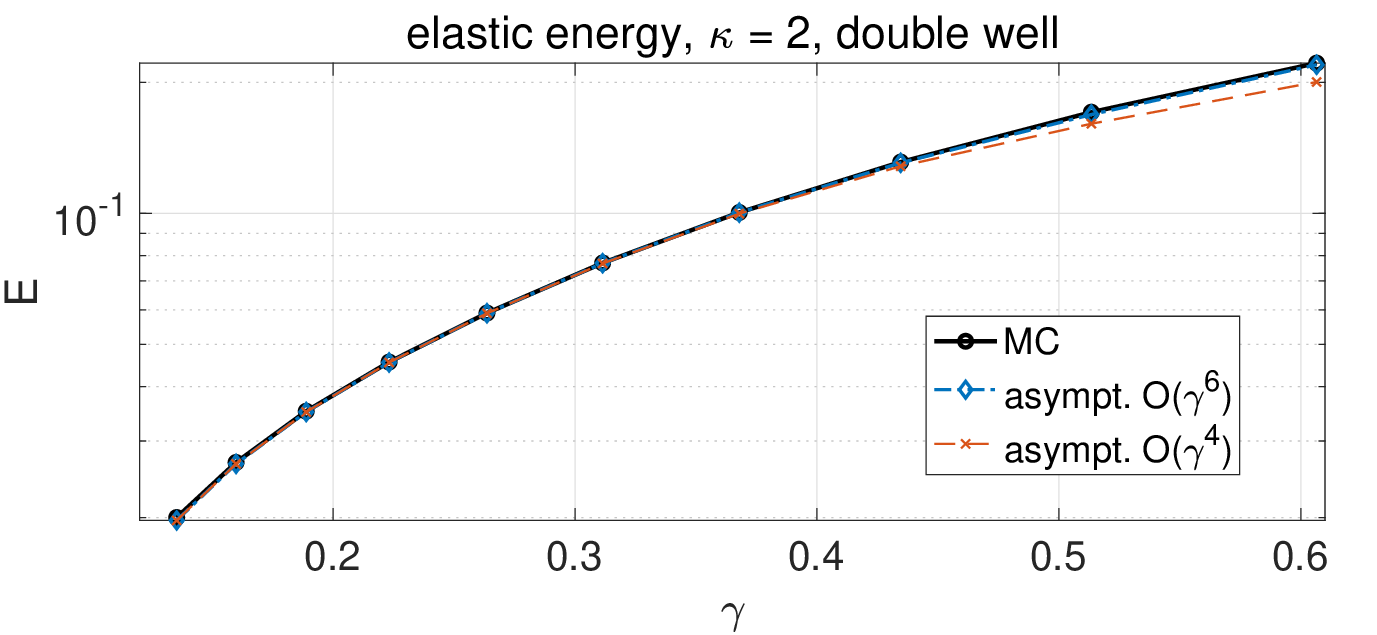}}
	
	\caption{Comparison of the polymer stress $\tau_{11}-\tau_{22}$ and elastic
		energy $E$ from the MC solution and asymptotic approximations in
		the double well model.\protect\label{fig:Comparison-stress-quad}}
\end{figure}

\begin{figure}
	\includegraphics[scale=0.31]{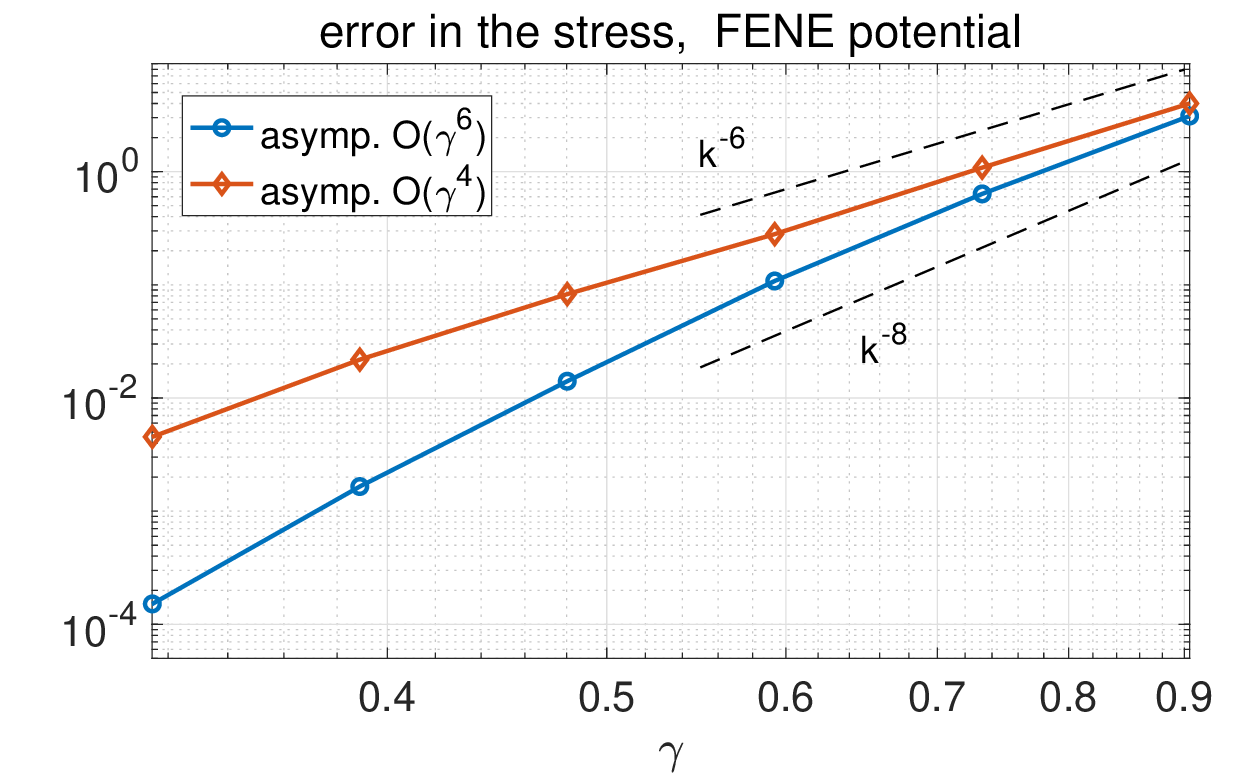}\includegraphics[scale=0.31]{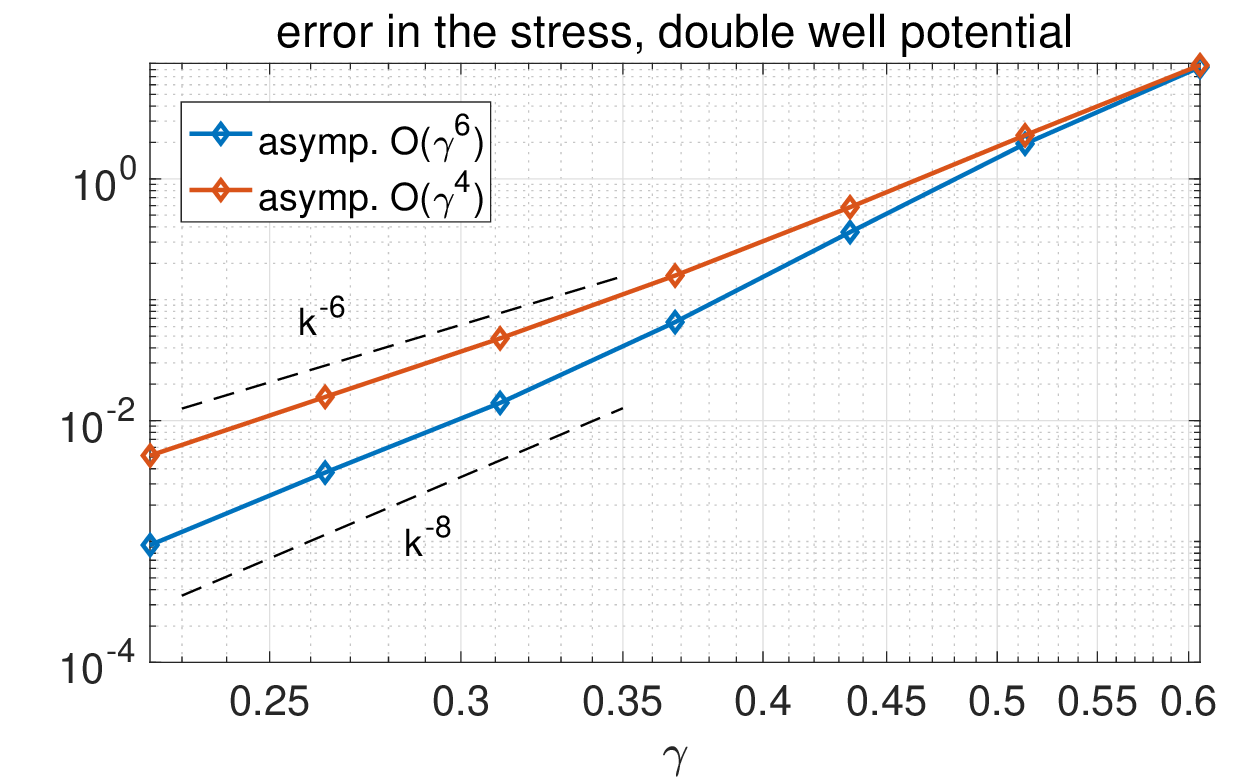}
	
	\caption{Absolute error in the polymer stress $\tau_{11}-\tau_{22}$ as a function
		of the parameter $\gamma$.\protect\label{fig:error-stress}}
	
\end{figure}

To have a more quantitative comparison of the approximation
skill, we compare the macroscopic quantities \eqref{eq:macro_quan}
captured by the approximation formulas in a range of different values
of $\gamma$ and $\kappa$ in Figure \ref{fig:Comparison-stress-fene} and \ref{fig:Comparison-stress-quad} for the two types of potentials. 
First, notice that with a small flow stress $\kappa=2$, the asymptotic
formula maintains accurate estimates for the stress and energy up to large values of $\gamma$ in both potential functions even only using a low-order approximation. This indicates
the robust performance of the closure model for the prediction of  macroscopic flow
fields. In the more challenging case with larger $\kappa=5$, 
the approximations remain accurate for a wide range of values in $\gamma$ far beyond the asymptotic limit.
As expected, the higher order approximation $O\left(\gamma^{6}\right)$
provides better estimates of the statistics. Especially, the double
well potential in this case becomes really stiff and will require a
very large ensemble size to achieve the accurate truth. Finally, to
further compare the order of accuracy in the approximation in the key macroscopic quantities, the absolute errors in the stress
approximation $\tau_{11}-\tau_{22}$ are plotted in Figure \ref{fig:error-stress}.
The precise scaling law with respect to $\gamma$ is observed, confirming the precise high-order  accuracy achieved from the asymptotic density expressions.

\subsection{Development of macroscopic flows from  microscopic elastic stress}\label{subsec:flow-test1}

Next, we examine the impact of the microscopic stress to the macroscopic
flow field through the closure model \eqref{eq:closure1}. Still, we consider a dominant steady potential flow $\tilde{u}$
in the leading order, and track the development of the vorticity flow
$\omega=\Delta\psi$ with the associated velocity field $v=\left(-\partial_{y}\psi,\partial_{x}\psi\right)$ and the stream function $\psi$ in smaller amplitude. Therefore,
we can find the simplified coupled equation for the macroscopic density
$f_{0}\left(x,t\right)$ and the vorticity field $\omega$ in the following form
\begin{equation}
	\begin{aligned}\partial_{t}\omega+\tilde{u}\cdot\nabla\omega+v\cdot\nabla\omega & =\mu\Delta\omega+\lambda\left[\left(\tau^{22}-\tau^{11}\right)\partial_{xy}f_{0}+\tau^{21}\partial_{xx}f_{0}-\tau^{12}\partial_{yy}f_{0}\right],\\
		\partial_{t}f_{0}+\tilde{u}\cdot\nabla f_{0}+v\cdot\nabla f_{0} & =0.
	\end{aligned}
	\label{eq:flow-macro}
\end{equation}
Above, we still adopt the steady potential flow $\tilde{u}=\left(\kappa x,-\kappa y\right)$
as in \eqref{eq:flow-potential}, and the vorticity flow $\omega=\left(\nabla\times\tilde{v}\right)\cdot\hat{k}$
is excited by the microscopic stress $\tau=\int\tilde{f}\nabla_{q}U\otimes q\mathrm{d}q$
where $\tilde{f}\left(q,t\right)$ is the equilibrium microscopic
density due to the potential flow $\tilde{u}$. The vorticity equation \eqref{eq:flow-macro} is solved by a pseudo-spectral code with a standard 4th order Runge-Kutta scheme for the time integration. For simplicity, periodic boundary condition is imposed in the vorticity flow solution. In the density equation, we
assume that the microscopic density quickly reached its equilibrium
with homogeneous macroscopic state, thus there is no need to run the
very expensive ensemble simulation at each macroscopic grid point in the closure models 
to get the fully resolved density function $f\left(x,q,t\right)$ as in the full model simulation. 
In this way, \eqref{eq:flow-macro} provides a simpler test case 
for checking the one-way coupling of the impact from microscopic stress
to the macroscopic vorticity flow structure.

We again test the performance in computing the density and vorticity field
in \eqref{eq:flow-macro} using both the FENE and double well potential
functions. The true elastic stress is generated by the full model through the MC simulations under the same setting in Section~\ref{subsec:density-simu}, while the
closure model uses the explicit formula \eqref{eq:stress_asymp} to
compute the stress term directly. We adopt a Gaussian distribution in
the initial density function $f_{0}=\frac{1}{2\pi\sqrt{\left(\sigma_{x}^{2}+\sigma_{y}^{2}\right)}}\exp\left(-\frac{x^{2}}{2\sigma_{x}^{2}}-\frac{y^{2}}{2\sigma_{y}^{2}}\right)$
with $\sigma_{x}=0.1,\sigma_{y}=0.75$. The initial flow field is
set to be purely potential with zero vorticity flow at the starting
time. Therefore, we track the development of vorticity flow due to
the microscopic elastic stress. 

\begin{figure}
	\subfloat{\includegraphics[scale=0.3]{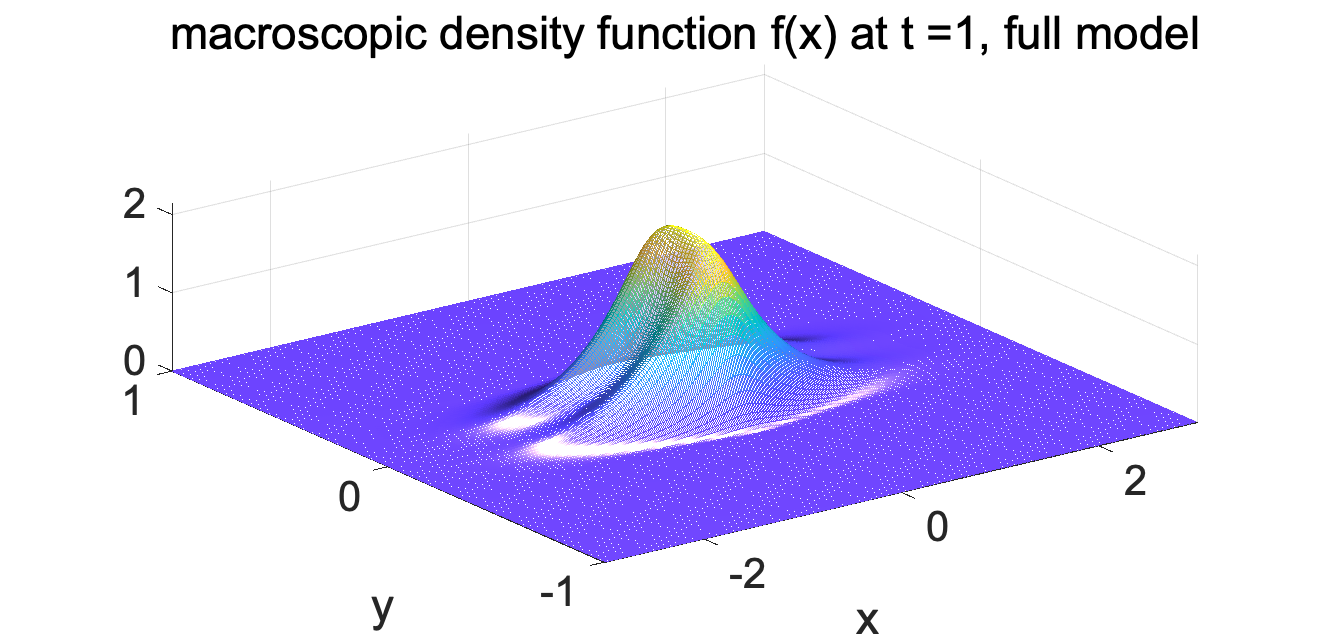}\includegraphics[scale=0.3]{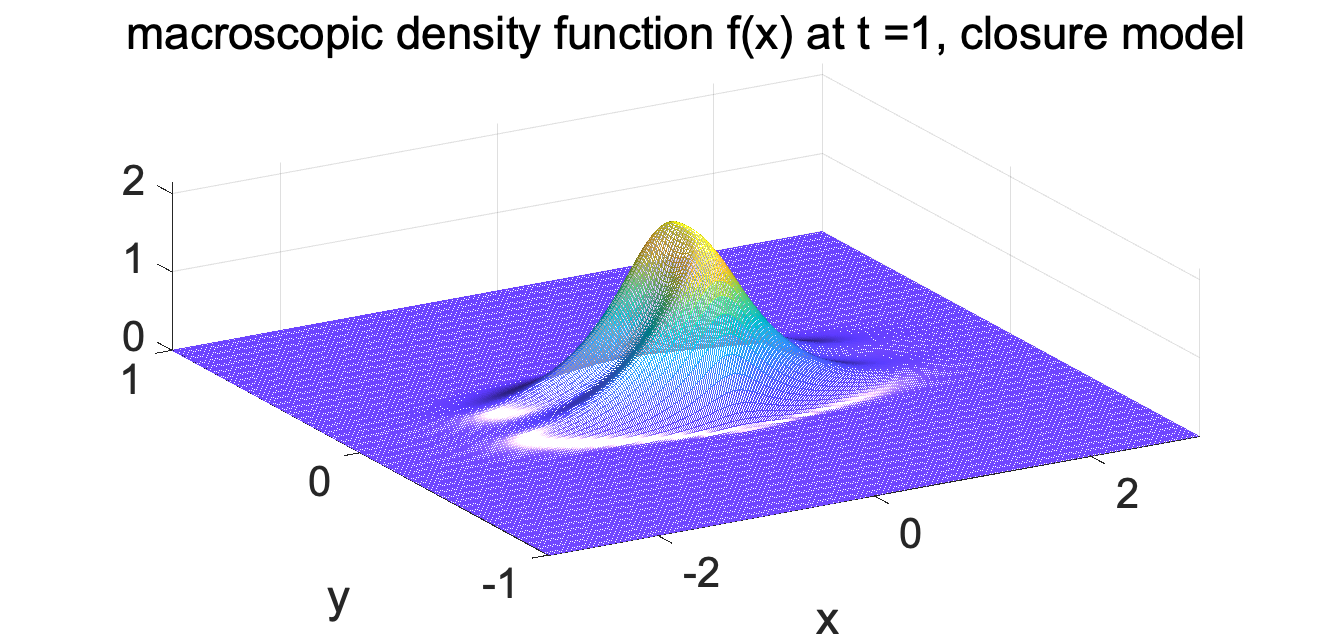}
		
	}
	
	\subfloat{\includegraphics[scale=0.3]{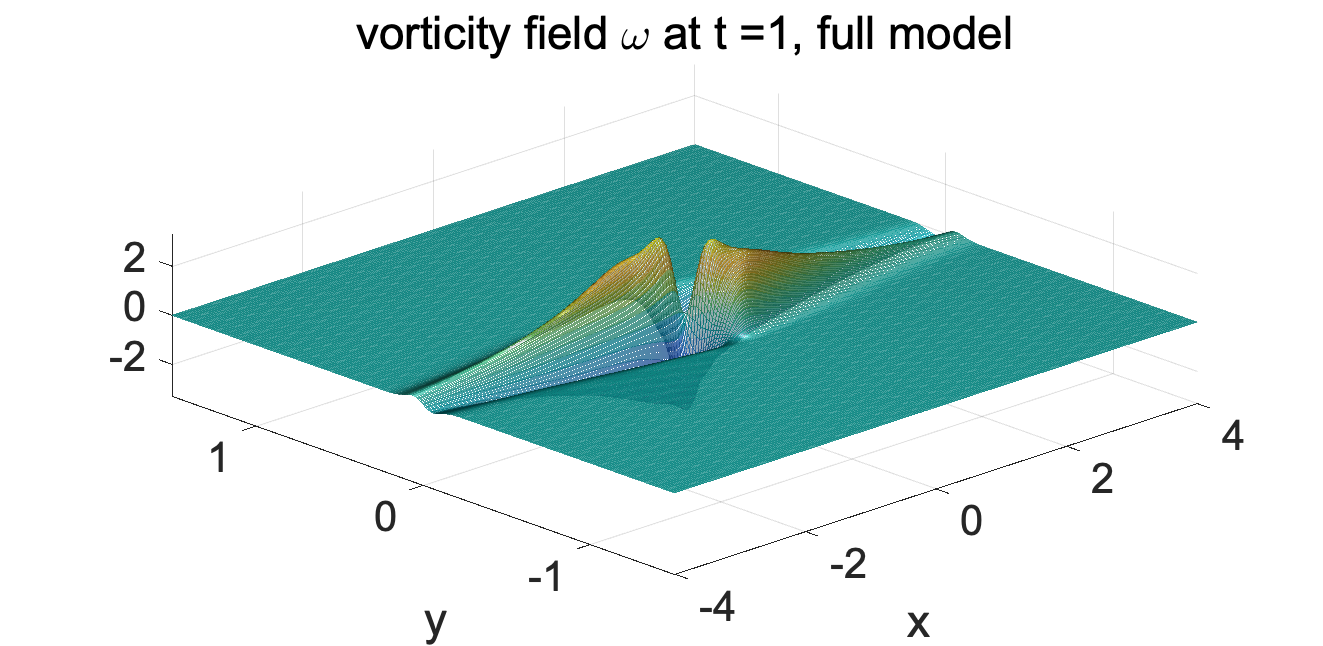}\includegraphics[scale=0.3]{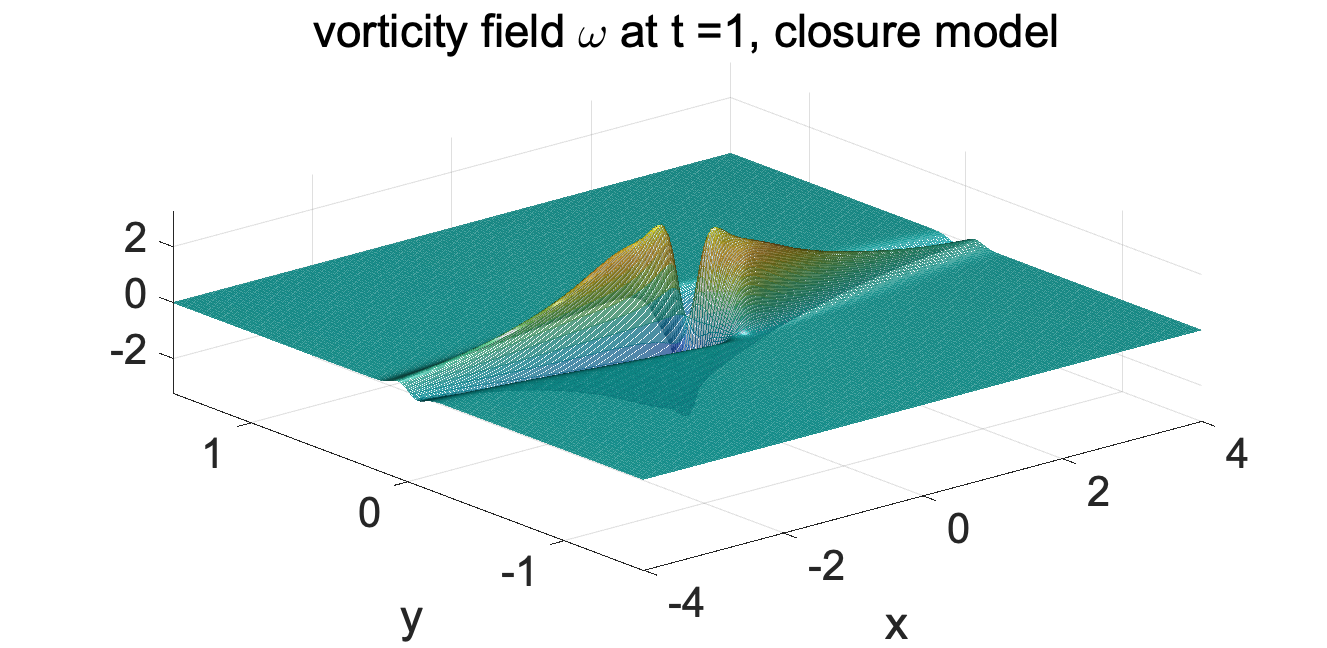}}
	
	\caption{Comparison of the macroscopic density and vorticity field from the
		full and closure model using the FENE potential.\protect\label{fig:Comparison-vort-fene}}
	
\end{figure}

\begin{figure}
	\subfloat{\includegraphics[scale=0.3]{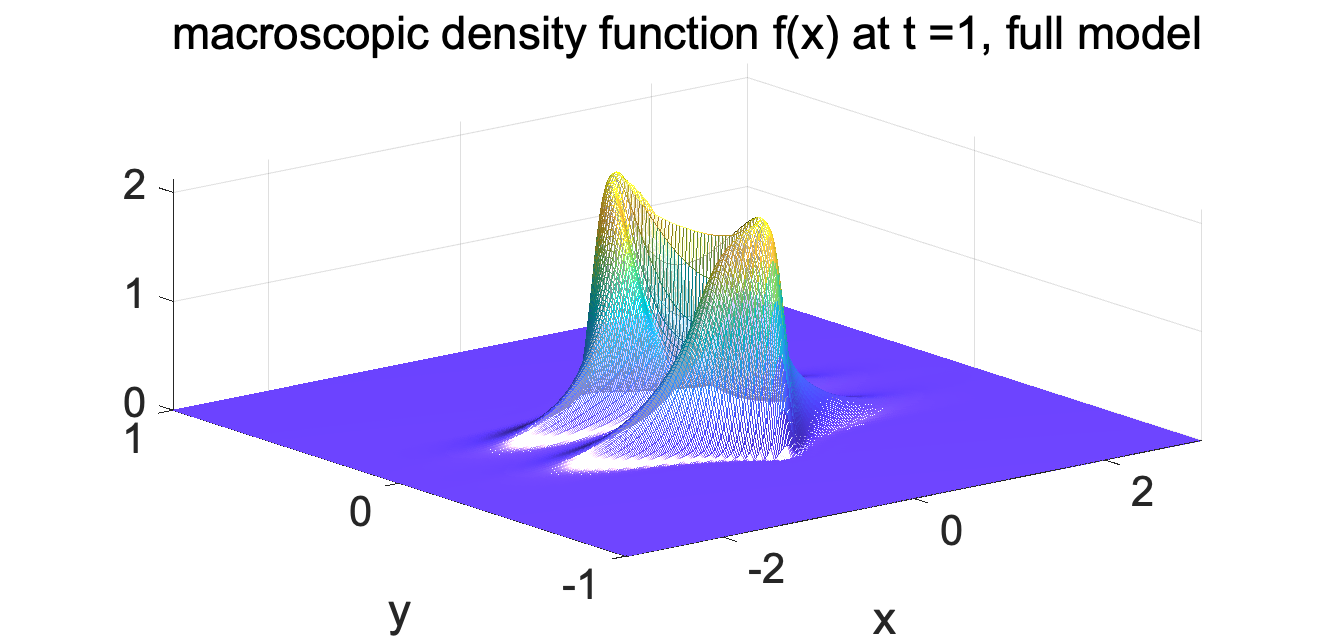}\includegraphics[scale=0.3]{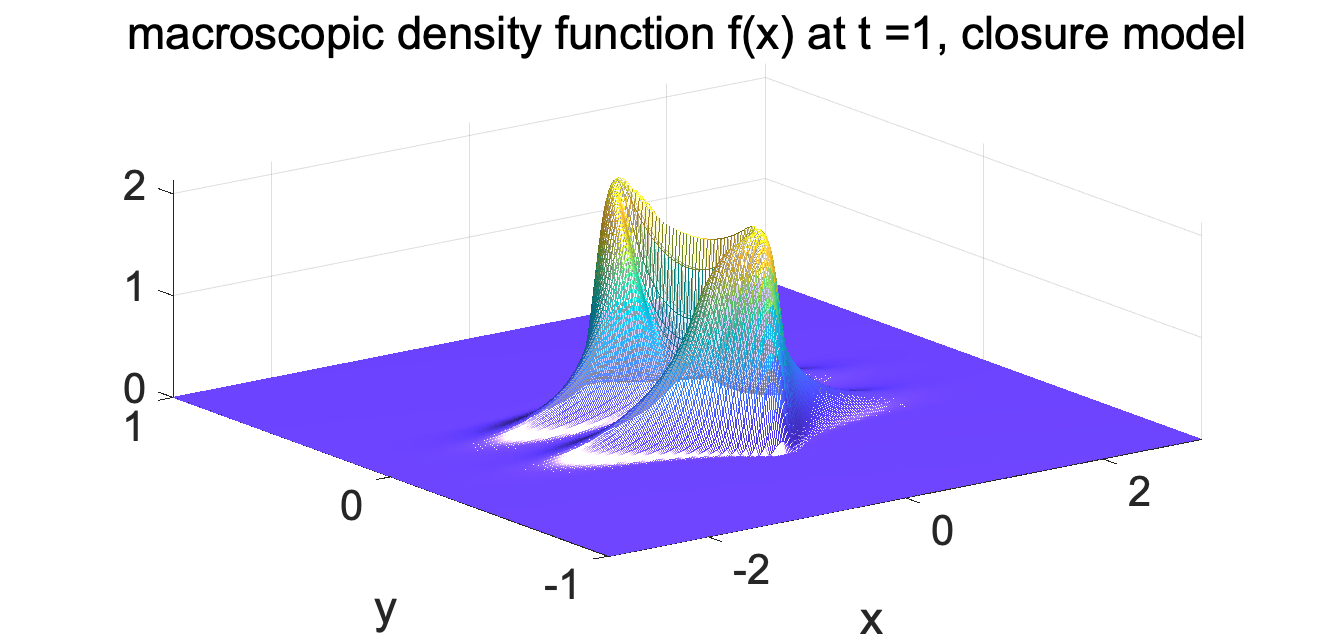}}
	
	\subfloat{\includegraphics[scale=0.3]{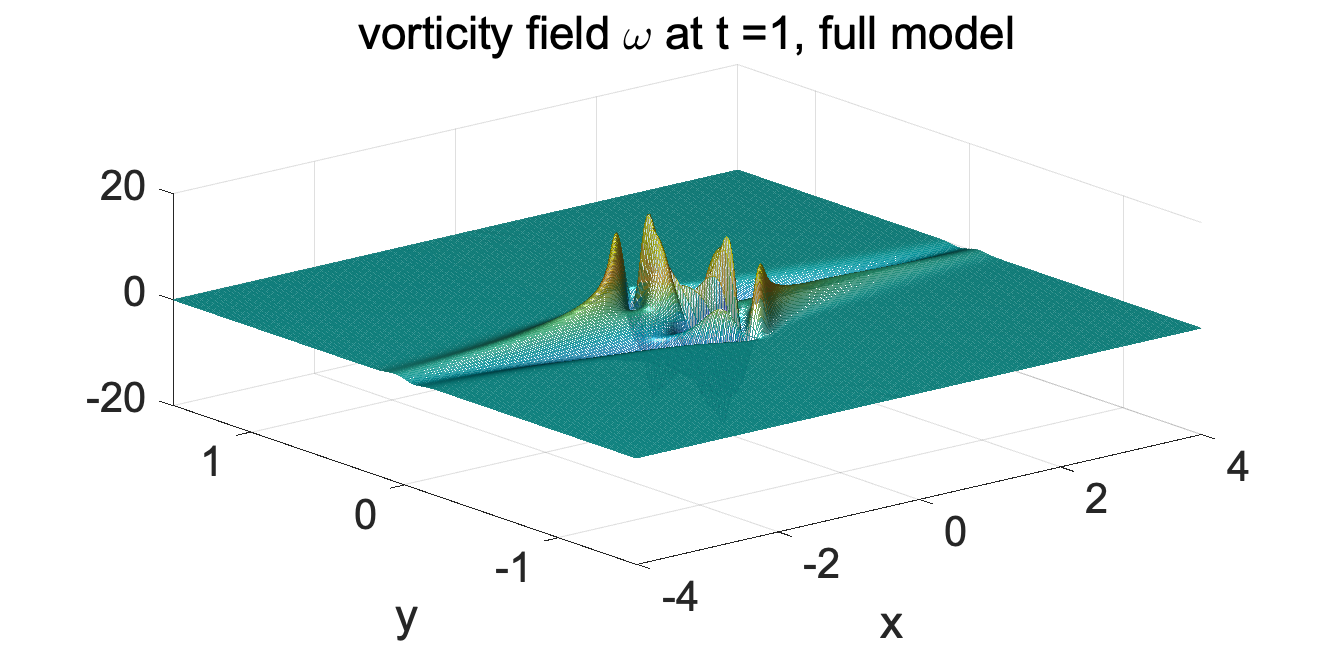}\includegraphics[scale=0.3]{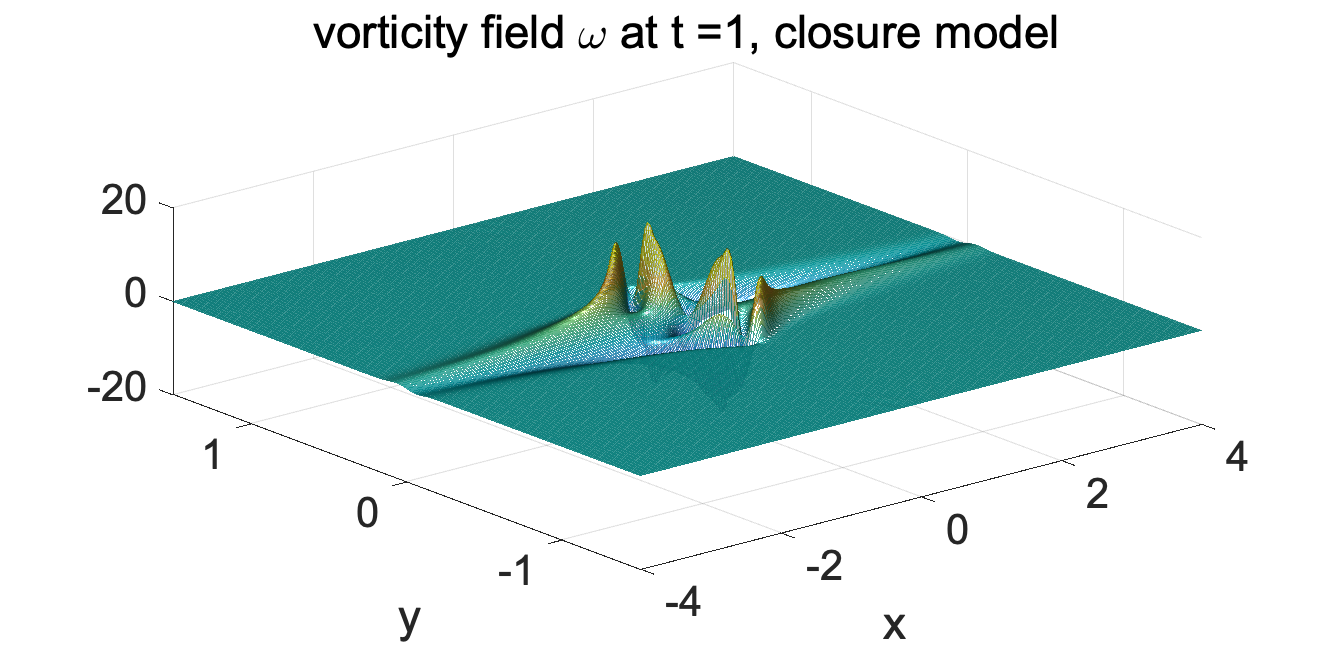}}
	
	\caption{Comparison of the macroscopic density and vorticity field from the
		full and closure model using the double well potential.\protect\label{fig:Comparison-vort-quad}}
\end{figure}

In Figure \ref{fig:Comparison-vort-fene} and \ref{fig:Comparison-vort-quad},
we compare the numerical results using the two potential
functions. We take a relatively strong potential flow strength $\kappa=2$
and $\gamma=0.5$ so that the the potential flow $u=\left(\kappa x,-\kappa y\right)$
is taking a dominant role in the starting time. The density, as well as the induced vorticity
field, then will be compressed along the $y$ direction and be extended
along the $x$ direction from the potential flow advection. In both test
cases, the finite-amplitude vorticity field is developed in time but remains in bounded amplitude.
The approximation model with the asymptotic stresses from the microscopic
potential gives accurate recovery for the macroscopic states in both density
and vorticity functions. Especially, it again shows that the asymptotic approximation
works even with relatively large values of $\gamma$ using the high-order
expansion. 
In addition, comparing the two potential functions, it is observed that
the FENE model with a centered symmetric microscopic potential will induce the vorticity flow in a smaller amplitude
compared to the double well potential under the same value of $\gamma$
and $\kappa$. This is consistent with the scaling law of the system found according to the leading-order term in the microscopic stress \eqref{eq:tensor_asympt}.

\begin{figure}
	\includegraphics[scale=0.32]{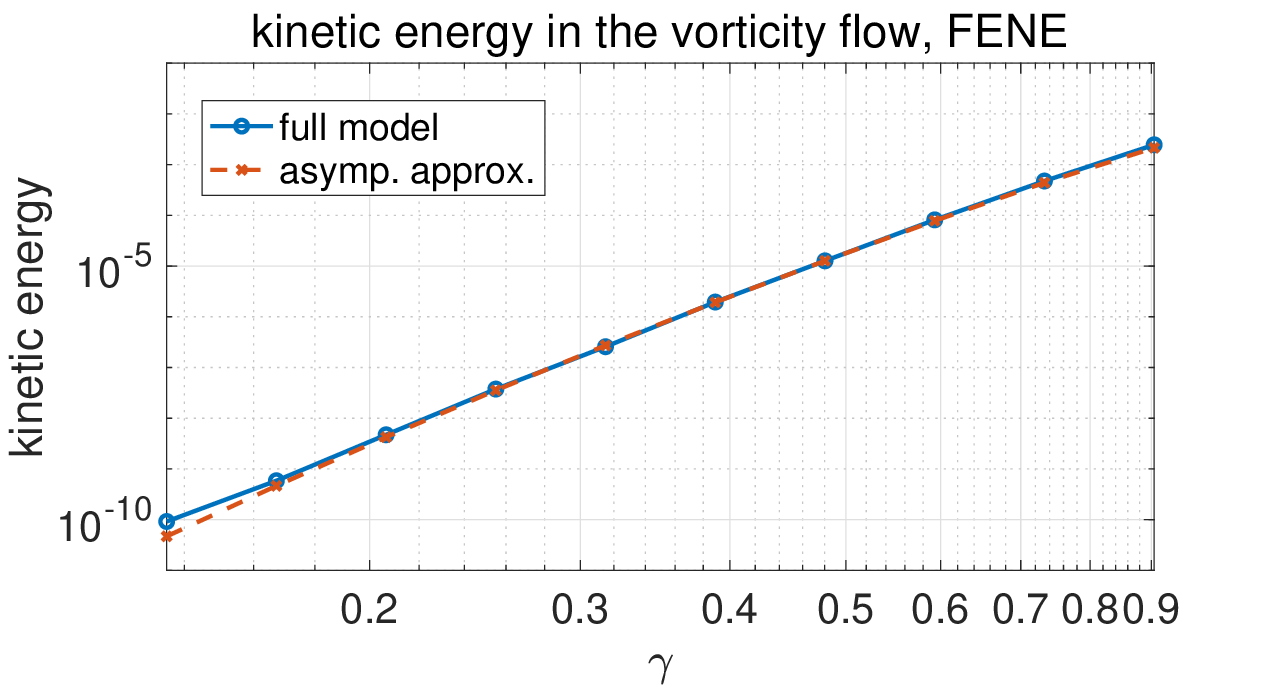}\includegraphics[scale=0.32]{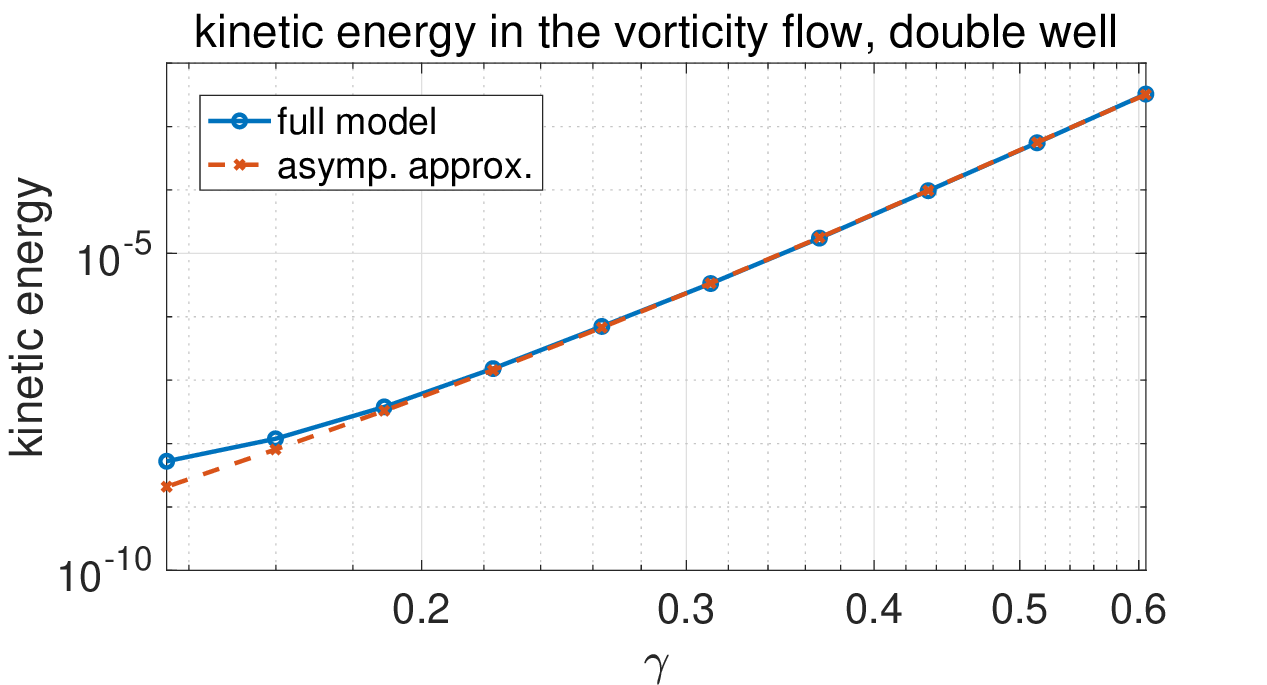}
	
	\caption{Total kinetic energy in the induced vorticity flow field $v$
		under two different potential functions.\protect\label{fig:Total-kinetic-energy}}
	
\end{figure}

Similarly, we show a more quantitative comparison of the total kinetic energy, $\mathrm{KE}=\int |v|^2\mathrm{d}x$, in the induced vorticity field with different values of $\gamma$ in Figure~\ref{fig:Total-kinetic-energy}. Again, the results confirm that the asymptotic closure models produce the same kinetic energy in solutions of the fluctuation flow field consistent up to large values of $\gamma$. 
Notice that in the FENE potential 
with the global minimum at the origin, the stress term has a higher
order feedback $O\left(\lambda\gamma^{6}\right)$, while the double
well potential with two critical points of minima gives a stress term
in the lower order $O\left(\lambda\gamma^{4}\right)$. This scaling law is also observed in the numerical results, where the FENE potential shows a slower growth with respect to $\gamma$ in comparison to the double well potential case in the fluctuation flow energy.
In addition, we would like to point out that the deviation at the small values of $\gamma$ is in fact not because the closure models are not accurate, but due to the errors in direct MC simulations where the equations become stiff in the regime with very large values of $D$. 

\subsection{Simulations of fully coupled micro-macro systems}\label{subsec:flow-test2}
\begin{figure}
	\includegraphics[scale=0.33]{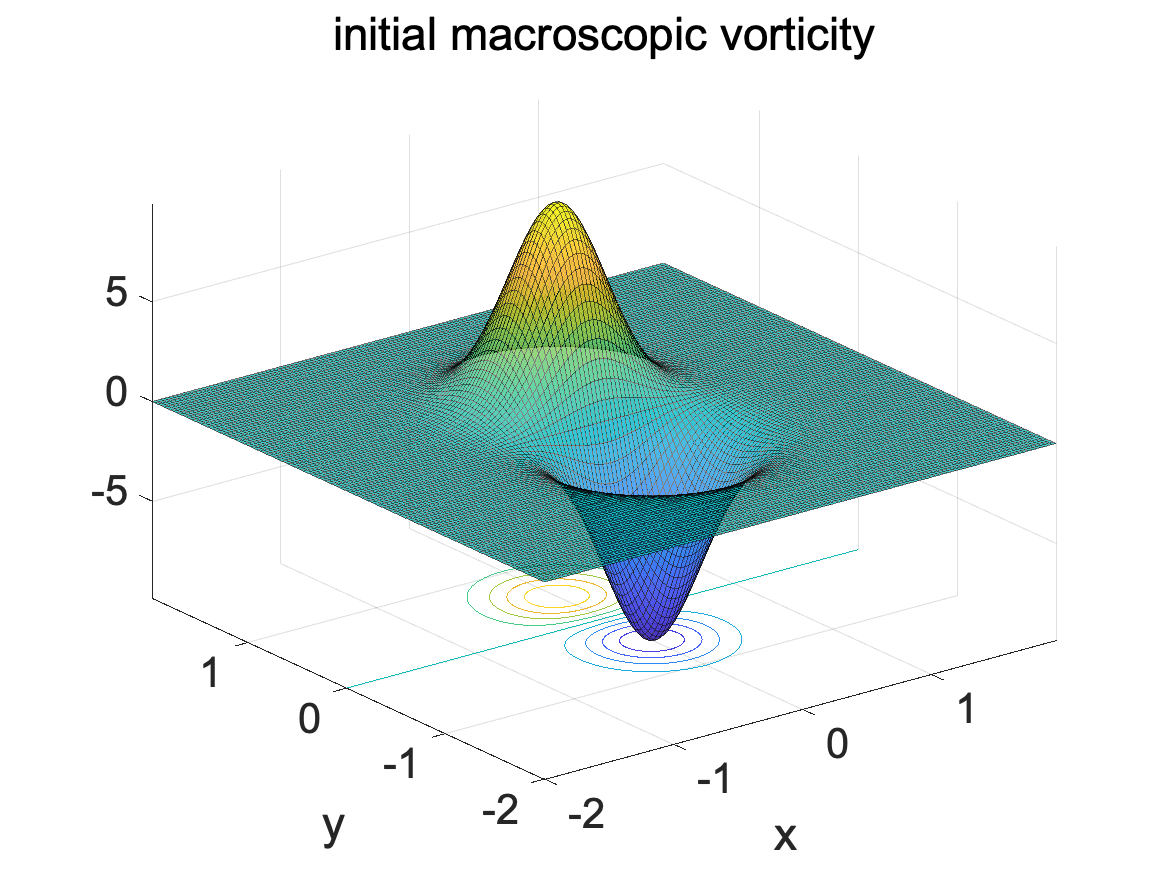}\includegraphics[scale=0.33]{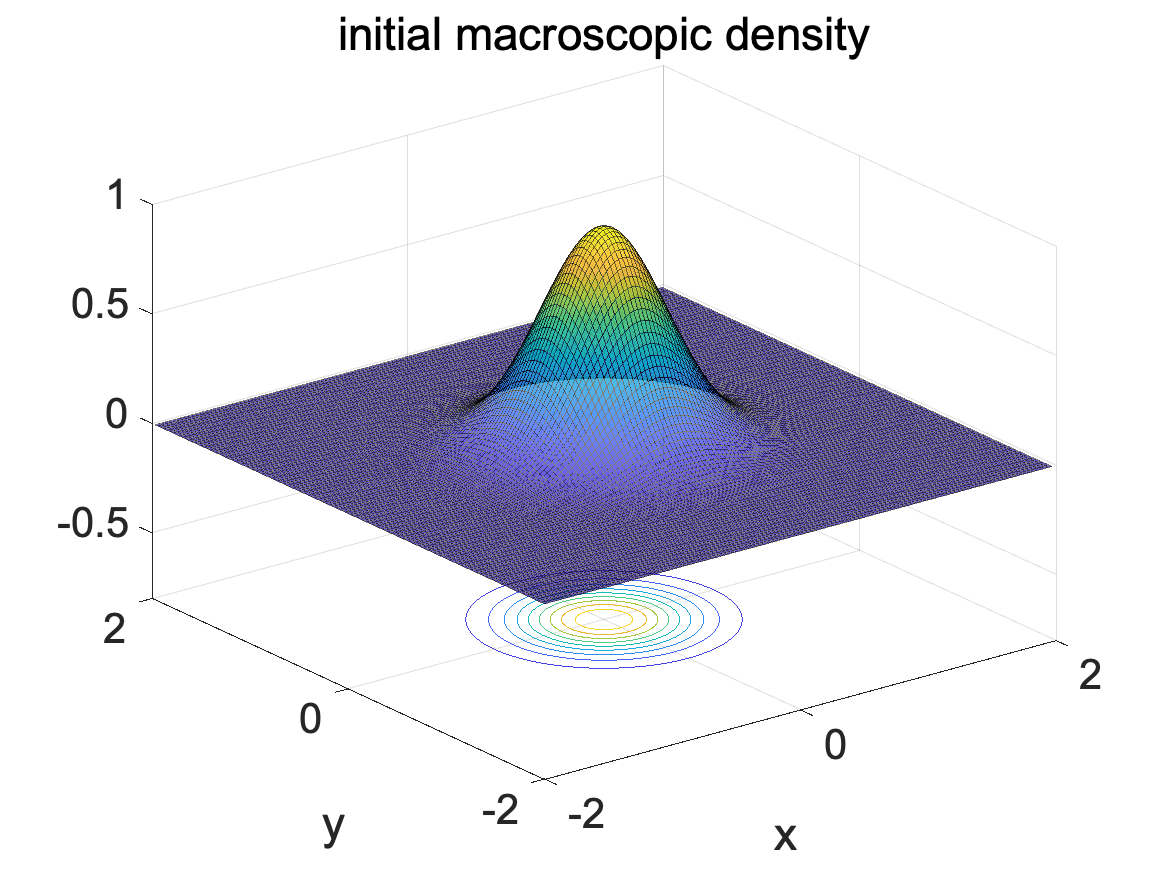}
	
	\caption{Initial states of the flow vorticity field and non-uniform macroscopic
		density.}\label{fig:initial-states}
\end{figure}

In the final test case, we show the performance of the asymptotic closure models on the fully coupled micro-macro system \eqref{eq:micro-macro} with the two-way interactions between the full density function $f(x,q,t)$ and the macroscopic flow field $u(x,t)$. In particular, we adopt the FENE potential satisfying the relation \eqref{eq:micro-potential} so that we have the equation for the flow vorticity field $\omega=\left(\nabla\times u\right)\cdot\hat{k}$ directly coupled with the full density distribution $f\left(x,q,t\right)$ as
\begin{equation}
	\partial_{t}\omega+u\cdot\nabla\omega=\mu\Delta\omega+\lambda\left(\nabla\times\nabla\cdot\tau\right)\cdot\hat{k}.
	\label{eq:flow-macro-full}
\end{equation}
Above, we consider the incompressible flow $u$ with the microscopic elastic stress as in \eqref{eq:macro-u}. The density field will be inversely subject to the macroscopic flow advection and deformation as in \eqref{eq:micro-f}. 
In addition, we will consider two initial distributions for the macroscopic density $N\left(x\right)=\int f\left(x,q,t\right)\mathrm{d}q$: i) uniform density $N\equiv 1$; and ii) a non-uniform initial density. The first case focuses on the impact of the microscopic stress, while the second case provides the most complicated test with combined contributions from both micro and macro interactions. We adopt the classic Lamb–Chaplygin dipole in \cite{batchelor2000introduction} as the initial state. The initial vorticity field and the non-uniform initial macro density are depicted in Figure~\ref{fig:initial-states}. 

The true solution is generated from the full density model through a direct MC simulation using $M=5\times 10^4$ samples by running independent ensemble simulation for $\left\{ Q^{i}\left(x,t\right)\right\} _{i=1}^{M}$
\begin{equation}
	\mathrm{d}Q^{i}+u\cdot\nabla Q^{i}\mathrm{d}t=\left(\nabla u Q^{i}-D\nabla_{q}U\right)\mathrm{d}t+\sqrt{2D}\gamma\mathrm{d} B_{t}^{i}.\label{eq:MC-micro}
\end{equation}
This is in general a very computationally demanding task to sufficiently sampling the full spatial and temporal space $(x,t)\in[-L,L]^2\times[0,T]$ using the Lagrangian samples $Q^i\left(x,t\right)\in\mathbb{R}^2$. The microscopic stress $\tau$ is then directly computed from the samples as
\begin{equation}
	\tau\left(x,t\right)=\int f\nabla_{q}U\otimes q\mathrm{d}q\approx\frac{N\left(x,t\right)}{M}\sum_{i=1}^{M}\nabla_{q}U\left(Q^{i}\right)\otimes Q^{i}.\label{eq:stress-MC}
\end{equation}
Above, the density function will be estimated from the empirical ensemble distribution, that is, $f\left(x,q,t\right)=\frac{N\left(x,t\right)}{M}\sum_{i}\delta\left(q-Q^{i}\left(x,t\right)\right)$. Especially, the total stress forcing $\nabla\times\nabla\cdot\tau$ in the macroscopic flow equation \eqref{eq:flow-macro-full} requires the second derivatives of $\tau$, thus the estimated ensemble averaged stress in \eqref{eq:stress-MC} needs to be smoothed under a proper Gaussian kernel. This makes it easy to miss the crucial small-scale structures when multiscale dynamics becomes dominant thus usually leads to large computational errors.

\begin{figure}
	\subfloat{\includegraphics[scale=0.3]{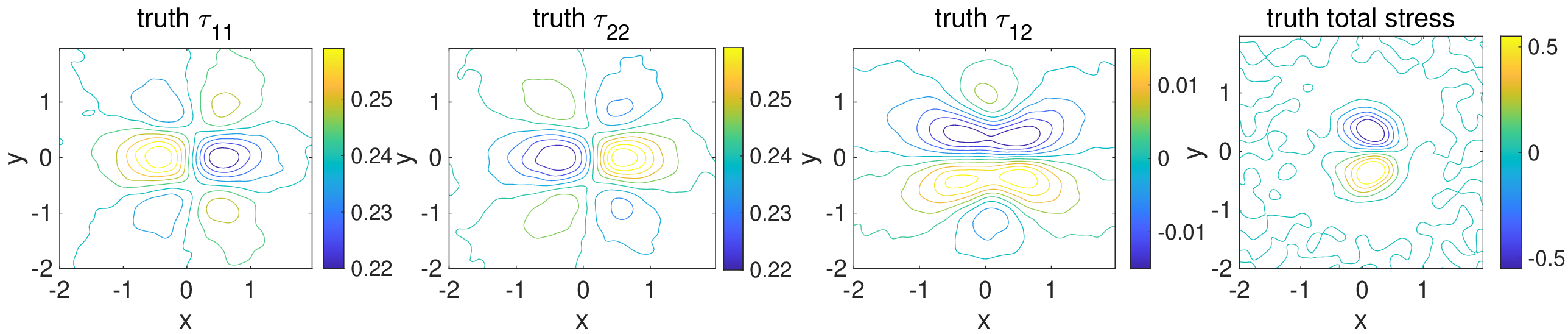}}
	
	\subfloat{\includegraphics[scale=0.3]{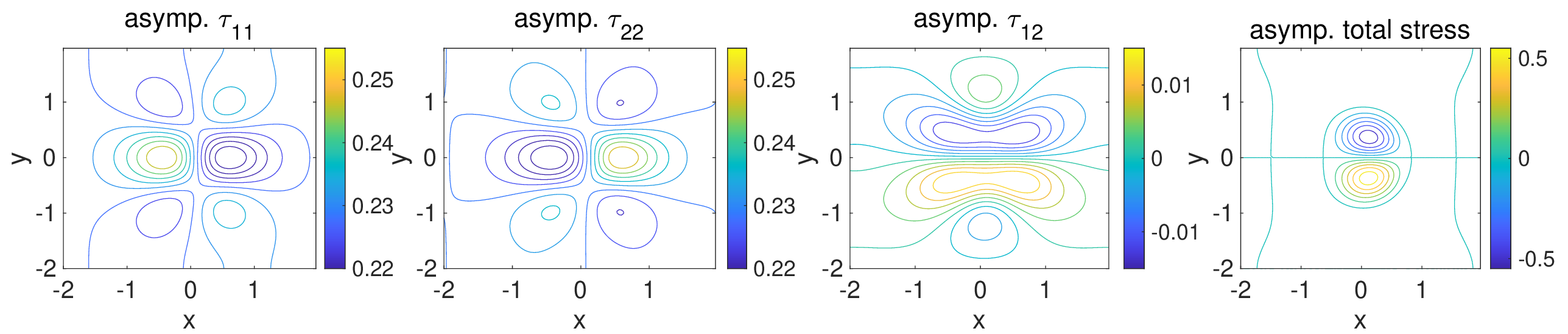}}
	
	\caption{Comparison of the stress terms at $t=0.5$ with a constant macroscopic
		density $N\equiv1$.\protect\label{fig:Comp-stress-const}}
	
\end{figure}

\begin{figure}
	\includegraphics[scale=0.26]{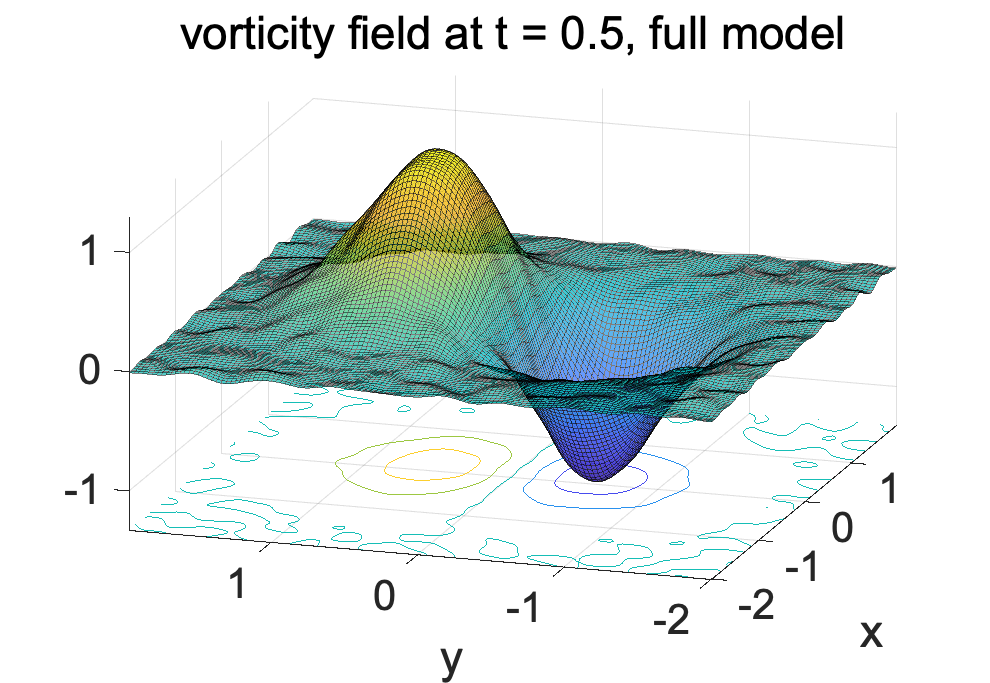}\includegraphics[scale=0.26]{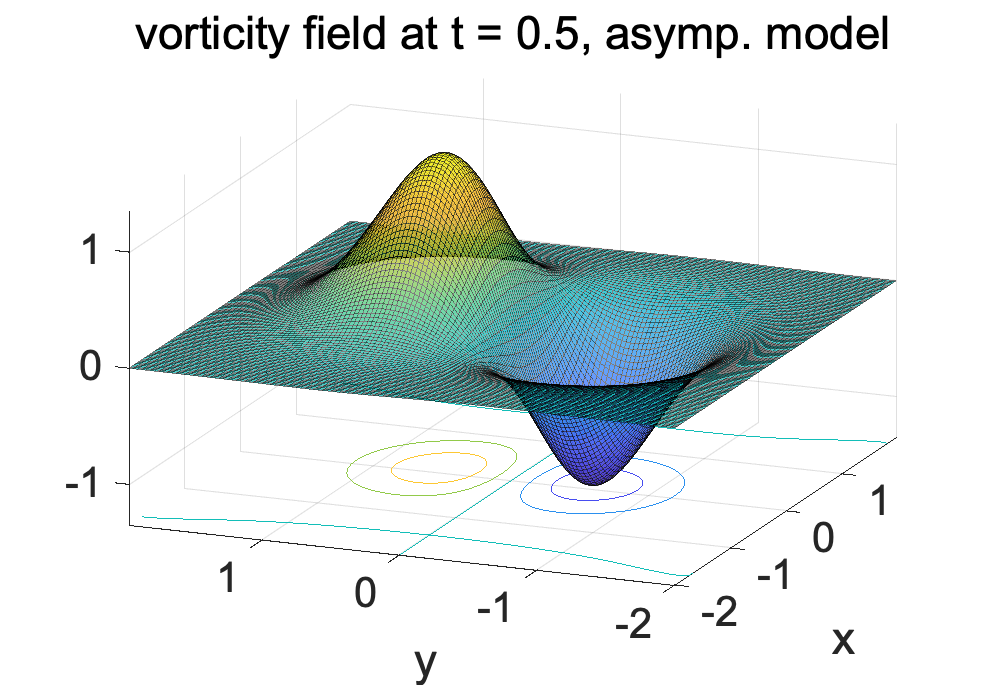}\includegraphics[scale=0.25]{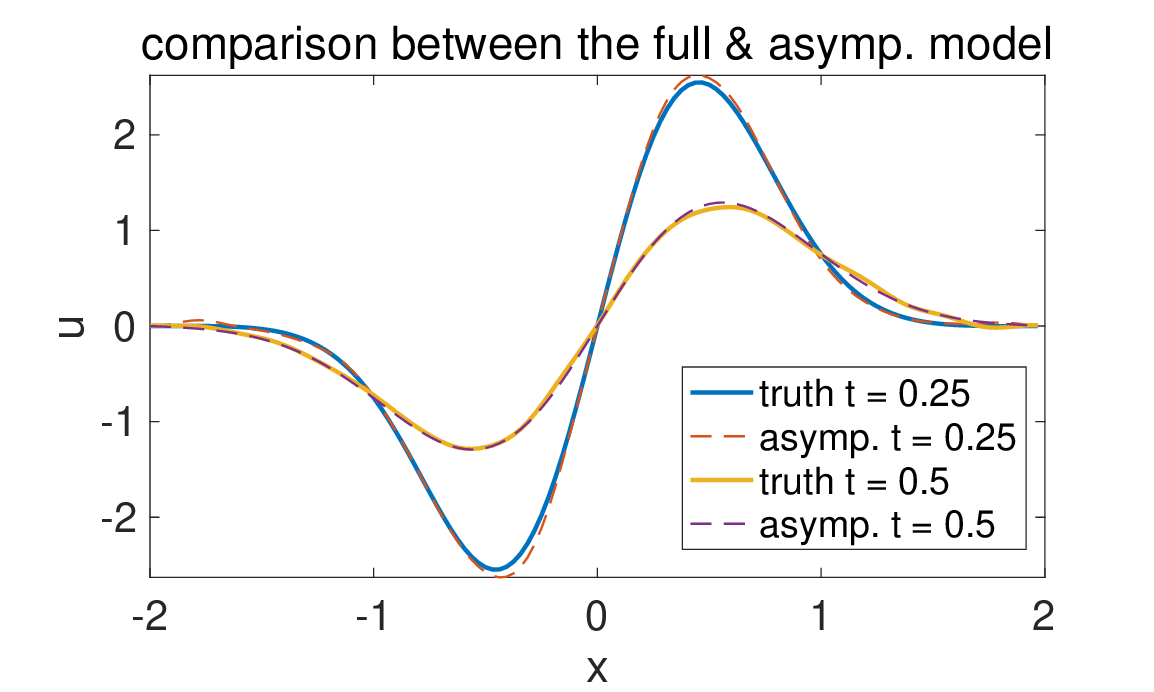}
	
	\caption{Comparison of the vorticity field with a constant macroscopic density
		$N\equiv1$. The right figure compares the cross-section of the solutions at $y=0$.\protect\label{fig:Comp-vort-const}}
	
\end{figure}

In the first test, we consider a constant macroscopic density field $N\equiv \mathrm{const.}$ under the FENE potential, then the elastic stress comes purely from the microscopic density distribution. The stress tensor and the total stress term from the direct MC simulation and the asymptotic model are compared in Figure~\ref{fig:Comp-stress-const}. The true stress from an ensemble estimate remains very noisy using even the very large sample size $M=5\times10^4$ and the results are already smoothed using a carefully tuned Gaussian kernel. In contrast, the asymptotic model captures the accurate stress solution without the need to run the very expensive large ensemble simulations. The corresponding vorticity flow solutions in the true and asymptotic closure models are compared in Figure~\ref{fig:Comp-vort-const}. Good agreements are observed in the vorticity flow structures. The asymptotic model demonstrates high skill in accurately capture the quantitative  flow structures during the entire time of the vorticity flow evolution.

\begin{figure}
	\subfloat{\includegraphics[scale=0.3]{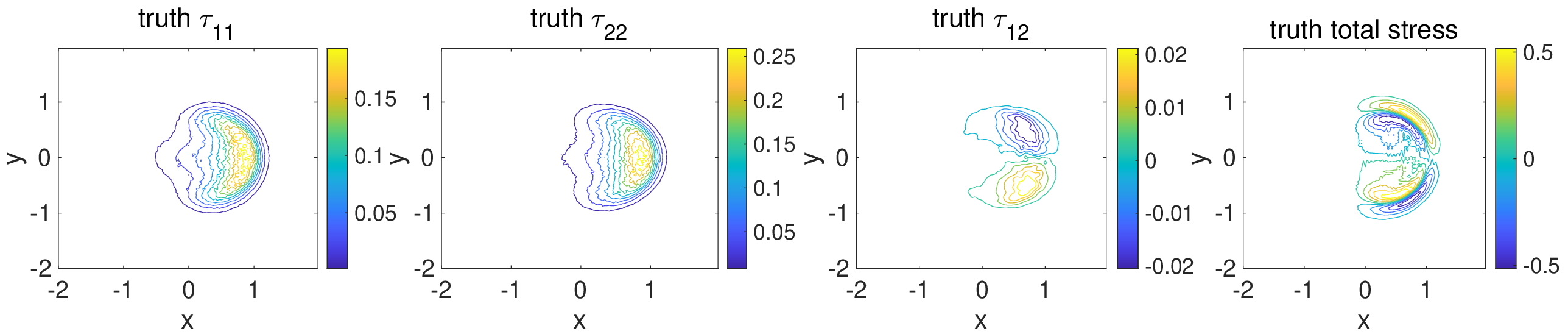}}
	
	\subfloat{\includegraphics[scale=0.3]{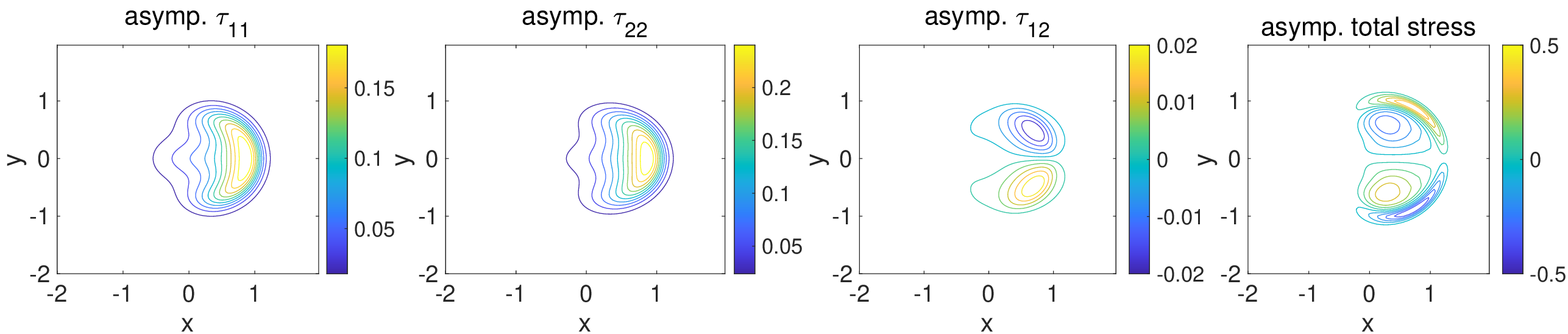}}
	
	\caption{Comparison of the stress terms at $t=0.5$ coupled with dynamical macroscopic
		density $N(x,t)$.\protect\label{fig:Comp-stress-full}}
	
\end{figure}

\begin{figure}
	\subfloat{\includegraphics[scale=0.38]{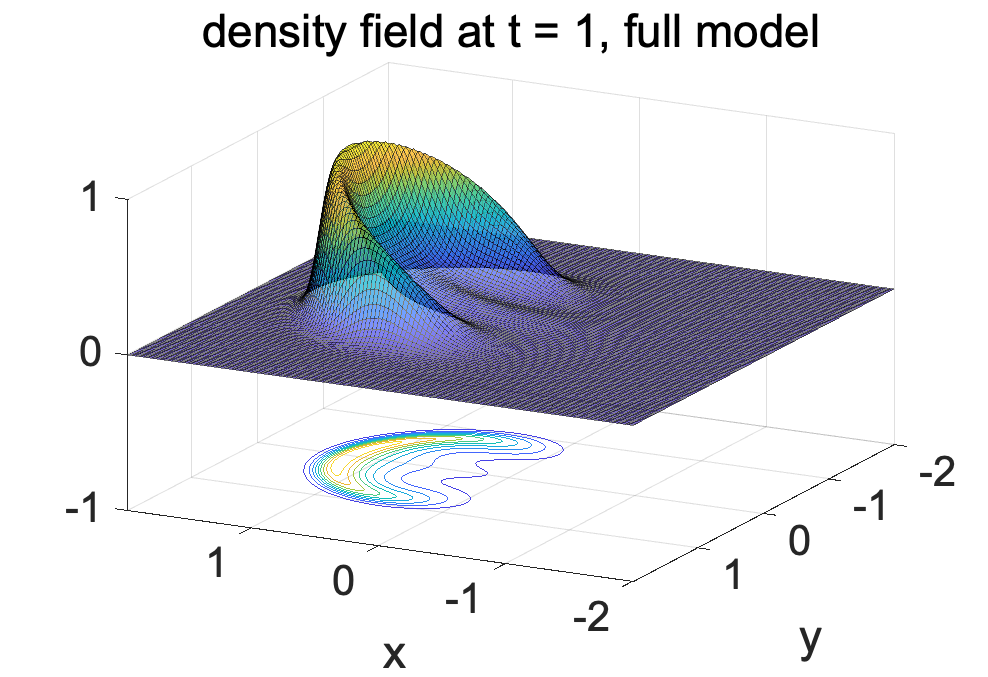}\includegraphics[scale=0.38]{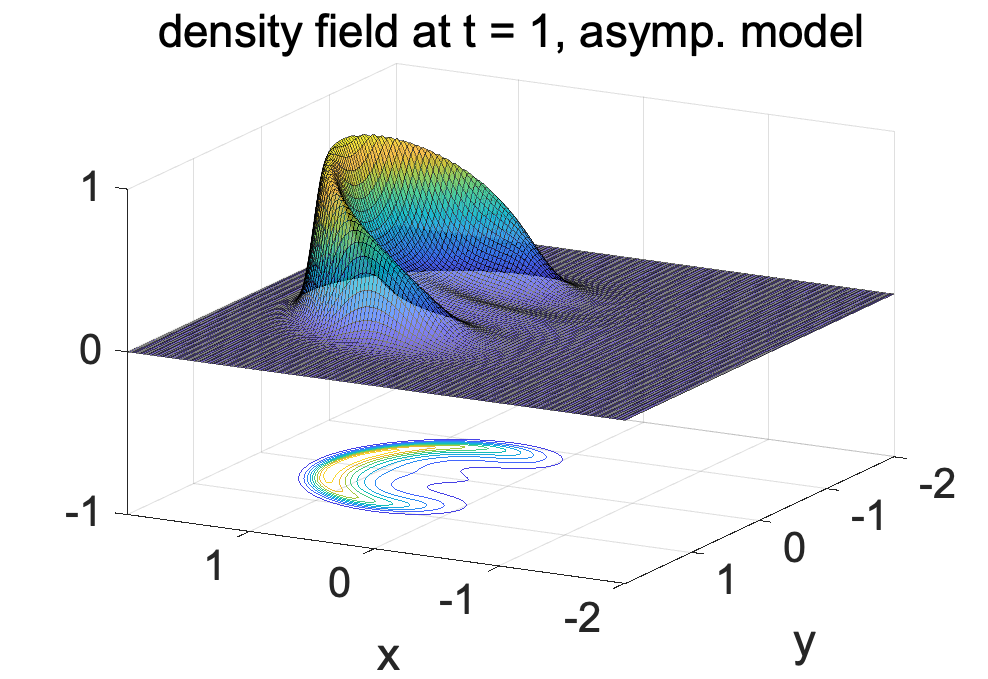}
	}
	
	\subfloat{\includegraphics[scale=0.38]{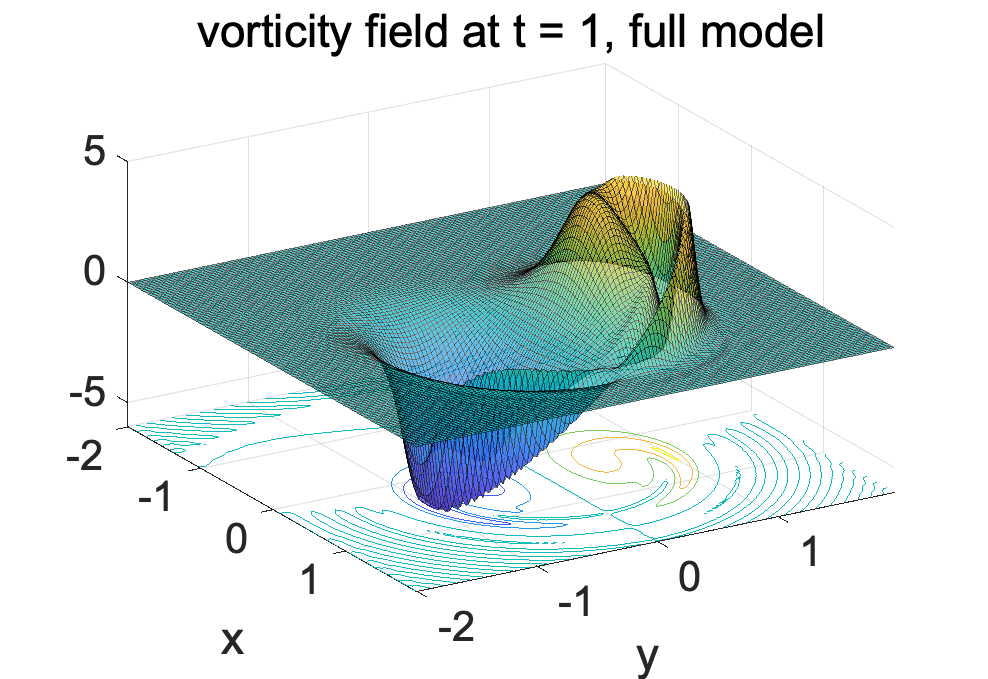}\includegraphics[scale=0.38]{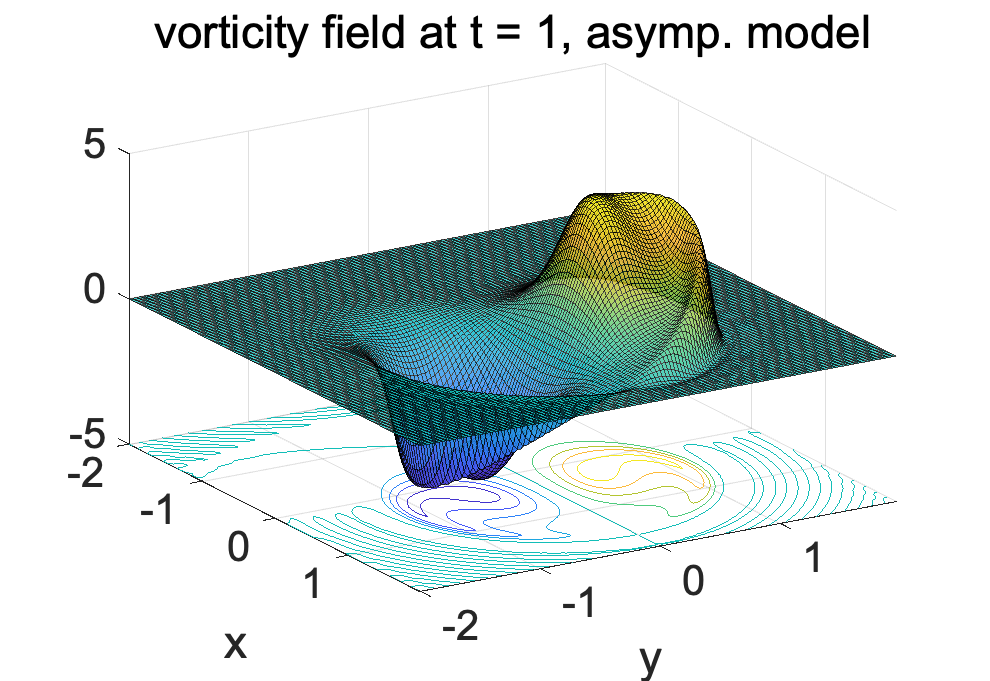}
	}
	
	\caption{Comparison of the vorticity field coupled with a dynamical macroscopic density function.\protect\label{fig:Comp-vort-full}}
\end{figure}

In the final test case, we consider the fully coupled model with also an evolution macroscopic density field for $N\left(x,t\right)$ transported by the flow field, 
$\partial_{t}N+u\cdot\nabla N=0$.  
A non-uniform initial density in Figure~\ref{fig:initial-states} is used in this case. As a result, the coupled model demonstrates more complicated multiscale structures. Similar to the previous uniform macro-density case, we compare the stress functions and the macroscopic flow and density fields captured by the full MC simulation with a large ensemble size and the efficient asymptotic model in Figure~\ref{fig:Comp-stress-full} and \ref{fig:Comp-vort-full} respectively. It can be seen that the macroscopic density is transported by the flow field which develops more complicated structures in the stress functions. Similar to the previous test case, the microscopic stresses become very noisy and need to be smoothed in the direct MC simulation, while the asymptotic model can accurately capture the structures with nice smooth solutions. Correspondingly, the macroscopic density and vorticity field solutions can be also precisely captured under the efficient asymptotic model. In this most complicated computational test, a larger difference appears in the vorticity flow solution between the full and asymptotic model. We suspect that this may due to the inaccurate ensemble estimate in the full model that is sensitive to small perturbations.

\section{Conclusion and future directions}\label{sec:conclusion}
In this paper, we present a new asymptotic closure approach for modeling multiscale viscoelastic fluids. The main ideas are based on the energetic variational formulation of the general micro-macro models \cite{giga2017variational} and the separation of scales in the microscopic density function as a Gibbs distribution near its fast relaxation equilibrium \cite{lin2007micro}. Efficient computational strategies are derived from the reduced-order modeling framework maintaining the detailed information in the microscopic structures from the kinetic theory. Especially, our new asymptotic closure models enjoy the  distinguished advantages of achieving a flexible high-order asymptotic expression for efficient computational implementation and maintaining a precise energy-dissipation equation for structure preserving properties. 

The performance of the new modeling strategy is validated on a series of detailed numerical experiments. It shows that both the multiscale density function and the macroscopic fluid field under different microscopic structures can be efficiently captured by the asymptotic approach, and the accuracy can be further improved by adding additional high-order terms. In particular, the numerical results show that the asymptotic representation maintains accuracy even up to large values of the scaling parameter $\gamma\sim 1$ far beyond the near-equilibrium regime at the fast relaxation limit. This implies  wider potential applications of the proposed strategy away from the fast relaxation region where the closure models are derived.\medskip

\noindent \emph{Future research directions.}\\
In the immediate follow-up research, the new asymptotic closure models can be directly applied to assist the efficient computation of fluid solutions involving  more complex microscopic structures and more general macroscopic fluid dynamics such as the lattice materials developed recently in \cite{silverberg2015origami,li2025nonlinear,li2023some}. Such materials can be created when sufficient chemical bridges are made between the entangled polymer chains, establishing richer properties intermediately between liquid and solid materials.
A new microscopic tensor $q \in \mathbb{R}^{n\times n}_{\text{sym}}$ is introduced, which indicates the effective deformation gradient along different directions of the lattice structure.  Following the same energetic variational approach according to the new energy-dissipation law, the macroscopic velocity equation can be derived according to the microscopic potential $U$ on the lattice
	\begin{equation}
		\partial_{t}u+u\cdot\nabla u+\nabla p=\eta\Delta u +\lambda\nabla_{x}\cdot\int_{\Omega_{q}}f\left[\left(\nabla_{q}U\right)q^{T}+\left(\nabla_{q}U\right)^{T}q\right]\mathrm{d}q.  \label{eq:pde_incompr}
	\end{equation}
This new model requires resolving the density function in an even higher-dimensional configuration space and dealing with microscopic potential demonstrating degenerating structures. We plan to have a detailed computational study of the lattice model \eqref{eq:pde_incompr} using the efficient new asymptotic formula.

In addition, the associated energy-dissipation equation of the asymptotic models implies the possibility of a detailed analytical study of the theoretical properties for the new approach. This will include the well-posedness of the closure models involving different levels of asymptotic truncations following the line of arguments in \cite{lin2007micro,yu2005micro}. Furthermore, the multiscale coupling structure of the micro-macro model fits into the modeling framework in \cite{qi2024coupled,qi2025data} for data assimilation schemes. It is interesting to investigate whether the new modeling equations can be combined with data assimilation approaches to further improve the prediction accuracy and recover unresolved microscopic structures.

%


%
\section*{Acknowledge}
The research of C.L. is partially supported by NSF grants DMS-2216926 and DMS-2410742. The research of D.Q. is partially supported by ONR grant N00014-24-1-2192 and NSF grant DMS-2407361.
%
%
%
%
%

\appendix
\section{Detailed derivations of the results in the main text}\label{sec:appen}
Here, we include the detailed derivations and computations of the results in the main text.

\subsection{Asymptotic expansion for the multiscale density function}\label{subsec:asymp-deri}
By substituting \eqref{eq:ansatz-spring} into \eqref{eq:asymp_den_spring}, we obtain with the scaling relation $D\gamma^4=c$
\begin{align}
	-\gamma^{-2} \overline{f}\nabla_qU \cdot (\nabla u q) +  \Big(\partial_t \overline{f} + 2(\nabla \cdot u) \overline{f} + u \cdot \nabla_x \overline{f}\Big) +\gamma^2 (\nabla u q)\cdot\nabla_{q}\overline{f}_1 \\ = -\gamma^{-2}c\nabla_q U \cdot \nabla_q \overline{f}_1 +  c\Delta_q \overline{f}_1. \label{eq:asymp-f0}
\end{align}
We seek the solution by comparing each order of the asymptotic dynamics \eqref{eq:asymp-f0}.

First, the leading order $O(\gamma^{-2})$ term in \eqref{eq:asymp-f0} gives
\begin{equation}
	(\nabla_q U) \cdot f_0(\nabla u q) = c(\nabla_q U) \cdot \nabla_q f_1, \label{eq:consistency-f0}
\end{equation}
which is given by the first order equation of \eqref{eq:f_order} with $n=1$. Notice that the condition i) $ f_0(\nabla u q) = c\nabla_q f_1,$ in \eqref{eq:f_order1} requires $\nabla u=(\nabla u)^T$, that is, $u$ needs to be a potential flow. In fact, the gradient field on the right hand side of \eqref{eq:consistency-f0} requires $\nabla u q$ to satisfy a curl-free condition with respect to $q$ so that
\begin{equation*}
	\frac{\partial u_i}{\partial x_j}  = \frac{\partial}{\partial q_j} \left(\sum_l \frac{\partial u_i}{\partial x_l} q_l\right) = \frac{\partial}{\partial q_i} \left(\sum_l \frac{\partial u_j}{\partial x_l}  q_l\right) =\frac{\partial u_j}{\partial x_i} .
\end{equation*}
A special explicit solution to the relations \eqref{eq:f_order1} then can be found in the following expression
\begin{equation}
	f_1\left(x,q,t\right) = \frac{1}{2c}f_0\left(x,t\right)\left[q^{T}\nabla u\left(x,t\right)q\right].\label{eq:special_f0}
\end{equation}
This can be checked by directly substituting \eqref{eq:special_f0} into the leading-order equation \eqref{eq:consistency-f0}.

Then, the going to the next order term $O\left(1\right)$ reaches the dynamical
equation for $f_0$
\begin{equation}
	\begin{aligned}
		\partial_t f_0 + \nabla \cdot (uf_0) = \partial_{t}f_{0}+\nabla\cdot uf_{0}+u\cdot\nabla_{}f_{0} &=-\nabla\cdot uf_{0} + c\Delta_{q}f_{1}\\
		& +\nabla_{q}U\cdot\left(\nabla uq\right) f_{1} -c\nabla_{q}U\cdot\nabla_{q}f_{2}.
	\end{aligned}\label{eq:asymp-f1}
\end{equation}
Similarly, the second line of \eqref{eq:asymp-f1} becomes zero using the second-order relation in \eqref{eq:f_order} with $n=1$. The first line leads to the governing dynamics for $f_{0}\left(x,t\right)$ as
\begin{equation}
	\partial_{t}f_{0}+\nabla\cdot \left(uf_{0}\right)=0,\label{eq:asymp2}
\end{equation}
where we use the relation \eqref{eq:consistency-f0} for the $f_1$ such that
\[
\nabla_q \cdot(c\nabla_q f_1) = \nabla_q\cdot(\nabla uq)f_0=(\nabla\cdot u) f_0.
\]
And the solution for $f_2\left(x,q,t\right)$ can be solved in a similar fashion with the explicit special solution as
\begin{equation}
	f_2\left(x,q,t\right) = \frac{1}{8c^2}f_0\left(x,t\right)\left[q^{T}\nabla u\left(x,t\right)q\right]^2.\label{eq:special_f1}
\end{equation}
The expression for $f_3\left(x,q,t\right)=\frac{\gamma^{6}}{48c^{3}}f_0\left[q^{T}\nabla u\left(x,t\right)q\right]^{3}$ follows in the same way from the next order equation.

Furthermore, the asymptotic expansion in \eqref{eq:density_asympt} can be continuously extended to higher orders. We may  consider matching the next order term $O\left(\gamma^2\right)$ satisfying the following equation
	equation for $f_0$
	\begin{equation}
		\begin{aligned}\partial_{t}f_{1}+2\nabla\cdot uf_{1}+u\cdot\nabla_{x}f_{1}= & \left(\nabla uq\right)\cdot\nabla_{q}f_{1}+c\Delta_{q}f_{2}\\
			& +\nabla_{q}U\cdot\left(\nabla uq\right)f_{2}-c\nabla_{q}U\cdot\nabla_{q}f_{3}
		\end{aligned}
		\label{eq:asymp-f2}
	\end{equation}
	Then, we seek the special solution of $f_3$ in the following decomposition
	\begin{equation}
		f_3\left(x,q,t\right)=\frac{\gamma^{6}}{48c^{3}}f_0\left[q^{T}\nabla u\left(x,t\right)q\right]^{3}+\tilde{f}_3\left(x,q,t\right).\label{eq:special_f2}
	\end{equation}
	Under the same strategy, the first term on the right hand side of \eqref{eq:special_f2} cancels the second line of the equation \eqref{eq:asymp-f2}, while the second term acts as the additional balancing effects to the first line of \eqref{eq:asymp-f2}. Substituting the special solutions \eqref{eq:density_asympt} and \eqref{eq:special_f2} to the next order equation \eqref{eq:asymp-f2} yields the relation
	\begin{equation}
		\partial_{t}\left(q^{T}\nabla uq\right)+u\cdot\nabla_{x}\left(q^{T}\nabla uq\right)+\frac{1}{2}q^{T}\left(\nabla u-\nabla u^{T}\right)\left(\nabla u+\nabla u^{T}\right)q=-\frac{2c^{2}}{f_{0}}\nabla_{q}U\cdot\nabla_{q}\tilde{f}_{3}.
	\end{equation}
	In addition, if we introduce the additional ansatz $\nabla_q U=\Psi(q)q$ and $\nabla_q \tilde{f}_3=\Psi(q)^{-1}K(x,t)q$, we find the explicit expression for the additional correction term to the order term $O(\gamma^6)$
	\begin{equation}
		\begin{aligned}\nabla_{q}U\cdot\nabla_{q}\tilde{f}_{3} & =\left(q^{T}\Psi\right)\Psi^{-1}Kq=q^{T}Kq,\\
			K\left(x,t\right) & =-\frac{f_{0}}{2c^{2}}\left[\partial_{t}\left(\nabla u\right)+u\cdot\nabla\left(\nabla u\right)+\frac{1}{2}\left(\nabla u-\nabla u^{T}\right)\left(\nabla u+\nabla u^{T}\right)\right].
		\end{aligned}
	\end{equation}
	Notice that in \eqref{eq:stress_asymp}, we only need $\nabla_{q}f_{3}\otimes q=\Psi^{-1}Kq\otimes q$ to compute the elastic stress $\tau$.

\subsection{Asymptotic expansion for the microscopic energy}\label{subsec:ene-variation}

\subsubsection{Closure models recovered from the energetic variational principle}\label{subsubsec:ene-vari-deri}
Here, we show that the closure models \eqref{eq:closure1} and \eqref{eq:closure2} can be derived from the energetic variational principle with respect to the total energy equation \eqref{eq:ene_asymp_leading}. In fact, we just need to apply the LAP and MDP to the corresponding free and dissipation energy due to the microscopic effects
\begin{equation}
	\begin{aligned}\mathcal{F}= & \frac{\tilde{\lambda}}{2c}\int_{\Omega_{x}}f_{0}\log\frac{f_{0}}{m_{0}}\left(\nabla u:\tilde{M}_{2}\right)\mathrm{d}x,\\
		\mathcal{D}_{2}= & \frac{\tilde{\lambda}}{8c}\int_{\Omega_{x}}f_{0}\mathrm{tr}\left[\left(\nabla u+\nabla u^{T}\right)\tilde{M}_{2}\left(\nabla u+\nabla u^{T}\right)\right]\mathrm{d}x.
	\end{aligned}
\end{equation}

First, the forcing can be recovered by taking the variation of the dissipation through (MDP), that is,
\begin{align*}
	\delta_{u}\mathcal{D}=&	\frac{\tilde{\lambda}}{8c}\delta_{u}\int f_{0}\mathrm{tr}\left[\left(\nabla u+\nabla u^{T}\right)\tilde{M}_{2}\left(\nabla u+\nabla u^{T}\right)\right]\mathrm{d}x\\
	=&	\frac{\tilde{\lambda}}{8c}\sum_{i,m,n}\left[\int f_{0}\left(\frac{\partial\delta u_{i}}{\partial x_{m}}+\frac{\partial\delta u_{m}}{\partial x_{i}}\right)\tilde{M}_{2}^{mn}\left(\frac{\partial u_{n}}{\partial x_{i}}+\frac{\partial u_{i}}{\partial x_{n}}\right)\mathrm{d}x\right.\\
	&+\left.\int f_{0}\left(\frac{\partial u_{i}}{\partial x_{m}}+\frac{\partial u_{m}}{\partial x_{i}}\right)\tilde{M}_{2}^{mn}\left(\frac{\partial\delta u_{n}}{\partial x_{i}}+\frac{\partial\delta u_{i}}{\partial x_{n}}\right)\mathrm{d}x\right]\\
	=	&-\frac{\tilde{\lambda}}{8c}\left\{ \int\delta u^{T}\nabla\cdot\left[f_{0}\left(\nabla u+\nabla u^{T}\right)\tilde{M}_{2}^{T}\right]\mathrm{d}x+\int\delta u^{T}\nabla\cdot\left[f_{0}\tilde{M}_{2}\left(\nabla u+\nabla u^{T}\right)\right]\mathrm{d}x\right.\\
	&+\left.\int\delta u^{T}\nabla\cdot\left[f_{0}\tilde{M}_{2}^{T}\left(\nabla u+\nabla u^{T}\right)\right]\mathrm{d}x+\int\delta u^{T}\nabla\cdot\left[f_{0}\left(\nabla u+\nabla u^{T}\right)\tilde{M}_{2}\right]\mathrm{d}x\right\} \\
	=	&-\frac{\tilde{\lambda}}{2c}\left\langle \nabla\cdot\left[f_{0}\left(\nabla u+\nabla u^{T}\right)\right]\tilde{M}_{2},\delta u\right\rangle _{L_{x}^{2}}.
\end{align*}
Above, in the last identity, we use the fact that $\tilde{M}_2$ is a symmetric matrix from the second moments. Therefore, the forcing from the elastic stress is recovered as
\begin{equation}
	\tau=-\frac{\delta\mathcal{D}}{\delta u}=\frac{\tilde{\lambda}}{2c}\nabla\cdot\left[f_{0}\left(\nabla u+\nabla u^{T}\right)\right]\tilde{M}_{2}.
\end{equation}

Then, the LAP computes the variation of the free energy under the Lagrangian coordinate as
\begin{align*}
	\delta_{x}\int_{0}^{T}\mathcal{F}\mathrm{d}t	=&\frac{\tilde{\lambda}}{2c}\delta_{x}\int_{0}^{T}\int f_{0}\log\frac{f_{0}}{m_{0}}\left(\nabla u:\tilde{M}_{2}\right)\mathrm{d}x\mathrm{d}t\\
	=&\frac{\tilde{\lambda}}{2c}\delta_{x}\int_{0}^{T}\sum_{i,j,l}\int\frac{f_{0}\left(X,0\right)}{\det F}\log\frac{f_{0}\left(X,0\right)}{m_{0}\det F}\left(\frac{\partial\dot{x}_{i}}{\partial X_{l}}\frac{\partial X_{l}}{\partial x_{j}}\tilde{M}_{2}^{ij}\right)\det F\mathrm{d}X\mathrm{d}t\\
	=&\frac{\tilde{\lambda}}{2c}\int_{0}^{T}\sum_{i,j,l}\int f_{0}\left(X,0\right)\log\frac{f_{0}\left(X,0\right)}{m_{0}}\delta_{x}\left[\partial_{t}F_{il}\left(F^{-1}\right)_{lj}\right]\tilde{M}_{2}^{ij}\mathrm{d}X\mathrm{d}t\\
	=&\frac{\tilde{\lambda}}{2c}\int_{0}^{T}\int f_{0}\left(X,0\right)\log\frac{f_{0}\left(X,0\right)}{m_{0}}\left[\partial_{t}\left(\delta_{x}F\right)F^{-1}+\partial_{t}F\delta_{x}\left(F^{-1}\right)\right]:\tilde{M}_{2}\mathrm{d}X\mathrm{d}t\\
	=&\frac{\tilde{\lambda}}{2c}\int_{0}^{T}\int f_{0}\left(X,0\right)\log\frac{f_{0}\left(X,0\right)}{m_{0}}\left[-\delta_{x}F\partial_{t}\left(F^{-1}\right)+\partial_{t}F\delta_{x}\left(F^{-1}\right)\right]:\tilde{M}_{2}\mathrm{d}X\mathrm{d}t\displaybreak[3]\\
	=&\frac{\tilde{\lambda}}{2c}\int_{0}^{T}\int f_{0}\left(X,0\right)\log\frac{f_{0}\left(X,0\right)}{m_{0}}\times\\
 &\qquad\qquad\left[\left(\delta_{x}F\right)F^{-1}\left(\partial_{t}F\right)F^{-1}-\left(\partial_{t}F\right)F^{-1}\left(\delta_{x}F\right)F^{-1}\right]:\tilde{M}_{2}\mathrm{d}X\mathrm{d}t\\
	=&\frac{\tilde{\lambda}}{2c}\int_{0}^{T}\int f_{0}\left(X,0\right)\log\frac{f_{0}\left(X,0\right)}{m_{0}}\times\\
    &\qquad\qquad\left[\left(\frac{\partial\delta x}{\partial X}\frac{\partial X}{\partial x}\right)\left(\frac{\partial\dot{x}}{\partial X}\frac{\partial X}{\partial x}\right)-\left(\frac{\partial\dot{x}}{\partial X}\frac{\partial X}{\partial x}\right)\left(\frac{\partial\delta x}{\partial X}\frac{\partial X}{\partial x}\right)\right]:\tilde{M}_{2}\mathrm{d}X\mathrm{d}t\\
	=&\frac{\tilde{\lambda}}{2c}\int_{0}^{T}\int f_{0}\log\frac{f_{0}}{m_{0}}\left(\nabla\delta x\nabla u-\nabla u\nabla\delta x\right):\tilde{M}_{2}\mathrm{d}x\mathrm{d}t\\
	=&\frac{\tilde{\lambda}}{2c}\left\langle \nabla\cdot\left[f_{0}\log\frac{f_{0}}{m_{0}}\left(-\tilde{M}_{2}\nabla u^{T}+\nabla u^{T}\tilde{M}_{2}\right)\right],\delta x\right\rangle _{L_{x,t}^{2}}.
\end{align*}
Above, for simplicity we consider the incompressible flow case with $\det F=1$. In the second equality, we use the definition $u=\dot{x}$ then the chain rule is applied. The third equality uses the definition of $F_{ij}=\frac{\partial x_i}{\partial X_j}$ and its inverse. Then, the derivative of the inverse matrix is used, $\delta (F^{-1})=-F^{-1}\delta FF^{-1}$ and $\partial_t (F^{-1})=-F^{-1}\partial_t FF^{-1}$. Finally, we convert the integration back to the Eulerian coordinate. 
Similarly, noticing that $\tilde{M}_2^{T}=\tilde{M}_2$ is symmetric, we can also get that
\begin{align*}
	\delta_{x}\int_{0}^{T}\mathcal{F}\mathrm{d}t	=&\frac{\tilde{\lambda}}{2c}\delta_{x}\int_{0}^{T}\int f_{0}\log\frac{f_{0}}{m_{0}}\left(\nabla u:\tilde{M}_{2}^{T}\right)\mathrm{d}x\mathrm{d}t\\
	=&\frac{\tilde{\lambda}}{2c}\left\langle \nabla\cdot\left[f_{0}\log\frac{f_{0}}{m_{0}}\left(-\nabla u\tilde{M}_{2}^{T}+\tilde{M}_{2}\nabla u\right)\right],\delta x\right\rangle _{L_{x,t}^{2}}.
\end{align*}
Therefore, we get the additional forcing term due to the free energy from microscopic coupling
\begin{equation}
	\frac{\delta\int_{0}^{T}\mathcal{F}\mathrm{d}t}{\delta x}=\frac{\tilde{\lambda}}{4c}\nabla\cdot\left\{ f_{0}\log\frac{f_{0}}{m_{0}}\left[\tilde{M}_{2}\left(\nabla u-\nabla u^{T}\right)-\left(\nabla u-\nabla u^{T}\right)\tilde{M}_{2}\right]\right\}.\label{eq:symm-variational}
\end{equation}
This implies the two additional structural assumptions indicated in \eqref{eq:flow-potential} and \eqref{eq:micro-potential}. 
In fact, under the leading-order potential flow assumption \eqref{eq:flow-potential}, there is $\nabla u =\nabla u^T$. 
Thus we direct get that the additional forcing in \eqref{eq:symm-variational} becomes zero. 
Furthermore, under the symmetric micro-potential assumption \eqref{eq:flow-potential}, the leading-order expansion in $\tilde{M}_2$ becomes an identity matrix, that is,
\[
\tilde{M}_{2}=\int q\otimes q\frac{e^{-U/\gamma^{2}}}{m_{0}}\approx e^{-U\left(0\right)/\gamma^{2}}\int q\otimes q\frac{e^{-\frac{1}{2}\Psi\left(0\right)\left|q\right|^{2}/\gamma^{2}}}{m_{0}}\approx cI.
\]
The forcing in \eqref{eq:symm-variational} also gets reduced to zero, that is
\[
\left(\delta_{x}FF^{-1}\partial_{t}FF^{-1}-\partial_{t}FF^{-1}\delta_{x}FF^{-1}\right):I=\mathrm{tr}\left(\delta_{x}FF^{-1}\partial_{t}FF^{-1}\right)-\mathrm{tr}\left(\partial_{t}FF^{-1}\delta_{x}FF^{-1}\right)=0.
\]

\subsubsection{Energy equations derived from asymptotic expansion}\label{subsubsec: ene-eqn-deri}
Here, we provide details on the computation of the energy in the closure models. By separating the equilibrium distribution in the density as in \eqref{eq:den-asymp}, we can rewrite the free energy and dissipation energy as
\begin{equation}
	\begin{aligned}\tilde{\mathcal{F}}= & \int_{\Omega_{q}}\left[\gamma^{2}\log f+U\left(q\right)\right]f\mathrm{d}q=\gamma^{2}\int_{\Omega_{q}}\overline{f}\log\overline{f}e^{-U(q)/\gamma^2}\mathrm{d}q,\\
		\tilde{\mathcal{D}}_{2}= & D\int_{\Omega_{q}}\left\vert \nabla_{q}\left(\gamma^{2}\log f+U(q)\right)\right\vert ^{2}f\mathrm{d}q=D\gamma^{4}\int_{\Omega_{q}}\overline{f}\left\vert \nabla_{q}\log\overline{f}\right\vert ^{2}e^{-U(q)/\gamma^2}\mathrm{d}q.
	\end{aligned}
	\label{eq:energy-asymp2}
\end{equation}

First, we find the asymptotic expression for the free energy $\tilde{\mathcal{F}}$. 
Using \eqref{eq:den-asymp} to write $\overline{f}=\frac{f_0}{m_0}\left(1+\gamma^2\tilde{f}_1+\gamma^4\tilde{f}_2\right)$, we have the expansion for the density function amplitude
\begin{equation}
	\log\overline{f}=\log \frac{f_{0}}{m_0}+\gamma^{2}\tilde{f}_{1}+\gamma^{4}\left(\tilde{f}_{2}-\frac{1}{2}\tilde{f}_{1}^{2}\right),
\end{equation}
where $\tilde{f}_1=f_1/f_0=\frac{1}{2c}q^T\nabla u q$ and $\tilde{f}_2=f_2/f_0=\frac{1}{8c^2}(q^T\nabla u q)^2$ according to \eqref{eq:special_f0} and \eqref{eq:special_f1}. This leads to the free energy expansion
\begin{align*}
    &\tilde{\mathcal{F}}=\gamma^{2} \times \\
    &\int_{\Omega_{q}}f_{0}\left\{ \log\frac{f_{0}}{m_{0}}+\gamma^{2}\left(1+\log\frac{f_{0}}{m_{0}}\right)\tilde{f}_{1}+\gamma^{4}\left[\frac{1}{2}\tilde{f}_{1}^{2}+\left(1+\log\frac{f_{0}}{m_{0}}\right)\tilde{f}_{2}\right]+O\left(\gamma^6\right)\right\} \frac{e^{-U/\gamma^{2}}}{m_{0}}\mathrm{d}q.
\end{align*}
Notice that the normalization of the density function requires 
\begin{equation*}
    \int f_{0}\left(1+\gamma^{2}\tilde{f}_{1}+\gamma^{4}\tilde{f}_{2}\right)\frac{e^{-U/\gamma^{2}}}{m_{0}}\mathrm{d}q\mathrm{d}x=1
\end{equation*}
and $\int f_0\mathrm{d}x=\mathrm{const}.$ Therefore, we can remove the constant terms and the microscopic free energy has the following  asymptotic representation
\begin{equation}
\begin{aligned}
\mathcal{F}&=\int_{\Omega_{x}}\tilde{\mathcal{F}}\mathrm{d}x\\
&=\gamma^{2}\int_{\Omega_{x}}\left[f_{0}\log f_{0}+\frac{\gamma^{2}}{2c}\frac{m_{1}}{m_{0}}f_{0}\log\frac{f_{0}}{m_{0}}+\frac{\gamma^{4}}{8c^{2}}\frac{m_{2}}{m_{0}}f_{0}\left(1+\log\frac{f_{0}}{m_{0}}\right)\right]\mathrm{d}x +O\left(\gamma^8\right),\label{eq:free_ene_asymp}
\end{aligned}
\end{equation}
where the notations for the moments in \eqref{eq:moments} are used. 
In particular, it can be shown that the leading-order term in the free energy \eqref{eq:free_ene_asymp} has no contribution in the total energy equation for the incompressible flow case. In fact, by direct computation using the continuity equation \eqref{eq:asymp2} for $f_0$ we have
\begin{equation}
	\begin{aligned}\frac{\mathrm{d}}{\mathrm{d}t}\int_{\Omega_{x}}f_{0}\log f_{0}\mathrm{d}x & =\int_{\Omega_{x}}\partial_{t}f_{0}\left(\log f_{0}+1\right)\mathrm{d}x\\
		& =-\int_{\Omega_{x}}\nabla\cdot\left(uf_{0}\right)\left(\log f_{0}+1\right)\mathrm{d}x\\
		& =\int_{\Omega_{x}}\tilde{f}_{0}u\cdot\nabla\log f_{0}\mathrm{d}x\\
		& =\int_{\Omega_{x}}u\cdot\nabla f_{0}\mathrm{d}x=-\int_{\Omega_{x}}\nabla\cdot uf_{0}\mathrm{d}x=0.
	\end{aligned}
\end{equation}

Next, we derive the  asymptotic expansion for the dissipation energy in \eqref{eq:energy-asymp2}. In a similar way, we find the leading-order terms according to \eqref{eq:den-asymp} and the scaling law $D\gamma^4=c$
\[
\tilde{\mathcal{D}}_2=c\int_{\Omega_{q}}f_{0}\left[\gamma^{4}\left|\nabla_{q}\tilde{f}_{1}\right|^{2}+\gamma^{6}\left(2\nabla_{q}\tilde{f}_{1}\cdot\nabla_{q}\tilde{f}_{2}-\tilde{f}_{1}\left|\nabla_{q}\tilde{f}_{1}\right|^{2}\right)+O\left(\gamma^8\right)\right]e^{-U/\gamma^{2}}\mathrm{d}q.
\]
According to \eqref{eq:special_f0} and \eqref{eq:special_f1}, we can find the explicit forms for the gradients of $\tilde{f}_1$ and $\tilde{f}_2$ as
\[
\nabla_{q}\tilde{f}_{1}=\frac{1}{2c}\left(\nabla u+\nabla u^T\right)q,\quad\nabla_{q}\tilde{f}_{2}=\frac{1}{c}\tilde{f}_{1}\nabla uq=\frac{1}{4c^{2}}\left(q^{T}\nabla uq\right)\left(\nabla u+\nabla u^T\right)q.
\]
This leads to the asymptotic expression for the microscopic dissipation energy
\begin{equation}
    \begin{aligned}\mathcal{D}_{2}=\int_{\Omega_{x}}\tilde{\mathcal{D}}_{2}\mathrm{d}x
    =&\int_{\Omega_{x}\times\Omega_{q}}f_0\left[\frac{\gamma^{4}}{4c}\left|\left(\nabla u+\nabla u^{T}\right)q\right|^2\right.\\
    &\left.+\frac{\gamma^{6}}{8c^{2}}\left(q^{T}\nabla uq\right)\left|\left(\nabla u+\nabla u^{T}\right)q\right|^{2}+O\left(\gamma^8\right)\right]\frac{e^{-U/\gamma^{2}}}{m_{0}}\mathrm{d}q\mathrm{d}x.
    \end{aligned}\label{eq:dissp_ene_asymp}
\end{equation}
For convenience, we also introduce the moments with respect to the equilibrium microscopic distribution
Above, we introduce the normalized moments with respect to the equilibrium microscopic distribution
\[
\begin{aligned}F_{1}\left(x,t\right)= & \frac{1}{m_{0}}\int_{\Omega_{q}}q^{T}\nabla uqe^{-\frac{U}{\gamma^{2}}}\mathrm{d}q=\sum_{i,j}\frac{\partial u_{i}}{\partial x_{j}}\tilde{M}_{2}^{ij},\\
	F_{2}\left(x,t\right)= & \frac{1}{m_{0}}\int_{\Omega_{q}}\left(q^{T}\nabla uq\right)^{2}e^{-\frac{U}{\gamma^{2}}}\mathrm{d}q=\sum_{i,j,k,l}\frac{\partial u_{i}}{\partial x_{j}}\frac{\partial u_{k}}{\partial x_{l}}\tilde{M}_{4}^{ijkl}.\\
	D_{1}\left(x,t\right)= & \frac{1}{m_{0}}\int_{\Omega_{q}}\left|\left(\nabla u+\nabla u^{T}\right)q\right|^{2}e^{-\frac{U}{\gamma^{2}}}\mathrm{d}q\\
    =&\sum_{m,i,j}\left(\frac{\partial u_{m}}{\partial x_{i}}+\frac{\partial u_{i}}{\partial x_{m}}\right)\left(\frac{\partial u_{m}}{\partial x_{j}}+\frac{\partial u_{j}}{\partial x_{m}}\right)\tilde{M}_{2}^{ij},\\
	D_{2}\left(x,t\right)= & \frac{1}{m_{0}}\int_{\Omega_{q}}\left(q^{T}\nabla uq\right)\left|\left(\nabla u+\nabla u^{T}\right)q\right|^{2}e^{-\frac{U}{\gamma^{2}}}\mathrm{d}q\\
    =&\sum_{m,i,j,k,l}\left(\frac{\partial u_{m}}{\partial x_{i}}+\frac{\partial u_{i}}{\partial x_{m}}\right)\left(\frac{\partial u_{m}}{\partial x_{j}}+\frac{\partial u_{j}}{\partial x_{m}}\right)\frac{\partial u_{k}}{\partial x_{l}}\tilde{M}_{4}^{ijkl},
\end{aligned}
\]
where the second and fourth-moments $\tilde{M}_2$ and $\tilde{M}_4$ are defined in the same way as in \eqref{eq:closure_mom}.


\bibliographystyle{plain}
\bibliography{refs-arxiv}

\end{document}